\documentclass[prd,superscriptaddress,nofootinbib,secnumarabic,preprintnumbers]{revtex4-2}[10pt]
\pdfoutput=1
\usepackage{latexsym}
\usepackage{mathrsfs, amsmath, amssymb}
\usepackage{diagbox, slashed, makecell}
\usepackage{hyperref, url}
\usepackage{color}
\usepackage{graphicx}
\usepackage{appendix}
\usepackage{multirow}
\usepackage{ucs}
\usepackage{indentfirst, layout, geometry}

\newcommand{\mr}{\mathrm}
\newcommand{\mc}{\mathcal}

\graphicspath{{figures/}}

\begin{document}

\begin{sloppypar}

\title{Leptoquark and vector-like quark extended models as the explanation of the muon $g-2$ anomaly}

\preprint{APCTP Pre2021-035}

\author{Shi-Ping He}
\email{shiping.he@apctp.org}
\affiliation{Asia Pacific Center for Theoretical Physics, Pohang 37673, Korea}

\date{\today}

\begin{abstract}
Leptoquark (LQ) models can be the solution to the $(g-2)_{\mu}$ anomaly because of the chiral enhancements. In this work, we consider the models extended by the LQ and vector-like quark (VLQ) simultaneously, which can be used to explain the anomaly because of the top quark and top partner (denoted as $T$) mixing. For the one LQ and one VLQ extended models, there are contributions from the top and $T$ quark. In the minimal LQ models, only the $R_2$ and $S_1$ representations can lead to the chiral enhancements. Here, we find one new $S_3$ solution to the anomaly in the presence of $(X,T,B)_{L,R}$ triplet. We also consider the one LQ and two VLQ extended models, which can solve the anomaly even in the absence of top interactions and mixings. Then, we propose new LQ search channels under the constraints of $(g-2)_{\mu}$. Besides the traditional $t\mu$ decay channel, the LQ can also decay into $T\mu$ final states, which will lead to the characteristic multi-top and multi-muon signals at hadron colliders.
\end{abstract}

\maketitle
%%%%%%%%%%%%%%%%%%%%%%%%%%%%%%%%%%%%%%%%%%%%%%%%%%%%%%%%%%%%%%%%%%%%%
%%%%%%%%%%%%%%%%%%%%%%%%%%%%%%%%%%%%%%%%%%%%%%%%%%%%%%%%%%%%%%%%%%%%%

%\tableofcontents
\clearpage

%%%%%%%%%%%%%%%%%%%%%%%%%%%%%%%%%%%%%%%%%%%%%%%%%%%%%%%%%%%%%%%%%%%%%
\section{Introduction}
The lepton magnetic moments play an important role in the developments of the standard model (SM) of elementary particle physics. One interesting topic of them is the muon anomalous magnetic dipole moment, which is labelled as $a_{\mu}\equiv(g-2)_{\mu}/2$. Its deviation from the SM prediction can give us some clues to the new physics \cite{Jegerlehner:2009ry, Jegerlehner:2017gek}. The so-called $(g-2)_{\mu}$ anomaly is first reported by the E821 experiment at BNL \cite{Muong-2:2006rrc}. Recently,  the FNAL muon $g-2$ experiment increases the tension between the SM prediction and experiment data \cite{Muong-2:2021ojo}. The average experimental result is $a_{\mu}^{\mr{Exp}}=116592061(41)\times10^{-11}$ after combining the BNL and FNAL data. The SM prediction is $a_{\mu}^{\mr{SM}}=116591810(43)\times10^{-11}$ \cite{Aoyama:2012wk, Aoyama:2019ryr, Czarnecki:2002nt, Gnendiger:2013pva, Davier:2017zfy, Keshavarzi:2018mgv, Colangelo:2018mtw, Hoferichter:2019mqg, Davier:2019can, Keshavarzi:2019abf, Kurz:2014wya, Melnikov:2003xd, Masjuan:2017tvw, Colangelo:2017fiz, Hoferichter:2018kwz, Gerardin:2019vio, Bijnens:2019ghy, Colangelo:2019uex, Blum:2019ugy, Colangelo:2014qya, Aoyama:2020ynm}, which results in the $4.2\sigma$ discrepancy with $\Delta a_{\mu}\equiv a_{\mu}^{\mr{Exp}}-a_{\mu}^{\mr{SM}}=(251\pm59)\times10^{-11}$. This excess may be pinned down with more data at FNAL and J-PARC \cite{Iinuma:2011zz}.

Although this anomaly can be caused by the experimental and theoretical uncertainties \cite{Cowan:2021sdy}, we are more interested in the new physics interpretations. Many new physics models are motivated to explain this anomaly \cite{Czarnecki:2001pv, Jegerlehner:2009ry, Queiroz:2014zfa, Lindner:2016bgg, Athron:2021iuf}, for example, supersymmetric models \cite{Moroi:1995yh, Martin:2001st, Heinemeyer:2004yq, Stockinger:2006zn, Yin:2021yqy}, scalar mediator models \cite{Ilisie:2015tra, Cherchiglia:2016eui, Davoudiasl:2018fbb, Sabatta:2019nfg, Jana:2020pxx, Botella:2020xzf, Barman:2021xeq}, new chiral lepton models \cite{Raby:2017igl, Crivellin:2021rbq}, vector-like lepton (VLL) models \cite{Kannike:2011ng, Freitas:2014pua, Kannike:2011ng, Dermisek:2013gta, Crivellin:2018qmi, Crivellin:2018qmi, Frank:2020smf, Dermisek:2021ajd, CarcamoHernandez:2019ydc, Navarro:2021sfb, Ko:2021lpx}, and so on. The LQs are well motivated in the grand unified theories \cite{Pati:1974yy, Georgi:1974sy, Fritzsch:1974nn}, and it can also be the alternative choice for the $(g-2)_{\mu}$ anomaly \cite{Buchmuller:1986zs, Djouadi:1989md, Dorsner:2016wpm, Crivellin:2020tsz, Dorsner:2019itg, Dorsner:2020aaz, Zhang:2021dgl} because of the significant chiral enhancement $m_t/m_\mu$. Although it can also be used to explain the anomalies in $B$ physics \cite{Dorsner:2016wpm, ColuccioLeskow:2016dox, Calibbi:2017qbu, Saad:2020ihm, Babu:2020hun, Marzocca:2021azj}, we will not discuss those here. On the other hand, vector-like fermions will occur in many new physics models \cite{Panico:2015jxa, Schmaltz:2005ky, Hewett:1988xc, Randall:1999ee}. However, most studies focus on the VLL extended models, and the VLQ models usually draw little attention. In this work, we will consider the simultaneous scalar LQ and VLQ extended models.
%%%%%%%%%%%%%%%%%%%%%%%%%%%%%%%%%%%%%%%%%%%%%%%%%%%%%%%%%%%%%%%%%%%%%
\section{Results in the LQ extended models}
\subsection{The minimal LQ models}
First of all, let us consider the minimal extension of SM with only one scalar LQ \cite{Buchmuller:1986zs, Chakraverty:2001yg, Cheung:2001ip, Queiroz:2014zfa, Dorsner:2016wpm}. There is a fermion number $F=3B+L$ for the LQ interactions, in which $B$ is the baryon number and $L$ is the lepton number. In fact, there are two types of interactions $\overline{\psi_R}\psi_L$ and $\overline{(\psi_L)^C}\psi_L$ depending on fermion number carried by the LQ. We name the first type as "$F=0$" (or "A") and the second type as "$F=2$" (or "B"). Then, the Lagrangian can be written as
\begin{align}\label{eqn:LQ:Lag}
&\mc{L}_{F=0}\supset\bar{\mu}(y_L^{S_A\mu q_A}\omega_-+y_R^{S_A\mu q_A}\omega_+)q_AS_A+\mathrm{h.c.}~,\nonumber\\
&\mc{L}_{F=2}\supset\bar{\mu}(y_L^{S_B\mu q_B}\omega_-+y_R^{S_B\mu q_B}\omega_+)q_B^CS_B+\mathrm{h.c.}.
\end{align}
In the above, $q_B^C$ is charge conjugate of the $q_B$ quark, which means $q_B^C=C\overline{q_B}^T$ with $C\equiv i\gamma^2\gamma^0$. \footnote{There can be different definitions for the $C$ operator, but this has no influence on the following discussions.} The chiral operators $\omega_{\pm}$ are defined as $(1\pm\gamma^5)/2$. In Tab. \ref{tab:LQ:couplings}, we list different scalar LQs and the coupling patterns between muon and up-type quark \cite{Dorsner:2016wpm}.
\begin{table}[!htb]
\begin{center}
\begin{tabular}{c|c|c|c|c}
\hline
\makecell{$SU(3)_C\times SU(2)_L\times U(1)_Y$ \\ representation} & label & $F$ & $y_L^{\mr{LQ}\mu t}$ & $y_R^{\mr{LQ}\mu t}$ \\
\hline
$(\bar{3},3,1/3)$ & $S_3$ & $-2$ & 0 & $\mc{O}(y)$ \\
\hline
$(3,2,7/6)$ & $R_2$ & 0 & $\mc{O}(y)$ & $\mc{O}(y)$ \\
\hline
$(3,2,1/6)$ & $\widetilde{R}_2$ & 0 & 0 & 0 \\
\hline
$(\bar{3},1,4/3)$ & $\widetilde{S}_1$ & $-2$ & 0 & 0 \\
\hline
$(\bar{3},1,1/3)$ & $S_1$ & $-2$ & $\mc{O}(y)$ & $\mc{O}(y)$ \\
\hline
\end{tabular}%}
\caption{Coupling patterns of the Yukawa couplings between muon and top quark in the minimal LQ models for different representations.} \label{tab:LQ:couplings}
\end{center}
\end{table}

\begin{figure}
\begin{center}
\includegraphics[scale=0.5]{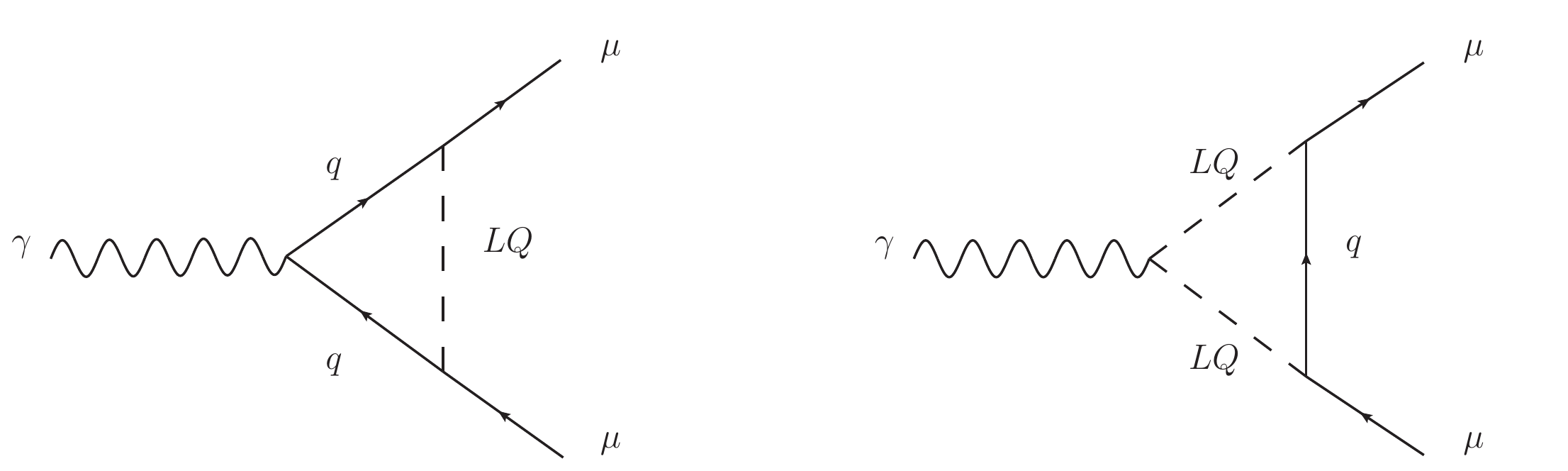}
\caption{The typical Feynman diagrams contributing to the $(g-2)_\mu$.}\label{fig:sim:FeynS}
\end{center}
\end{figure}
When involving the LQ, we need to consider two types of Feynman diagrams in Fig. \ref{fig:sim:FeynS} at the same time. The contributions to $(g-2)_\mu$ are slightly different for the $F=0$ and $F=2$ type interactions \cite{Cheung:2001ip}. The one-loop results have been calculated in the previous studies \cite{Leveille:1977rc, Dorsner:2016wpm}, and note that there should be a color factor $N_C=3$. Here, we consider the muon coupled to the LQ and top quark. Thus, $m_\mu\ll m_q,m_S$ is a good approximation. For convenience, we define the following four integrals:
\begin{align}
f_{LL}^q(x)=\frac{2+3x-6x^2+x^3+6x\log x}{12(1-x)^4},\nonumber\\
f_{LR}^q(x)=-\frac{3-4x+x^2+2\log x}{4(1-x)^3},\nonumber\\
f_{LL}^S(x)=-\frac{1-6x+3x^2+2x^3-6x^2\log x}{12(1-x)^4},\nonumber\\
f_{LR}^S(x)=-\frac{1-x^2+2x\log x}{4(1-x)^3}.\nonumber\\
\end{align}
For $F=0$ or A type interactions, the contributions to $(g-2)_\mu$ are given as
\begin{align}\label{eqn:g-2:LQA}
&\Delta a_{\mu}^{F=0}=-\frac{N_Cm_{\mu}^2}{8\pi^2m_{S_A}^2}\Bigg[(|y_L^{S_A\mu q_A}|^2+|y_R^{S_A\mu q_A}|^2)\big(Q_{q_A}f_{LL}^q(x_A)+Q_{S_A}f_{LL}^S(x_A)\big)\nonumber\\
	&+\frac{m_{q_A}}{m_{\mu}}\left(y_L^{S_A\mu q_A}(y_R^{S_A\mu q_A})^\ast+y_R^{S_A\mu q_A}(y_L^{S_A\mu q_A})^\ast\right)\big(Q_{q_A}f_{LR}^q(x_A)+Q_{S_A}f_{LR}^S(x_A)\big)\Bigg],
\end{align}
where $x_A$ is defined as $m_{q_A}^2/m_{S_A}^2$, $Q_{q_A}$ is the electric charge of $q_A$ quark and $Q_{S_A}$ is the electric charge of scalar $S_A$ in Eq. \eqref{eqn:LQ:Lag}. For $F=2$ or B type interactions, we need to perform the transformations $y_{L,R}^{S_A\mu q_A}\rightarrow y_{L,R}^{S_B\mu q_B},Q_{q_A}\rightarrow-Q_{q_B},Q_{S_A}\rightarrow Q_{S_B},m_{q_A}\rightarrow m_{q_B},x_A\rightarrow x_B$ in Eq. \eqref{eqn:g-2:LQA}. Because the quark field is $q_B^C$ in the type B interactions, it should be $Q_{q_B^C}=-Q_{q_B}$.\footnote{Here, we only change the sign of quark electric charge from the $F=0$ to $F=2$ case. But, it says that we need to change both the signs of quark electric charge and mass in Kingman's paper \cite{Cheung:2001ip}. We think the sign of quark mass should stay the same, because the mass term $\bar{\psi}\psi$ is invariant under the $C$ transformation. The sign of electric charge should be changed, because the electro-magnetic current term $\bar{\psi}\gamma^{\mu}\psi$ changes the sign under the $C$ transformation.} More specifically, the contributions to $(g-2)_\mu$ are given as
\begin{align}\label{eqn:g-2:LQB}
&\Delta a_{\mu}^{F=2}=-\frac{N_Cm_{\mu}^2}{8\pi^2m_{S_B}^2}\Bigg[(|y_L^{S_B\mu q_B}|^2+|y_R^{S_B\mu q_B}|^2)\big(-Q_{q_B}f_{LL}^q(x_B)+Q_{S_B}f_{LL}^S(x_B)\big)\nonumber\\
	&+\frac{m_{q_B}}{m_{\mu}}\left(y_L^{S_B\mu q_B}(y_R^{S_B\mu q_B})^\ast+y_R^{S_B\mu q_B}(y_L^{S_B\mu q_B})^\ast\right)\big(-Q_{q_B}f_{LR}^q(x_B)+Q_{S_B}f_{LR}^S(x_B)\big)\Bigg],
\end{align}
where $x_B$ is defined as $m_{q_B}^2/m_{S_B}^2$.

As we can see in Tab. \ref{tab:LQ:couplings}, only the $R_2,S_1$ representations can give both the left-handed and right-handed (non-chiral) couplings to muons, which means the chiral enhancement is possible. The $R_2$ induces the $F=0$ type interactions, and $S_1$ induces the $F=2$ type interactions. Now, let us make the conventions. We use $Q_L^i,u_R^i,d_R^i,L_L^i,e_R^i$ to label the SM quark and lepton fields, where $i$ means the generation index and ranges from 1 to 3. Especially, $Q_L^3,u_R^3,d_R^3,L_L^2,e_R^2$ are also marked as $t_L,t_R,b_L,b_R,\mu_L,\mu_R$. For the doublet and following triplet fields, the labels $a,b,c$ are the weak isospin indices. The $\tau^i(i=1,2,3)$ are the Pauli matrices, and $\epsilon$ is defined as $i\tau^2$. Then, we have the following gauge eigenstate interactions \cite{Dorsner:2016wpm}
\begin{align}
&\mc{L}_{R_2}\supset-x_{ij}\overline{u_R}^iR_2^a\epsilon^{ab}L_L^{j,b}+y_{ij}\overline{e_R}^i(R_2^a)^\ast Q_L^{j,a}+\mathrm{h.c.}~,\nonumber\\
&\mc{L}_{S_1}\supset v_{ij}\overline{(Q_L)^C}^{i,a}S_1\epsilon^{ab}L_L^{j,b}+x_{ij}\overline{(u_R)^C}^iS_1e_R^j+\mathrm{h.c.}.
\end{align}
In the mass eigenstates, they can be written as
\begin{align}
&\mc{L}_{R_2}\supset\bar{\mu}(y_L^{R_2\mu t}\omega_-+y_R^{R_2\mu t}\omega_+)t(R_2^{5/3})^\ast+y_L^{R_2\mu t}\bar{\mu}~\omega_-~b(R_2^{2/3})^\ast+\mathrm{h.c.}~,\nonumber\\
&\mc{L}_{S_1}\supset\bar{\mu}(y_L^{S_1\mu t}\omega_-+y_R^{S_1\mu t}\omega_+)t^C(S_1)^\ast+\mathrm{h.c.}.
\end{align}
Before obtaining the explicit results of $R_2$ and $S_1$ contributions, let us define the following four integrals related to the $\mr{LQ}\mu t(T)$ interactions:
\begin{align}\label{eqn:LQ:R2S1}
&f_{LL}^{R_2}(x)\equiv-2f_{LL}^q(x)+5f_{LL}^S(x)=-\frac{3-8x+x^2+4x^3+2x(2-5x)\log x}{4(1-x)^4},\nonumber\\
&f_{LR}^{R_2}(x)\equiv-2f_{LR}^q(x)+5f_{LR}^S(x)=\frac{1-8x+7x^2+(4-10x)\log x}{4(1-x)^3},\nonumber\\
&f_{LL}^{S_1}(x)\equiv2f_{LL}^q(x)+f_{LL}^S(x)=\frac{1+4x-5x^2+2x(2+x)\log x}{4(1-x)^4},\nonumber\\
&f_{LR}^{S_1}(x)\equiv2f_{LR}^q(x)+f_{LR}^S(x)=-\frac{7-8x+x^2+(4+2x)\log x}{4(1-x)^3}.
\end{align}
For the $S_3$ interactions in the following context, we have the results $f_{LL}^{S_3}(x)=f_{LL}^{S_1}(x)$ and $f_{LR}^{S_3}(x)=f_{LR}^{S_1}(x)$ because of the similar interaction types. Besides, the $\mr{LQ}\mu b(B)$ interactions will be involved for some cases later. Thus, we also define the following four integrals in advance:
\begin{align}\label{eqn:LQ:R2S3b}
&\widetilde{f}_{LL}^{R_2}(x)\equiv f_{LL}^q(x)+2f_{LL}^S(x)=\frac{x[5-4x-x^2+(2+4x)\log x]}{4(1-x)^4},\nonumber\\
&\widetilde{f}_{LR}^{R_2}(x)\equiv f_{LR}^q(x)+2f_{LR}^S(x)=-\frac{5-4x-x^2+(2+4x)\log x}{4(1-x)^3},\nonumber\\
&\widetilde{f}_{LL}^{S_3}(x)\equiv-f_{LL}^q(x)+4f_{LL}^S(x)=-\frac{2-7x+2x^2+3x^3+2x(1-4x)\log x}{4(1-x)^4},\nonumber\\
&\widetilde{f}_{LR}^{S_3}(x)\equiv-f_{LR}^q(x)+4f_{LR}^S(x)=-\frac{1+4x-5x^2-(2-8x)\log x}{4(1-x)^3}.
\end{align}
In the above, the functions $\widetilde{f}^{\mr{LQ}}(x)$ are related with bottom quark type interactions, while the previous functions $f^{\mr{LQ}}(x)$ are for the top quark type interactions.

By setting $Q_{q_A}=Q_t=2/3$ and $Q_{S_A}=Q_{(R_2^{5/3})^\ast}=-5/3$ in Eq. \eqref{eqn:g-2:LQA}, we have the following results for the $R_2$ case:
\begin{align}\label{eqn:LQmut:R2}
&\Delta a_{\mu}^{R_2}=\frac{m_{\mu}^2}{8\pi^2m_{R_2}^2}\Bigg[(|y_L^{R_2\mu t}|^2+|y_R^{R_2\mu t}|^2)f_{LL}^{R_2}(m_t^2/m_{R_2}^2)+|y_L^{R_2\mu t}|^2\widetilde{f}_{LL}^{R_2}(m_b^2/m_{R_2}^2)\nonumber\\
&+\frac{m_t}{m_{\mu}}\left(y_L^{R_2\mu t}(y_R^{R_2\mu t})^\ast+y_R^{R_2\mu t}(y_L^{R_2\mu t})^\ast\right)f_{LR}^{R_2}(m_t^2/m_{R_2}^2)\Bigg].
\end{align}
Considering $m_t\ll m_{R_2}$, it can be approximated as
\begin{align}
&\Delta a_{\mu}^{R_2}\approx\frac{m_{\mu}^2}{32\pi^2m_{R_2}^2}\Bigg[-3(|y_L^{R_2\mu t}|^2+|y_R^{R_2\mu t}|^2)+\frac{m_t}{m_{\mu}}(1+4\log \frac{m_t^2}{m_{R_2}^2})\left(y_L^{R_2\mu t}(y_R^{R_2\mu t})^\ast+y_R^{R_2\mu t}(y_L^{R_2\mu t})^\ast\right)\Bigg].
\end{align}
By setting $Q_{q_B}=Q_t=2/3$ and $Q_{S_B}=Q_{(S_1)^\ast}=-1/3$ in Eq. \eqref{eqn:g-2:LQB}, we have the following results for the $S_1$ case:
\begin{align}\label{eqn:LQmut:S1}
&\Delta a_{\mu}^{S_1}=\frac{m_{\mu}^2}{8\pi^2m_{S_1}^2}\Bigg[(|y_L^{S_1\mu t}|^2+|y_R^{S_1\mu t}|^2)f_{LL}^{S_1}(m_t^2/m_{S_1}^2)+\frac{m_t}{m_{\mu}}\left(y_L^{S_1\mu t}(y_R^{S_1\mu t})^\ast+y_R^{S_1\mu t}(y_L^{S_1\mu t})^\ast\right)f_{LR}^{S_1}(m_t^2/m_{S_1}^2)\Bigg].
\end{align}
Considering $m_t\ll m_{S_1}$, it can be approximated as
\begin{align}
&\Delta a_{\mu}^{S_1}\approx\frac{m_{\mu}^2}{32\pi^2m_{S_1}^2}\Bigg[(|y_L^{S_1\mu t}|^2+|y_R^{S_1\mu t}|^2)-\frac{m_t}{m_{\mu}}(7+4\log \frac{m_t^2}{m_{S_1}^2})\left(y_L^{S_1\mu t}(y_R^{S_1\mu t})^\ast+y_R^{S_1\mu t}(y_L^{S_1\mu t})^\ast\right)\Bigg].
\end{align}
Considering the electric conservation condition and the notations, our results agree with those in Refs. \cite{Cheung:2001ip, Dorsner:2016wpm, Aebischer:2021uvt}. For LQ with mass above TeV, we have the rough estimation $m_{\mu}^2/(8\pi^2m_{\mr{LQ}}^2)\lesssim10^{-10}$. Because the LQ Yukawa couplings are bounded by the perturbative unitarity, the chiral enhancements are required to explain the $(g-2)_{\mu}$.

\subsection{The one LQ and one VLQ extended models}
There are totally seven type VLQs, while only five of them contain the top partner $T$ \cite{Aguilar-Saavedra:2013qpa}. We will consider the five cases: one singlet $T_{L,R}$, two doublets $(X,T)_{L,R},(T,B)_{L,R}$, and two triplets $(X,T,B)_{L,R},(T,B,Y)_{L,R}$. Here, $X,T,B,Y$ carry the $5/3,2/3,-1/3,-4/3$ electric charge, respectively. For the five scalar type LQs and five type interesting VLQs, it can be totally twenty-five combinations if we consider one LQ and one VLQ at the same time, which will be dubbed as "$\mr{LQ}+\mr{VLQ}$" in the following context. After some attempts, we find that only the combinations with specific LQ representations can lead to the up-type quark chiral enhancements. Then, we will study these combinations.

\begin{enumerate}
\item For the $T_{L,R}$ singlet case, the LQ should be $R_2$ or $S_1$.
\begin{itemize}
\item For the $R_2+T_{L,R}$ model, there are following gauge eigenstate interactions:
\begin{align}
\mc{L}_{R_2+T_{L,R}}\supset-x_{ij}\overline{u_R}^iR_2^a\epsilon^{ab}L_L^{j,b}-x_{Tj}\overline{T_R}R_2^a\epsilon^{ab}L_L^{j,b}+y_{ij}\overline{e_R}^i(R_2^a)^\ast Q_L^{j,a}+\mathrm{h.c.}.
\end{align}
\item For the $S_1+T_{L,R}$ model, there are following gauge eigenstate interactions:
\begin{align}
\mc{L}_{S_1+T_{L,R}}\supset v_{ij}\overline{(Q_L)^C}^{i,a}S_1\epsilon^{ab}L_L^{j,b}+x_{ij}\overline{(u_R)^C}^iS_1e_R^j+x_{Tj}\overline{(T_R)^C}S_1e_R^j+\mathrm{h.c.}.
\end{align}
\end{itemize}

\item For the $(X,T)_{L,R}$ doublet case, the LQ should be $R_2$ or $S_1$.
\begin{itemize}
\item For the $R_2+(X,T)_{L,R}$ model, there are following gauge eigenstate interactions:
\begin{align}
\mc{L}_{R_2+(X,T)_{L,R}}\supset-x_{ij}\overline{u_R}^iR_2^a\epsilon^{ab}L_L^{j,b}+y_{ij}\overline{e_R}^i(R_2^a)^\ast Q_L^{j,a}+\mathrm{h.c.}.
\end{align}

\item For the $S_1+(X,T)_{L,R}$ model, there are following gauge eigenstate interactions:
\begin{align}
\mc{L}_{S_1+(X,T)_{L,R}}\supset v_{ij}\overline{(Q_L)^C}^{i,a}S_1\epsilon^{ab}L_L^{j,b}+x_{ij}\overline{(u_R)^C}^iS_1e_R^j+\mathrm{h.c.}.
\end{align}
\end{itemize}

\item For the $(T,B)_{L,R}$ doublet case, the LQ should be $R_2$ or $S_1$.
\begin{itemize}
\item For the $R_2+(T,B)_{L,R}$ model, there are following gauge eigenstate interactions:
\begin{align}
\mc{L}_{R_2+(T,B)_{L,R}}\supset-x_{ij}\overline{u_R}^iR_2^a\epsilon^{ab}L_L^{j,b}+y_{ij}\overline{e_R}^i(R_2^a)^\ast Q_L^{j,a}+y_{iT}\overline{e_R}^i(R_2^a)^\ast \left(\begin{array}{c}T_L\\B_L\end{array}\right)^a+\mathrm{h.c.}.
\end{align}
\item For the $S_1+(T,B)_{L,R}$ model, there are following gauge eigenstate interactions:
\begin{align}
\mc{L}_{S_1+(T,B)_{L,R}}\supset v_{ij}\overline{(Q_L)^C}^{i,a}S_1\epsilon^{ab}L_L^{j,b}+v_{Tj}\left(\overline{(T_L)^C},\overline{(B_L)^C}\right)^aS_1\epsilon^{ab}L_L^{j,b}+x_{ij}\overline{(u_R)^C}^iS_1e_R^j+\mathrm{h.c.}.
\end{align}
\end{itemize}

\item For the $(X,T,B)_{L,R}$ triplet case, the LQ can be $R_2$, $S_1$ or $S_3$.
\begin{itemize}
\item For the $R_2+(X,T,B)_{L,R}$ model, there are following gauge eigenstate interactions:
\begin{align}
\mc{L}_{R_2+(X,T,B)_{L,R}}\supset-x_{ij}\overline{u_R}^iR_2^a\epsilon^{ab}L_L^{j,b}-x_{iT}\overline{L_L}^{i,a}(\Psi_R)^{ab}\epsilon^{bc}(R_2^c)^{\ast}+y_{ij}\overline{e_R}^i(R_2^a)^\ast Q_L^{j,a}+\mathrm{h.c.}.
\end{align}
\item For the $S_1+(X,T,B)_{L,R}$ model, there are following gauge eigenstate interactions:
\begin{align}
\mc{L}_{S_1+(X,T,B)_{L,R}}\supset v_{ij}\overline{(Q_L)^C}^{i,a}S_1\epsilon^{ab}L_L^{j,b}+x_{ij}\overline{(u_R)^C}^iS_1e_R^j+\mathrm{h.c.}.
\end{align}
\item For the $S_3+(X,T,B)_{L,R}$ model, there are following gauge eigenstate interactions:
\begin{align}
\mc{L}_{S_3+(X,T,B)_{L,R}}\supset x_{ij}\overline{(Q_L)^C}^{i,a}\epsilon^{ab}(S_3)^{bc}L_L^{j,c}+y_{Tj}\mr{Tr}[\overline{(\Psi_R)^C}S_3]e_R^j+\mathrm{h.c.}.
\end{align}
In the above, we define the $(X,T,B)_{L,R}$ and $S_3$ triplets in the following forms:
\begin{align}
\Psi_{L,R}\equiv\Big(\begin{array}{cc}T_{L,R}&\sqrt{2}X_{L,R}\\ \sqrt{2}B_{L,R}&-T_{L,R}\end{array}\Big),\qquad S_3\equiv\Big(\begin{array}{cc}S_3^{1/3}&\sqrt{2}S_3^{4/3}\\ \sqrt{2}S_3^{-2/3}&-S_3^{1/3}\end{array}\Big).
\end{align}
Then, the matrix elements $\big(\overline{(\Psi_R)^C}\big)_{ij}$ are defined as $\overline{\big((\Psi_R)_{ij}\big)^C}$. 
\end{itemize}

\item For the $(T,B,Y)_{L,R}$ triplet case, the LQ should be $R_2$ or $S_1$.
\begin{itemize}
\item For the $R_2+(T,B,Y)_{L,R}$ model, there are following gauge eigenstate interactions:
\begin{align}
\mc{L}_{R_2+(T,B,Y)_{L,R}}\supset-x_{ij}\overline{u_R}^iR_2^a\epsilon^{ab}L_L^{j,b}+y_{ij}\overline{e_R}^i(R_2^a)^\ast Q_L^{j,a}+\mathrm{h.c.}.
\end{align}
\item For the $S_1+(T,B,Y)_{L,R}$ model, there are following gauge eigenstate interactions:
\begin{align}
\mc{L}_{S_1+(T,B,Y)_{L,R}}\supset v_{ij}\overline{(Q_L)^C}^{i,a}S_1\epsilon^{ab}L_L^{j,b}+x_{ij}\overline{(u_R)^C}^iS_1e_R^j+\mathrm{h.c.}.
\end{align}
\end{itemize}

\end{enumerate}

After the electro-weak symmetry breaking (EWSB), we can obtain the $\mr{LQ}\mu q$ interactions. Then, we give the explicit forms in the following.
\begin{itemize}
\item For the $R_2+T_{L,R}$ model, they can be parametrized as
\begin{align}
\mc{L}_{R_2+T_{L,R}}\supset\bar{\mu}(y_L^{R_2\mu t}\omega_-+y_R^{R_2\mu t}\omega_+)t(R_2^{5/3})^\ast+y_R^{R_2\mu T}\bar{\mu}~\omega_+~T(R_2^{5/3})^\ast+y_L^{R_2\mu t}\bar{\mu}~\omega_-~b(R_2^{2/3})^\ast+\mathrm{h.c.}.
\end{align}
\item For the $S_1+T_{L,R}$ model, they can be parametrized as
\begin{align}
\mc{L}_{S_1+T_{L,R}}\supset\bar{\mu}(y_L^{S_1\mu t}\omega_-+y_R^{S_1\mu t}\omega_+)t^C(S_1)^\ast+y_L^{S_1\mu T}\bar{\mu}~\omega_-~T^C(S_1)^\ast+\mathrm{h.c.}.
\end{align}
\item For the $R_2+(X,T)_{L,R}$ model, they can be parametrized as
\begin{align}
\mc{L}_{R_2+(X,T)_{L,R}}\supset\bar{\mu}(y_L^{R_2\mu t}\omega_-+y_R^{R_2\mu t}\omega_+)t(R_2^{5/3})^\ast+y_L^{R_2\mu t}\bar{\mu}~\omega_-~b(R_2^{2/3})^\ast+\mathrm{h.c.}.
\end{align}
\item For the $S_1+(X,T)_{L,R}$ model, they can be parametrized as
\begin{align}
\mc{L}_{S_1+(X,T)_{L,R}}\supset\bar{\mu}(y_L^{S_1\mu t}\omega_-+y_R^{S_1\mu t}\omega_+)t^C(S_1)^\ast+\mathrm{h.c.}.
\end{align}
\item For the $R_2+(T,B)_{L,R}$ model, they can be parametrized as
\begin{align}
&\mc{L}_{R_2+(T,B)_{L,R}}\supset\bar{\mu}(y_L^{R_2\mu t}\omega_-+y_R^{R_2\mu t}\omega_+)t(R_2^{5/3})^\ast+y_L^{R_2\mu T}\bar{\mu}~\omega_-~T(R_2^{5/3})^\ast\nonumber\\
&+y_L^{R_2\mu t}\bar{\mu}~\omega_-~b(R_2^{2/3})^\ast+y_L^{R_2\mu T}\bar{\mu}~\omega_-~B(R_2^{2/3})^\ast+\mathrm{h.c.}.
\end{align}
\item For the $S_1+(T,B)_{L,R}$ model, they can be parametrized as
\begin{align}
\mc{L}_{S_1+(T,B)_{L,R}}\supset\bar{\mu}(y_L^{S_1\mu t}\omega_-+y_R^{S_1\mu t}\omega_+)t^C(S_1)^\ast+y_R^{S_1\mu T}\bar{\mu}~\omega_+~T^C(S_1)^\ast+\mathrm{h.c.}.
\end{align}
\item For the $R_2+(X,T,B)_{L,R}$ model, they can be parametrized as
\begin{align}
&\mc{L}_{R_2+(X,T,B)_{L,R}}\supset\bar{\mu}(y_L^{R_2\mu t}\omega_-+y_R^{R_2\mu t}\omega_+)t(R_2^{5/3})^\ast+y_R^{R_2\mu T}\bar{\mu}~\omega_+~T(R_2^{5/3})^\ast\nonumber\\
&+y_L^{R_2\mu t}\bar{\mu}~\omega_-~b(R_2^{2/3})^\ast+\sqrt{2}y_R^{R_2\mu T}\bar{\mu}~\omega_+~B(R_2^{2/3})^\ast+\mathrm{h.c.}.
\end{align}
\item For the $S_1+(X,T,B)_{L,R}$ model, they can be parametrized as
\begin{align}
\mc{L}_{S_1+(X,T,B)_{L,R}}\supset\bar{\mu}(y_L^{S_1\mu t}\omega_-+y_R^{S_1\mu t}\omega_+)t^C(S_1)^\ast+\mathrm{h.c.}.
\end{align}
\item For the $S_3+(X,T,B)_{L,R}$ model, they can be parametrized as
\begin{align}
&\mc{L}_{S_3+(X,T,B)_{L,R}}\supset y_L^{S_3\mu T}\bar{\mu}~\omega_-~T^C(S_3^{1/3})^\ast+y_R^{S_3\mu t}\bar{\mu}~\omega_+~t^C(S_3^{1/3})^\ast\nonumber\\
&+y_L^{S_3\mu T}\bar{\mu}~\omega_-~B^C(S_3^{4/3})^\ast+\sqrt{2}y_R^{S_3\mu t}\bar{\mu}~\omega_+~b^C(S_3^{4/3})^\ast+y_L^{S_3\mu T}\bar{\mu}~\omega_-~X^C(S_3^{-2/3})^\ast+\mathrm{h.c.}.
\end{align}
\item For the $R_2+(T,B,Y)_{L,R}$ model, they can be parametrized as
\begin{align}
&\mc{L}_{R_2+(T,B,Y)_{L,R}}\supset\bar{\mu}(y_L^{R_2\mu t}\omega_-+y_R^{R_2\mu t}\omega_+)t(R_2^{5/3})^\ast+y_L^{R_2\mu t}\bar{\mu}~\omega_-~b(R_2^{2/3})^\ast+\mathrm{h.c.}.
\end{align}
\item For the $S_1+(T,B,Y)_{L,R}$ model, they can be parametrized as
\begin{align}
\mc{L}_{S_1+(T,B,Y)_{L,R}}\supset\bar{\mu}(y_L^{S_1\mu t}\omega_-+y_R^{S_1\mu t}\omega_+)t^C(S_1)^\ast+\mathrm{h.c.}.
\end{align}
\end{itemize}

Now, let us consider the mixing between top quark and $T$ quark. Based on the Refs. \cite{Aguilar-Saavedra:2013qpa, He:2020suf, He:2020fqj}, the $t$ and $T$ quarks can be rotated into mass eigenstates by the following transformations:
\begin{align}\label{eqn:quarktT:rotation}
\left[\begin{array}{c}t_L\\T_L\end{array}\right]\rightarrow
	\left[\begin{array}{cc}\cos\theta_L&\sin\theta_L\\-\sin\theta_L&\cos\theta_L\end{array}\right]
	\left[\begin{array}{c}t_L\\T_L\end{array}\right],\quad
%%%%%%%%%%%%%%%%%%%%%%%%%%%%%%%%%%%%%%%%%%%%%%%%%%%%%%%%%%%%%%%%%%%%%%%%
\left[\begin{array}{c}t_R\\T_R\end{array}\right]\rightarrow
	\left[\begin{array}{cc}\cos\theta_R&\sin\theta_R\\-\sin\theta_R&\cos\theta_R\end{array}\right]
	\left[\begin{array}{c}t_R\\T_R\end{array}\right].
\end{align}
Hereafter, $\sin\theta_L,\cos\theta_L,\sin\theta_R,\cos\theta_R$ will be abbreviated as $s_L,c_L,s_R,c_R$ (also $s_L^t,c_L^t,s_R^t,c_R^t$), respectively. If the VLQ contains the $B$ component, things are a little complicated. For the $R_2+(T,B)_{L,R}/(X,T,B)_{L,R}/(T,B,Y)_{L,R}$ and $S_3+(X,T,B)_{L,R}$ models, there are also $b$ and $B$ quark mixings. Similarly, their transformations can be written as
\begin{align}\label{eqn:quarkbB:rotation}
\left[\begin{array}{c}b_L\\B_L\end{array}\right]\rightarrow
	\left[\begin{array}{cc}c_L^b&s_L^b\\-s_L^b&c_L^b\end{array}\right]
	\left[\begin{array}{c}b_L\\B_L\end{array}\right],\quad
%%%%%%%%%%%%%%%%%%%%%%%%%%%%%%%%%%%%%%%%%%%%%%%%%%%%%%%%%%%%%%%%%%%%%%%%
\left[\begin{array}{c}b_R\\B_R\end{array}\right]\rightarrow
	\left[\begin{array}{cc}c_R^b&s_R^b\\-s_R^b&c_R^b\end{array}\right]
	\left[\begin{array}{c}b_R\\B_R\end{array}\right].
\end{align}
In the above, the superscript ``$b$" is added to distinguish from the top quark case. For the $S_1+(T,B)_{L,R}/(X,T,B)_{L,R}/(T,B,Y)_{L,R}$ models, it is not necessary to perform the rotations for the $b$ and $B$ quarks. Although the involved $B$ quark can mixing with the $b$ quark, it does not affect the $(g-2)_{\mu}$ because of the absence of $S_1\mu b(B)$ type interactions.

In fact, the left-handed and right-handed field mixing angles can be related through the following relations \cite{Aguilar-Saavedra:2013qpa}:
\begin{align}
&\tan\theta_L^{t(b)}=\frac{m_{t(b)}}{m_{T(B)}}\tan\theta_R^{t(b)}, \qquad\mr{for~doublet~VLQs},\nonumber\\
&\tan\theta_R^{t(b)}=\frac{m_{t(b)}}{m_{T(B)}}\tan\theta_L^{t(b)}, \qquad\mr{for~singlet~and~triplet~VLQs}.
\end{align}
Besides, there are two independent mixing angles $\theta_R^t$ and $\theta_R^b$ for the $(T,B)_{L,R}$ extended models. While the mixing angle $\theta_L^b$ is not free in the $(X,T,B)_{L,R}$ and $(T,B,Y)_{L,R}$ extended models, because we have the following identities \footnote{For the $(T,B,Y)_{L,R}$, there is a sign difference from the relation in Ref. \cite{Aguilar-Saavedra:2013qpa}. The reason is that we adopt the similar parametrization convention as the $(X,T,B)_{L,R}$ case, which can be removed through redefinition of gauge eigenstate $T$ quark. While, this notation difference does not lead to physical difference.}:
\begin{align}
\sin2\theta_L^b=\frac{\sqrt{2}(m_T^2-m_t^2)}{m_B^2-m_b^2}\sin2\theta_L^t, \qquad\mr{for~}(X,T,B)_{L,R},\nonumber\\
\sin2\theta_L^b=-\frac{m_T^2-m_t^2}{\sqrt{2}(m_B^2-m_b^2)}\sin2\theta_L^t, \qquad\mr{for~}(T,B,Y)_{L,R}.
\end{align}
In the limit of $m_{t(b)}\ll m_T=m_B$ and $\theta_L^{t(b)}\ll1$, they will lead to the simple approximations $\theta_L^b\approx\sqrt{2}\theta_L^t$ for the $(X,T,B)_{L,R}$ and $\theta_L^b\approx-\theta_L^t/\sqrt{2}$ for the $(T,B,Y)_{L,R}$.

Then, we collect the mass eigenstate interactions in the following.
\begin{itemize}

\item For the $R_2+T_{L,R}$ model, they can be written as
\begin{align}
&\mc{L}_{R_2+T_{L,R}}\supset\bar{\mu}[y_L^{R_2\mu t}c_L\omega_-+(y_R^{R_2\mu t}c_R-y_R^{R_2\mu T}s_R)\omega_+]t(R_2^{5/3})^\ast\nonumber\\
	&+\bar{\mu}[y_L^{R_2\mu t}s_L\omega_-+(y_R^{R_2\mu t}s_R+y_R^{R_2\mu T}c_R)\omega_+]T(R_2^{5/3})^\ast+y_L^{R_2\mu t}\bar{\mu}~\omega_-~b(R_2^{2/3})^\ast+\mathrm{h.c.}.
\end{align}
\item For the $S_1+T_{L,R}$ model, they can be written as
\begin{align}
&\mc{L}_{S_1+T_{L,R}}\supset\bar{\mu}[(y_L^{S_1\mu t}c_R-y_L^{S_1\mu T}s_R)\omega_-+y_R^{S_1\mu t}c_L\omega_+]t^C(S_1)^\ast\nonumber\\
	&+\bar{\mu}[(y_L^{S_1\mu t}s_R+y_L^{S_1\mu T}c_R)\omega_-+y_R^{S_1\mu t}s_L\omega_+]T^C(S_1)^\ast+\mathrm{h.c.}.
\end{align}

\item For the $R_2+(X,T)_{L,R}$ model, they can be written as
\begin{align}
&\mc{L}_{R_2+(X,T)_{L,R}}\supset\bar{\mu}(y_L^{R_2\mu t}c_L\omega_-+y_R^{R_2\mu t}c_R\omega_+)t(R_2^{5/3})^\ast+\bar{\mu}(y_L^{R_2\mu t}s_L\omega_-+y_R^{R_2\mu t}s_R\omega_+)T(R_2^{5/3})^\ast\nonumber\\
&+y_L^{R_2\mu t}\bar{\mu}~\omega_-~b(R_2^{2/3})^\ast+\mathrm{h.c.}.
\end{align}
\item For the $S_1+(X,T)_{L,R}$ model, they can be written as
\begin{align}
&\mc{L}_{S_1+(X,T)_{L,R}}\supset\bar{\mu}(y_L^{S_1\mu t}c_R\omega_-+y_R^{S_1\mu t}c_L\omega_+)t^C(S_1)^\ast+\bar{\mu}(y_L^{S_1\mu t}s_R\omega_-+y_R^{S_1\mu t}s_L\omega_+)T^C(S_1)^\ast+\mathrm{h.c.}.
\end{align}

\item For the $R_2+(T,B)_{L,R}$ model, they can be written as
\begin{align}
&\mc{L}_{R_2+(T,B)_{L,R}}\supset\bar{\mu}[(y_L^{R_2\mu t}c_L^t-y_L^{R_2\mu T}s_L^t)\omega_-+y_R^{R_2\mu t}c_R^t\omega_+]t(R_2^{5/3})^\ast\nonumber\\
	&+\bar{\mu}[(y_L^{R_2\mu t}s_L^t+y_L^{R_2\mu T}c_L^t)\omega_-+y_R^{R_2\mu t}s_R^t\omega_+]T(R_2^{5/3})^\ast\nonumber\\
&+(y_L^{R_2\mu t}c_L^b-y_L^{R_2\mu T}s_L^b)\bar{\mu}~\omega_-~b(R_2^{2/3})^\ast+(y_L^{R_2\mu T}c_L^b+y_L^{R_2\mu t}s_L^b)\bar{\mu}~\omega_-~B(R_2^{2/3})^\ast+\mathrm{h.c.}.
\end{align}
\item For the $S_1+(T,B)_{L,R}$ model, they can be written as
\begin{align}
&\mc{L}_{S_1+(T,B)_{L,R}}\supset\bar{\mu}[y_L^{S_1\mu t}c_R\omega_-+(y_R^{S_1\mu t}c_L-y_R^{S_1\mu T}s_L)\omega_+]t^C(S_1)^\ast\nonumber\\
	&+\bar{\mu}[y_L^{S_1\mu t}s_R\omega_-+(y_R^{S_1\mu t}s_L+y_R^{S_1\mu T}c_L)\omega_+]T^C(S_1)^\ast+\mathrm{h.c.}.
\end{align}

\item For the $R_2+(X,T,B)_{L,R}$ model, they can be written as
\begin{align}
&\mc{L}_{R_2+(X,T,B)_{L,R}}\supset\bar{\mu}[y_L^{R_2\mu t}c_L^t\omega_-+(y_R^{R_2\mu t}c_R^t-y_R^{R_2\mu T}s_R^t)\omega_+]t(R_2^{5/3})^\ast\nonumber\\
	&+\bar{\mu}[y_L^{R_2\mu t}s_L^t\omega_-+(y_R^{R_2\mu T}c_R^t+y_R^{R_2\mu t}s_R^t)\omega_+]T(R_2^{5/3})^\ast\nonumber\\
&+\bar{\mu}[y_L^{R_2\mu t}c_L^b\omega_--\sqrt{2}y_R^{R_2\mu T}s_R^b\omega_+]b(R_2^{2/3})^\ast+\bar{\mu}[y_L^{R_2\mu t}s_L^b\omega_-+\sqrt{2}y_R^{R_2\mu T}c_R^b\omega_+]B(R_2^{2/3})^\ast+\mathrm{h.c.}.
\end{align}
\item For the $S_1+(X,T,B)_{L,R}$ model, they can be written as
\begin{align}
&\mc{L}_{S_1+(X,T,B)_{L,R}}\supset\bar{\mu}(y_L^{S_1\mu t}c_R\omega_-+y_R^{S_1\mu t}c_L\omega_+)t^C(S_1)^\ast+\bar{\mu}(y_L^{S_1\mu t}s_R\omega_-+y_R^{S_1\mu t}s_L\omega_+)T^C(S_1)^\ast+\mathrm{h.c.}.
\end{align}
\item For the $S_3+(X,T,B)_{L,R}$ model, they can be written as
\begin{align}
&\mc{L}_{S_3+(X,T,B)_{L,R}}\supset\bar{\mu}(-y_L^{S_3\mu T}s_R^t\omega_-+y_R^{S_3\mu t}c_L^t\omega_+)t^C(S_3^{1/3})^\ast+\bar{\mu}(y_L^{S_3\mu T}c_R^t\omega_-+y_R^{S_3\mu t}s_L^t\omega_+)T^C(S_3^{1/3})^\ast\nonumber\\
&+\bar{\mu}(-y_L^{S_3\mu T}s_R^b\omega_-+\sqrt{2}y_R^{S_3\mu t}c_L^b\omega_+)b^C(S_3^{4/3})^\ast+\bar{\mu}(y_L^{S_3\mu T}c_R^b\omega_-+\sqrt{2}y_R^{S_3\mu t}s_L^b\omega_+)B^C(S_3^{4/3})^\ast\nonumber\\
&+y_L^{S_3\mu T}\bar{\mu}~\omega_-~X^C(S_3^{-2/3})^\ast+\mathrm{h.c.}.
\end{align}

\item For the $R_2+(T,B,Y)_{L,R}$ model, they can be written as
\begin{align}
&\mc{L}_{R_2+(T,B,Y)_{L,R}}\supset\bar{\mu}(y_L^{R_2\mu t}c_L\omega_-+y_R^{R_2\mu t}c_R\omega_+)t(R_2^{5/3})^\ast+\bar{\mu}(y_L^{R_2\mu t}s_L\omega_-+y_R^{R_2\mu t}s_R\omega_+)T(R_2^{5/3})^\ast\nonumber\\
&+y_L^{R_2\mu t}c_L^b\bar{\mu}~\omega_-~b(R_2^{2/3})^\ast+y_L^{R_2\mu t}s_L^b\bar{\mu}~\omega_-~B(R_2^{2/3})^\ast+\mathrm{h.c.}.
\end{align}
\item For the $S_1+(T,B,Y)_{L,R}$ model, they can be written as
\begin{align}
&\mc{L}_{S_1+(T,B,Y)_{L,R}}\supset\bar{\mu}(y_L^{S_1\mu t}c_R\omega_-+y_R^{S_1\mu t}c_L\omega_+)t^C(S_1)^\ast+\bar{\mu}(y_L^{S_1\mu t}s_R\omega_-+y_R^{S_1\mu t}s_L\omega_+)T^C(S_1)^\ast+\mathrm{h.c.}.
\end{align}
\end{itemize}

From the above mass eigenstate interactions, we can obtain the top and $T$ contributions to the $(g-2)_{\mu}$. For the $R_2+\mr{VLQ}$ models, we can derive the $t$ and $T$ quark contributions from Eq. \eqref{eqn:LQmut:R2}. For the $S_1/S_3+\mr{VLQ}$ models, we can also derive the $t$ and $T$ quark contributions from Eq. \eqref{eqn:LQmut:S1}. As to the $b$ and $B$ quark contributions, we need to adopt the functions $\widetilde{f}^{\mr{LQ}}(x)$ defined in Eq. \eqref{eqn:LQ:R2S3b}. The contribution from $X$ quark only occurs in the $S_3+(X,T,B)_{L,R}$ model, which is negligible because of the absence of chiral enhancement. Besides, there are no $Y$ quark contributions in the $R_2/S_1+(T,B,Y)_{L,R}$ models. In the following, we only show the chirally enhanced parts for simplicity.
\begin{itemize}

\item For the $R_2+T_{L,R}$ model, they are calculated as
\begin{align}
&\Delta a_{\mu}^{R_2+T_{L,R}}\approx\frac{m_{\mu}m_t}{4\pi^2m_{R_2}^2}\Bigg\{\frac{m_T}{m_t}f_{LR}^{R_2}(m_T^2/m_{R_2}^2)\mr{Re}[(y_R^{R_2\mu t}s_R+y_R^{R_2\mu T}c_R)(y_L^{R_2\mu t})^\ast]s_L\nonumber\\
	&+(\frac{1}{4}+\log \frac{m_t^2}{m_{R_2}^2})\mr{Re}[(y_R^{R_2\mu t}c_R-y_R^{R_2\mu T}s_R)(y_L^{R_2\mu t})^\ast]c_L\Bigg\}.
\end{align}
\item For the $S_1+T_{L,R}$ model, they are calculated as
\begin{align}
&\Delta a_{\mu}^{S_1+T_{L,R}}\approx\frac{m_{\mu}m_t}{4\pi^2m_{S_1}^2}\Bigg\{\frac{m_T}{m_t}f_{LR}^{S_1}(m_T^2/m_{S_1}^2)\mr{Re}[(y_L^{S_1\mu T}c_R+y_L^{S_1\mu t}s_R)(y_R^{S_1\mu t})^\ast]s_L\nonumber\\
	&-(\frac{7}{4}+\log \frac{m_t^2}{m_{S_1}^2})\mr{Re}[(y_L^{S_1\mu t}c_R-y_L^{S_1\mu T}s_R)(y_R^{S_1\mu t})^\ast]c_L\Bigg\}.
\end{align}

\item For the $R_2+(X,T)_{L,R}$ model, they are calculated as
\begin{align}
&\Delta a_{\mu}^{R_2+(X,T)_{L,R}}\approx\frac{m_{\mu}m_t}{4\pi^2m_{R_2}^2}\Bigg\{\frac{m_T}{m_t}f_{LR}^{R_2}(m_T^2/m_{R_2}^2)\mr{Re}[y_R^{R_2\mu t}(y_L^{R_2\mu t})^\ast]s_Ls_R\nonumber\\
	&+(\frac{1}{4}+\log \frac{m_t^2}{m_{R_2}^2})\mr{Re}[y_R^{R_2\mu t}(y_L^{R_2\mu t})^\ast]c_Lc_R\Bigg\}.
\end{align}
\item For the $S_1+(X,T)_{L,R}$ model, they are calculated as
\begin{align}
&\Delta a_{\mu}^{S_1+(X,T)_{L,R}}\approx\frac{m_{\mu}m_t}{4\pi^2m_{S_1}^2}\Bigg\{\frac{m_T}{m_t}f_{LR}^{S_1}(m_T^2/m_{S_1}^2)\mr{Re}[y_L^{S_1\mu t}(y_R^{S_1\mu t})^\ast]s_Ls_R\nonumber\\
	&-(\frac{7}{4}+\log \frac{m_t^2}{m_{S_1}^2})\mr{Re}[(y_L^{S_1\mu t}(y_R^{S_1\mu t})^\ast]c_Lc_R\Bigg\}.
\end{align}

\item For the $R_2+(T,B)_{L,R}$ model, they are calculated as
\begin{align}
&\Delta a_{\mu}^{R_2+(T,B)_{L,R}}\approx\frac{m_{\mu}m_t}{4\pi^2m_{R_2}^2}\Bigg\{\frac{m_T}{m_t}f_{LR}^{R_2}(m_T^2/m_{R_2}^2)\mr{Re}[(y_L^{R_2\mu t}s_L+y_L^{R_2\mu T}c_L)(y_R^{R_2\mu t})^\ast]s_R\nonumber\\
	&+(\frac{1}{4}+\log \frac{m_t^2}{m_{R_2}^2})\mr{Re}[(y_L^{R_2\mu t}c_L-y_L^{R_2\mu T}s_L)(y_R^{R_2\mu t})^\ast]c_R\Bigg\}.
\end{align}
\item For the $S_1+(T,B)_{L,R}$ model, they are calculated as
\begin{align}
&\Delta a_{\mu}^{S_1+(T,B)_{L,R}}\approx\frac{m_{\mu}m_t}{4\pi^2m_{S_1}^2}\Bigg\{\frac{m_T}{m_t}f_{LR}^{S_1}(m_T^2/m_{S_1}^2)\mr{Re}[(y_R^{S_1\mu T}c_L+y_R^{S_1\mu t}s_L)(y_L^{S_1\mu t})^\ast]s_R\nonumber\\
	&-(\frac{7}{4}+\log \frac{m_t^2}{m_{S_1}^2})\mr{Re}[(y_R^{S_1\mu t}c_L-y_R^{S_1\mu T}s_L)(y_L^{S_1\mu t})^\ast]c_R\Bigg\}.
\end{align}

\item For the $R_2+(X,T,B)_{L,R}$ model, they are calculated as
\begin{align}
&\Delta a_{\mu}^{R_2+(X,T,B)_{L,R}}\approx\frac{m_{\mu}m_t}{4\pi^2m_{R_2}^2}\Bigg\{\frac{m_T}{m_t}f_{LR}^{R_2}(\frac{m_T^2}{m_{R_2}^2})\mr{Re}[(y_R^{R_2\mu t}s_R^t+y_R^{R_2\mu T}c_R^t)(y_L^{R_2\mu t})^\ast]s_L^t\nonumber\\
	&+(\frac{1}{4}+\log\frac{m_t^2}{m_{R_2}^2})\mr{Re}[(y_R^{R_2\mu t}c_R^t-y_R^{R_2\mu T}s_R^t)(y_L^{R_2\mu t})^\ast]c_L^t\nonumber\\
&+[\frac{\sqrt{2}m_B}{m_t}\widetilde{f}_{LR}^{R_2}(\frac{m_B^2}{m_{R_2}^2})s_L^bc_R^b-\frac{\sqrt{2}m_b}{m_t}\widetilde{f}_{LR}^{R_2}(\frac{m_b^2}{m_{R_2}^2})c_L^bs_R^b]\cdot\mr{Re}[y_R^{R_2\mu T}(y_L^{R_2\mu t})^\ast]\Bigg\}.
\end{align}
\item For the $S_1+(X,T,B)_{L,R}$ model, they are calculated as
\begin{align}
&\Delta a_{\mu}^{S_1+(X,T,B)_{L,R}}\approx\frac{m_{\mu}m_t}{4\pi^2m_{S_1}^2}\Bigg\{\frac{m_T}{m_t}f_{LR}^{S_1}(m_T^2/m_{S_1}^2)\mr{Re}[y_L^{S_1\mu t}(y_R^{S_1\mu t})^\ast]s_Ls_R\nonumber\\
	&-(\frac{7}{4}+\log \frac{m_t^2}{m_{S_1}^2})\mr{Re}[y_L^{S_1\mu t}(y_R^{S_1\mu t})^\ast]c_Lc_R\Bigg\}.
\end{align}
\item For the $S_3+(X,T,B)_{L,R}$ model, they are calculated as
\begin{align}
&\Delta a_{\mu}^{S_3+(X,T,B)_{L,R}}\approx\frac{m_{\mu}m_t}{4\pi^2m_{S_3}^2}\Bigg\{\frac{m_T}{m_t}f_{LR}^{S_3}(\frac{m_T^2}{m_{S_3}^2})s_L^tc_R^t+(\frac{7}{4}+\log \frac{m_t^2}{m_{S_3}^2})c_L^ts_R^t\nonumber\\
&+\frac{\sqrt{2}m_B}{m_t}\widetilde{f}_{LR}^{S_3}(\frac{m_B^2}{m_{S_3}^2})s_L^bc_R^b-\frac{\sqrt{2}m_b}{m_t}\widetilde{f}_{LR}^{S_3}(\frac{m_b^2}{m_{S_3}^2})s_R^bc_L^b\Bigg\}\mr{Re}[y_L^{S_3\mu T}(y_R^{S_3\mu t})^\ast].
\end{align}

\item For the $R_2+(T,B,Y)_{L,R}$ model, they are calculated as
\begin{align}
&\Delta a_{\mu}^{R_2+(T,B,Y)_{L,R}}\approx\frac{m_{\mu}m_t}{4\pi^2m_{R_2}^2}\Bigg\{\frac{m_T}{m_t}f_{LR}^{R_2}(m_T^2/m_{R_2}^2)\mr{Re}[y_R^{R_2\mu t}(y_L^{R_2\mu t})^\ast]s_Ls_R\nonumber\\
	&+(\frac{1}{4}+\log \frac{m_t^2}{m_{R_2}^2})\mr{Re}[y_R^{R_2\mu t}(y_L^{R_2\mu t})^\ast]c_Lc_R\Bigg\}.
\end{align}
\item For the $S_1+(T,B,Y)_{L,R}$ model, they are calculated as
\begin{align}
&\Delta a_{\mu}^{S_1+(T,B,Y)_{L,R}}\approx\frac{m_{\mu}m_t}{4\pi^2m_{S_1}^2}\Bigg\{\frac{m_T}{m_t}f_{LR}^{S_1}(m_T^2/m_{S_1}^2)\mr{Re}[y_L^{S_1\mu t}(y_R^{S_1\mu t})^\ast]s_Ls_R\nonumber\\
	&-(\frac{7}{4}+\log \frac{m_t^2}{m_{S_1}^2})\mr{Re}[y_L^{S_1\mu t}(y_R^{S_1\mu t})^\ast]c_Lc_R\Bigg\}.
\end{align}

\end{itemize}

Now, let us make a summary for the one LQ and one VLQ models. In Tab. \ref{tab:LQ+VLQ:inputs}, we list the related input parameters and mixing angle identities in the previously mentioned one LQ and one VLQ models. For the singlet and triplet VLQs, we choose $\theta_L$ as the input mixing angle. For the doublet VLQs, we choose $\theta_R$ as the input mixing angle \cite{Aguilar-Saavedra:2013qpa}. In Tab. \ref{tab:LQ+VLQ:LQmut}, we list the coupling patterns of $\mr{LQ}\mu t$ and $\mr{LQ}\mu T$ interactions. Besides, we also show the coupling patterns of $\mr{LQ}\mu b$ and $\mr{LQ}\mu B$ interactions in Tab. \ref{tab:LQ+VLQ:LQmub}.
\begin{table}[!htb]
\begin{center}
\begin{tabular}{c|c|c|c}
\hline
LQ & VLQ & related input parameters & mixing angle relation \\
\hline
\multirow{5}{*}{$R_2$} & $T_{L,R}$ & $y_L^{R_2\mu t},~y_R^{R_2\mu t},~y_R^{R_2\mu T},~\theta_L,~m_T$ & $\tan\theta_R=\frac{m_t}{m_T}\tan\theta_L$ \\
\cline{2-4}
& $(X,T)_{L,R}$ & $y_L^{R_2\mu t},~y_R^{R_2\mu t},~\theta_R,~m_T$ & $\tan\theta_L=\frac{m_t}{m_T}\tan\theta_R$ \\
\cline{2-4}
& $(T,B)_{L,R}$ & $y_L^{R_2\mu t},~y_R^{R_2\mu t},~y_L^{R_2\mu T},~\theta_R^t,~\theta_R^b,~m_T$ & $\tan\theta_L^{t(b)}=\frac{m_{t(b)}}{m_{T(B)}}\tan\theta_R^{t(b)}$ \\
\cline{2-4}
\rule{0pt}{15pt} & $(X,T,B)_{L,R}$ & $y_L^{R_2\mu t},~y_R^{R_2\mu t},~y_R^{R_2\mu T},~\theta_L,~m_T$ & $\tan\theta_R^{t(b)}=\frac{m_{t(b)}}{m_{T(B)}}\tan\theta_L^{t(b)},~\sin2\theta_L^b=\frac{\sqrt{2}(m_T^2-m_t^2)}{m_B^2-m_b^2}\sin2\theta_L^t$ \\
\cline{2-4}
\rule{0pt}{15pt} & $(T,B,Y)_{L,R}$ & $y_L^{R_2\mu t},~y_R^{R_2\mu t},~\theta_L,~m_T$ & $\tan\theta_R^{t(b)}=\frac{m_{t(b)}}{m_{T(B)}}\tan\theta_L^{t(b)},~\sin2\theta_L^b=-\frac{m_T^2-m_t^2}{\sqrt{2}(m_B^2-m_b^2)}\sin2\theta_L^t$ \\
\hline\hline
\multirow{5}{*}{$S_1$} & $T_{L,R}$ & $y_L^{S_1\mu t},~y_R^{S_1\mu t},~y_L^{S_1\mu T},~\theta_L,~m_T$ & $\tan\theta_R=\frac{m_t}{m_T}\tan\theta_L$ \\
\cline{2-4}
& $(X,T)_{L,R}$ & $y_L^{S_1\mu t},~y_R^{S_1\mu t},~\theta_R,~m_T$ & $\tan\theta_L=\frac{m_t}{m_T}\tan\theta_R$ \\
\cline{2-4}
& $(T,B)_{L,R}$ & $y_L^{S_1\mu t},~y_R^{S_1\mu t},~y_R^{S_1\mu T},~\theta_R^t,~\theta_R^b,~m_T$ & $\tan\theta_L^{t(b)}=\frac{m_{t(b)}}{m_{T(B)}}\tan\theta_R^{t(b)}$ \\
\cline{2-4}
\rule{0pt}{15pt} & $(X,T,B)_{L,R}$ & $y_L^{S_1\mu t},~y_R^{S_1\mu t},~\theta_L,~m_T$ & $\tan\theta_R^{t(b)}=\frac{m_{t(b)}}{m_{T(B)}}\tan\theta_L^{t(b)},~\sin2\theta_L^b=\frac{\sqrt{2}(m_T^2-m_t^2)}{m_B^2-m_b^2}\sin2\theta_L^t$ \\
\cline{2-4}
\rule{0pt}{15pt} & $(T,B,Y)_{L,R}$ & $y_L^{S_1\mu t},~y_R^{S_1\mu t},~\theta_L,~m_T$ & $\tan\theta_R^{t(b)}=\frac{m_{t(b)}}{m_{T(B)}}\tan\theta_L^{t(b)},~\sin2\theta_L^b=-\frac{m_T^2-m_t^2}{\sqrt{2}(m_B^2-m_b^2)}\sin2\theta_L^t$ \\
\hline
\rule{0pt}{15pt} $S_3$ & $(X,T,B)_{L,R}$ & $y_L^{S_3\mu T},~y_R^{S_3\mu t},~\theta_L,~m_T$ & $\tan\theta_R^{t(b)}=\frac{m_{t(b)}}{m_{T(B)}}\tan\theta_L^{t(b)},~\sin2\theta_L^b=\frac{\sqrt{2}(m_T^2-m_t^2)}{m_B^2-m_b^2}\sin2\theta_L^t$ \\ \hline
\end{tabular}
\caption{The input parameters and mixing angle relations in the LQ+VLQ models for different representations. In the above, the $\theta_{L,R}$ means $\theta_{L,R}^t$ by default.} \label{tab:LQ+VLQ:inputs} 
\end{center}
\end{table}

\begin{table}[!htb]\small
\begin{center}
%\resizebox{168mm}{18mm}{
\begin{tabular}{c|c|c|c|c|c}
\hline
LQ & VLQ & $\overline{\mu_R}t_L$ & $\overline{\mu_L}t_R$ & $\overline{\mu_R}T_L$ & $\overline{\mu_L}T_R$ \\
\hline
\multirow{5}{*}{$R_2$} & $T_{L,R}$ & $y_L^{R_2\mu t}c_L$ & $y_R^{R_2\mu t}c_R-y_R^{R_2\mu T}s_R$ & $y_L^{R_2\mu t}s_L$ & $y_R^{R_2\mu t}s_R+y_R^{R_2\mu T}c_R$ \\
\cline{2-6}
& $(X,T)_{L,R}$ & $y_L^{R_2\mu t}c_L$ & $y_R^{R_2\mu t}c_R$ & $y_L^{R_2\mu t}s_L$ & $y_R^{R_2\mu t}s_R$ \\
\cline{2-6}
& $(T,B)_{L,R}$ & $y_L^{R_2\mu t}c_L-y_L^{R_2\mu T}s_L$ & $y_R^{R_2\mu t}c_R$ & $y_L^{R_2\mu t}s_L+y_L^{R_2\mu T}c_L$ & $y_R^{R_2\mu t}s_R$ \\
\cline{2-6}
& $(X,T,B)_{L,R}$ & $y_L^{R_2\mu t}c_L$ & $y_R^{R_2\mu t}c_R-y_R^{R_2\mu T}s_R$ & $y_L^{R_2\mu t}s_L$ & $y_R^{R_2\mu t}s_R+y_R^{R_2\mu T}c_R$ \\
\cline{2-6}
& $(T,B,Y)_{L,R}$ & $y_L^{R_2\mu t}c_L$ & $y_R^{R_2\mu t}c_R$ & $y_L^{R_2\mu t}s_L$ & $y_R^{R_2\mu t}s_R$ \\
\hline\hline
LQ & VLQ & $\overline{\mu_R}(t_R)^C$ & $\overline{\mu_L}(t_L)^C$ & $\overline{\mu_R}(T_R)^C$ & $\overline{\mu_L}(T_L)^C$\\
\hline
\multirow{5}{*}{$S_1$} & $T_{L,R}$ & $y_L^{S_1\mu t}c_R-y_L^{S_1\mu T}s_R$ & $y_R^{S_1\mu t}c_L$ & $y_L^{S_1\mu t}s_R+y_L^{S_1\mu T}c_R$ & $y_R^{S_1\mu t}s_L$ \\
\cline{2-6}
& $(X,T)_{L,R}$ & $y_L^{S_1\mu t}c_R$ & $y_R^{S_1\mu t}c_L$ & $y_L^{S_1\mu t}s_R$ & $y_R^{S_1\mu t}s_L$ \\
\cline{2-6}
& $(T,B)_{L,R}$ & $y_L^{S_1\mu t}c_R$ & $y_R^{S_1\mu t}c_L-y_R^{S_1\mu T}s_L$ & $y_L^{S_1\mu t}s_R$ & $y_R^{S_1\mu t}s_L+y_R^{S_1\mu T}c_L$ \\
\cline{2-6}
& $(X,T,B)_{L,R}$ & $y_L^{S_1\mu t}c_R$ & $y_R^{S_1\mu t}c_L$ & $y_L^{S_1\mu t}s_R$ & $y_R^{S_1\mu t}s_L$ \\
\cline{2-6}
& $(T,B,Y)_{L,R}$ & $y_L^{S_1\mu t}c_R$ & $y_R^{S_1\mu t}c_L$ & $y_L^{S_1\mu t}s_R$ & $y_R^{S_1\mu t}s_L$ \\
\hline
$S_3$ & $(X,T,B)_{L,R}$ & $-y_L^{S_3\mu T}s_R$ & $y_R^{S_3\mu t}c_L$ & $y_L^{S_3\mu T}c_R$ & $y_R^{S_3\mu t}s_L$ \\ \hline
\end{tabular}%}
\caption{Coupling patterns of the Yukawa couplings between muon and $t/T$ quarks in the LQ+VLQ models for different representations.}\label{tab:LQ+VLQ:LQmut}
\end{center}
\end{table}

\begin{table}[!htb]\small
\begin{center}
%\resizebox{168mm}{18mm}{
\begin{tabular}{c|c|c|c|c|c}
\hline
LQ & VLQ & $\overline{\mu_R}b_L$ & $\overline{\mu_L}b_R$ & $\overline{\mu_R}B_L$ & $\overline{\mu_L}B_R$ \\
\hline
\multirow{5}{*}{$R_2$} & $T_{L,R}$ & $y_L^{R_2\mu t}$ & 0 & $\times$ & $\times$ \\
\cline{2-6}
& $(X,T)_{L,R}$ & $y_L^{R_2\mu t}$ & 0 & $\times$ & $\times$ \\
\cline{2-6}
& $(T,B)_{L,R}$ & $y_L^{R_2\mu t}c_L^b-y_L^{R_2\mu T}s_L^b$ & 0 & $y_L^{R_2\mu T}c_L^b+y_L^{R_2\mu t}s_L^b$ & 0 \\
\cline{2-6}
& $(X,T,B)_{L,R}$ & $y_L^{R_2\mu t}c_L^b$ & $-\sqrt{2}y_R^{R_2\mu T}s_R^b$ & $y_L^{R_2\mu t}s_L^b$ & $\sqrt{2}y_R^{R_2\mu T}c_R^b$ \\
\cline{2-6}
& $(T,B,Y)_{L,R}$ & $y_L^{R_2\mu t}c_L^b$ & 0 & $y_L^{R_2\mu t}s_L^b$ & 0 \\
\hline\hline
LQ & VLQ & $\overline{\mu_R}(b_R)^C$ & $\overline{\mu_L}(b_L)^C$ & $\overline{\mu_R}(B_R)^C$ & $\overline{\mu_L}(B_L)^C$\\
\hline
\multirow{5}{*}{$S_1$} & $T_{L,R}$ & 0 & 0 & $\times$ & $\times$ \\
\cline{2-6}
& $(X,T)_{L,R}$ & 0 & 0 & $\times$ & $\times$ \\
\cline{2-6}
& $(T,B)_{L,R}$ & 0 & 0 & 0 & 0 \\
\cline{2-6}
& $(X,T,B)_{L,R}$ & 0 & 0 & 0 & 0 \\
\cline{2-6}
& $(T,B,Y)_{L,R}$ & 0 & 0 & 0 & 0 \\
\hline
$S_3$ & $(X,T,B)_{L,R}$ & $-y_L^{S_3\mu T}s_R^b$ & $\sqrt{2}y_R^{S_3\mu t}c_L^b$ & $y_L^{S_3\mu T}c_R^b$ & $\sqrt{2}y_R^{S_3\mu t}s_L^b$ \\ \hline
\end{tabular}%}
\caption{Coupling patterns of the Yukawa couplings between muon and $b/B$ quarks in the LQ+VLQ models for different representations. The symbol ``$\times$" means no such interactions.}\label{tab:LQ+VLQ:LQmub}
\end{center}
\end{table}

In Tab. \ref{tab:LQ+VLQ:g-2}, we show the approximate formulae of $\Delta a_{\mu}$ in different LQ+VLQ models. Assuming $y_{L,R}^{\mr{LQ}\mu t(T)}$ to be the same order, $m_b\ll m_t\ll m_T\approx m_B$, and $s_{L,R}\ll1$, we also show the ratio order of the product of left-handed and right-handed $\mr{LQ}\mu T$ couplings to the product of top quark ones. From the behaviours in Tab. \ref{tab:LQ+VLQ:g-2}, we find that the $T$ contributions are strongly suppressed by the factor $m_ts_{L,R}^2/m_T$ for the $R_2+(X,T)_{L,R}/(T,B,Y)_{L,R},~S_1+(X,T)_{L,R}/(X,T,B)_{L,R}/(T,B,Y)_{L,R}$ models. For the $R_2+T_{L,R}/(T,B)_{L,R}/(X,T,B)_{L,R}$ and $S_1+T_{L,R}/(T,B)_{L,R}$ models, the $T$ contributions are suppressed by the factor $s_{L,R}$. While, the $T$ and $B$ quark contributions are dominated for the $S_3+(X,T,B)_{L,R}$ model because of the $s_L/s_R\approx m_T/m_t$ factor. As $s_L$ and $s_R$ go to zeros, there will be no chiral enhancements from the $T$ quark in all the one LQ and one VLQ extended models. Here, we just capture the leading behaviors. We do not consider all the subleading corrections for simplicity, but we can include the full contributions in the following numerical calculations.
\begin{table}[!htb]\footnotesize
\begin{center}
%\resizebox{168mm}{18mm}{
\begin{tabular}{>{\scriptsize}p{0.25cm}<{\centering}|>{\scriptsize}p{1.5cm}<{\centering}|c|c}
\hline
LQ & VLQ & the approximate expressions of $\Delta\overline{a_{\mu}}$ & \makecell{coupling product order \\ of $T$ compared to $t$}\\
\hline
\multirow{5}{*}{$R_2$} & $T_{L,R}$ & $\frac{m_T}{m_t}f_{LR}^{R_2}(m_T^2/m_{R_2}^2)\mr{Re}[y_R^{R_2\mu T}(y_L^{R_2\mu t})^\ast]s_L+(\frac{1}{4}+\log \frac{m_t^2}{m_{R_2}^2})\mr{Re}[y_R^{R_2\mu t}(y_L^{R_2\mu t})^\ast]$ & $s_L$ \\
\cline{2-4}
& $(X,T)_{L,R}$ & $\frac{m_T}{m_t}f_{LR}^{R_2}(m_T^2/m_{R_2}^2)\mr{Re}[y_R^{R_2\mu t}(y_L^{R_2\mu t})^\ast]s_Ls_R+(\frac{1}{4}+\log \frac{m_t^2}{m_{R_2}^2})\mr{Re}[y_R^{R_2\mu t}(y_L^{R_2\mu t})^\ast]$ & $m_ts_R^2/m_T$\\
\cline{2-4}
& $(T,B)_{L,R}$ & $\frac{m_T}{m_t}f_{LR}^{R_2}(m_T^2/m_{R_2}^2)\mr{Re}[y_R^{R_2\mu t}(y_L^{R_2\mu T})^\ast]s_R+(\frac{1}{4}+\log \frac{m_t^2}{m_{R_2}^2})\mr{Re}[y_R^{R_2\mu t}(y_L^{R_2\mu t})^\ast]$ & $s_R$\\
\cline{2-4}
\rule{0pt}{20pt}& $(X,T,B)_{L,R}$ & \makecell{$\frac{m_T}{m_t}[f_{LR}^{R_2}(m_T^2/m_{R_2}^2)+2\widetilde{f}_{LR}^{R_2}(m_T^2/m_{R_2}^2)]\cdot\mr{Re}[y_R^{R_2\mu T}(y_L^{R_2\mu t})^\ast]s_L$\\$+(\frac{1}{4}+\log \frac{m_t^2}{m_{R_2}^2})\mr{Re}[y_R^{R_2\mu t}(y_L^{R_2\mu t})^\ast]$} & $s_L$\\
\cline{2-4}
& $(T,B,Y)_{L,R}$ & $\frac{m_T}{m_t}f_{LR}^{R_2}(m_T^2/m_{R_2}^2)\mr{Re}[y_R^{R_2\mu t}(y_L^{R_2\mu t})^\ast]s_Ls_R+(\frac{1}{4}+\log \frac{m_t^2}{m_{R_2}^2})\mr{Re}[y_R^{R_2\mu t}(y_L^{R_2\mu t})^\ast]$ & $m_ts_L^2/m_T$\\
\hline\hline
\multirow{5}{*}{$S_1$} & $T_{L,R}$ & $\frac{m_T}{m_t}f_{LR}^{S_1}(m_T^2/m_{S_1}^2)\mr{Re}[y_L^{S_1\mu T}(y_R^{S_1\mu t})^\ast]s_L-(\frac{7}{4}+\log \frac{m_t^2}{m_{S_1}^2})\mr{Re}[y_L^{S_1\mu t}(y_R^{S_1\mu t})^\ast]$ & $s_L$\\
\cline{2-4}
& $(X,T)_{L,R}$ & $\frac{m_T}{m_t}f_{LR}^{S_1}(m_T^2/m_{S_1}^2)\mr{Re}[y_L^{S_1\mu t}(y_R^{S_1\mu t})^\ast]s_Ls_R-(\frac{7}{4}+\log \frac{m_t^2}{m_{S_1}^2})\mr{Re}[y_L^{S_1\mu t}(y_R^{S_1\mu t})^\ast]$ & $m_ts_R^2/m_T$\\
\cline{2-4}
& $(T,B)_{L,R}$ & $\frac{m_T}{m_t}f_{LR}^{S_1}(m_T^2/m_{S_1}^2)\mr{Re}[y_L^{S_1\mu t}(y_R^{S_1\mu T})^\ast]s_R-(\frac{7}{4}+\log \frac{m_t^2}{m_{S_1}^2})\mr{Re}[y_L^{S_1\mu t}(y_R^{S_1\mu t})^\ast]$ & $s_R$\\
\cline{2-4}
& $(X,T,B)_{L,R}$ & $\frac{m_T}{m_t}f_{LR}^{S_1}(m_T^2/m_{S_1}^2)\mr{Re}[y_L^{S_1\mu t}(y_R^{S_1\mu t})^\ast]s_Ls_R-(\frac{7}{4}+\log \frac{m_t^2}{m_{S_1}^2})\mr{Re}[y_L^{S_1\mu t}(y_R^{S_1\mu t})^\ast]$ & $m_ts_L^2/m_T$\\
\cline{2-4}
& $(T,B,Y)_{L,R}$ & $\frac{m_T}{m_t}f_{LR}^{S_1}(m_T^2/m_{S_1}^2)\mr{Re}[y_L^{S_1\mu t}(y_R^{S_1\mu t})^\ast]s_Ls_R-(\frac{7}{4}+\log \frac{m_t^2}{m_{S_1}^2})\mr{Re}[y_L^{S_1\mu t}(y_R^{S_1\mu t})^\ast]$ & $m_ts_L^2/m_T$\\
\hline
\rule{0pt}{20pt}$S_3$ & $(X,T,B)_{L,R}$ & \makecell{$\frac{m_T}{m_t}[f_{LR}^{S_3}(m_T^2/m_{S_3}^2)+2\widetilde{f}_{LR}^{S_3}(m_T^2/m_{S_3}^2)]\cdot\mr{Re}[y_L^{S_3\mu T}(y_R^{S_3\mu t})^\ast]s_L$\\$+(\frac{7}{4}+\log \frac{m_t^2}{m_{S_3}^2})\mr{Re}[y_L^{S_3\mu T}(y_R^{S_3\mu t})^\ast]s_R$} & $m_T/m_t$\\ \hline
\end{tabular}%}
\caption{The third column shows the approximate formulae of the $\Delta a_{\mu}$ in the LQ+VLQ models for different representations, and the fourth column shows the multiplication order of $T$ left-handed and right-handed LQ Yukawa couplings with respect to the top quark ones. In the above, we have extracted the common factor, which means $\Delta a_{\mu}$ is redefined as $m_{\mu}m_t\Delta\overline{a_{\mu}}/(4\pi^2m_{\mr{LQ}}^2)$.} \label{tab:LQ+VLQ:g-2}
\end{center}
\end{table}

\subsection{The one LQ and two VLQs extended models}
Here, we will not enumerate all the possible combinations. Instead, we just take singlet $T_{1L,R}$ and doublet $(T_2,B_2)_{L,R}$ extended models as the examples, which will be named as "$\mr{LQ}+T_{L,R}+(T,B)_{L,R}$" with $\mr{LQ}$ to be $R_2$ or $S_1$. For simplicity, we just consider the mixing between two VLQs. Then, we have the following mass term and Yukawa interactions:
\begin{align}
\mc{L}\supset-M_1\overline{T_{1L}}T_{1R}-M_2\overline{(T_2,B_2)_{L}}\left(\begin{array}{c}T_2\\B_2\end{array}\right)_{R}-y_L^{TB}\overline{(T_2,B_2)_{L}}T_{1R}\widetilde{\phi}+y_R^{TB}\phi^T\epsilon\overline{T_{1L}}\left(\begin{array}{c}T_2\\B_2\end{array}\right)_{R}+\mathrm{h.c.}~,
\end{align}
where $\widetilde{\phi}\equiv\epsilon\phi^{\ast}$ and $\phi$ is the SM Higgs doublet. Thus, the mass terms should be
\begin{align}
-\overline{(T_1,T_2)_{L}}\left(\begin{array}{cc}M_1&\frac{1}{\sqrt{2}}y_R^{TB}v\\\frac{1}{\sqrt{2}}y_L^{TB}v&M_2\end{array}\right)
\left(\begin{array}{c}T_1\\T_2\end{array}\right)_{R}+\mathrm{h.c.}~,
\end{align}
which can be diagonalized through the following transformations:
\begin{align}\label{eqn:quarkXT:rotation}
\left[\begin{array}{c}T_{1L}\\T_{2L}\end{array}\right]\rightarrow
	\left[\begin{array}{cc}\cos\theta_L^T&\sin\theta_L^T\\-\sin\theta_L^T&\cos\theta_L^T\end{array}\right]
	\left[\begin{array}{c}T_L\\T'_L\end{array}\right],\quad
%%%%%%%%%%%%%%%%%%%%%%%%%%%%%%%%%%%%%%%%%%%%%%%%%%%%%%%%%%%%%%%%%%%%%%%%
\left[\begin{array}{c}T_{1R}\\T_{2R}\end{array}\right]\rightarrow
	\left[\begin{array}{cc}\cos\theta_R^T&\sin\theta_R^T\\-\sin\theta_R^T&\cos\theta_R^T\end{array}\right]
	\left[\begin{array}{c}T_R\\T'_R\end{array}\right].
\end{align}
In the following, $\sin\theta_L^T,\cos\theta_L^T,\sin\theta_R^T,\cos\theta_R^T$ will be abbreviated as $s_L^T,c_L^T,s_R^T,c_R^T$, respectively. Besides, we have the following relations:
\begin{align}
&M_1=m_Tc_L^Tc_R^T+m_{T'}s_L^Ts_R^T,M_2=m_{T'}c_L^Tc_R^T+m_Ts_L^Ts_R^T,
\end{align}
and
\begin{align}
&c_L^Ts_R^T=\frac{v(y_R^{TB}m_T+y_L^{TB}m_{T'})}{\sqrt{2}(m_{T'}^2-m_T^2)},s_L^Tc_R^T=\frac{v(y_L^{TB}m_T+y_R^{TB}m_{T'})}{\sqrt{2}(m_{T'}^2-m_T^2)}.
\end{align}
In the above, $m_T$ and $m_{T'}$ are the physical masses of $T_{L,R}$ and $T'_{L,R}$. Different from the single VLQ cases, the left-handed and right-handed field mixing angles can not be related with each other. 

Let us consider the case where only the VLQs can couple to muon through one LQ, and we have the following gauge eigenstate LQ interactions:
\begin{align}
&\mc{L}_{R_2+T_{L,R}+(T,B)_{L,R}}\supset-x_{Ti}\overline{T_{1R}}R_2^a\epsilon^{ab}L_L^{i,b}+y_{iT}\overline{e_R}^i(R_2^a)^\ast \left(\begin{array}{c}T_2\\B_2\end{array}\right)_L^a+\mathrm{h.c.}~,\nonumber\\
&\mc{L}_{S_1+T_{L,R}+(T,B)_{L,R}}\supset x_{Ti}\overline{(T_{1R})^C}S_1e_R^i+v_{Ti}\left(\overline{(T_{2L})^C},\overline{(B_{2L})^C}\right)^aS_1\epsilon^{ab}L_L^{i,b}+\mathrm{h.c.}.
\end{align}
After the EWSB, they can be written as
\begin{align}
&\mc{L}_{R_2+T_{L,R}+(T,B)_{L,R}}\supset y_L^{R_2\mu T_2}\bar{\mu}~\omega_-~T_{2L}(R_2^{\frac{5}{3}})^\ast+y_R^{R_2\mu T_1}\bar{\mu}~\omega_+~T_{1R}(R_2^{\frac{5}{3}})^\ast+y_L^{R_2\mu T_2}\bar{\mu}~\omega_-~B_{2L}(R_2^{\frac{2}{3}})^\ast+\mathrm{h.c.}~,\nonumber\\
&\mc{L}_{S_1+T_{L,R}+(T,B)_{L,R}}\supset y_L^{S_1\mu T_1}\bar{\mu}~\omega_-~(T_{1R})^C(S_1)^\ast+y_R^{S_1\mu T_2}\bar{\mu}~\omega_+~(T_{2L})^C(S_1)^\ast+\mathrm{h.c.}.
\end{align}
When transforming the fields through the rotation in Eq. \eqref{eqn:quarkXT:rotation}, we can get the following mass eigenstate interactions:
\begin{align}\label{eqn:R2+TTB:Lagmass}
&\mc{L}_{R_2+T_{L,R}+(T,B)_{L,R}}\supset\bar{\mu}(-y_L^{R_2\mu T_2}s_L^T\omega_-+y_R^{R_2\mu T_1}c_R^T\omega_+)T(R_2^{\frac{5}{3}})^\ast\nonumber\\
	&+\bar{\mu}(y_L^{R_2\mu T_2}c_L^T\omega_-+y_R^{R_2\mu T_1}s_R^T\omega_+)T'(R_2^{\frac{5}{3}})^\ast+y_L^{R_2\mu T_2}\bar{\mu}~\omega_-~B_{2L}(R_2^{\frac{2}{3}})^\ast+\mathrm{h.c.}~,
\end{align}
and
\begin{align}\label{eqn:S1+TTB:Lagmass}
&\mc{L}_{S_1+T_{L,R}+(T,B)_{L,R}}\supset\bar{\mu}(y_L^{S_1\mu T_1}c_R^T\omega_--y_R^{S_1\mu T_2}s_L^T\omega_+)T^C(S_1)^\ast\nonumber\\
	&+\bar{\mu}(y_L^{S_1\mu T_1}s_R^T\omega_-+y_R^{S_1\mu T_2}c_L^T\omega_+)T'^C(S_1)^\ast+\mathrm{h.c.}.
\end{align}
For the $R_2$ leptoquark, the contributions can be approximated as
\begin{align}\label{eqn:LQ+TTB:g-2appR2}
&\Delta a_{\mu}^{R_2+T_{L,R}+(T,B)_{L,R}}\approx\frac{m_{\mu}^2}{4\pi^2m_{R_2}^2}\mr{Re}[y_R^{R_2\mu T_1}(y_L^{R_2\mu T_2})^\ast][-\frac{m_T}{m_\mu}f_{LR}^{R_2}(\frac{m_T^2}{m_{R_2}^2})s_L^Tc_R^T+\frac{m_{T'}}{m_\mu}f_{LR}^{R_2}(\frac{m_{T'}^2}{m_{R_2}^2})s_R^Tc_L^T].
\end{align}
For the $S_1$ leptoquark, the contributions can be approximated as
\begin{align}\label{eqn:LQ+TTB:g-2appS1}
&\Delta a_{\mu}^{S_1+T_{L,R}+(T,B)_{L,R}}\approx\frac{m_{\mu}^2}{4\pi^2m_{S_1}^2}\mr{Re}[y_L^{S_1\mu T_1}(y_R^{S_1\mu T_2})^\ast][-\frac{m_T}{m_\mu}f_{LR}^{S_1}(\frac{m_T^2}{m_{S_1}^2})s_L^Tc_R^T+\frac{m_{T'}}{m_\mu}f_{LR}^{S_1}(\frac{m_{T'}^2}{m_{S_1}^2})s_R^Tc_L^T].
\end{align}
We find that there can be cancellation for both $R_2$ and $S_1$ in the case of $m_T=m_T'$ and $\theta_L^T=\theta_R^T$. While, it is constructive interference for the case of $\theta_L^T=-\theta_R^T$.
%%%%%%%%%%%%%%%%%%%%%%%%%%%%%%%%%%%%%%%%%%%%%%%%%%%%%%%%%%%%%%%%%%%%%%
\section{Numerical analysis}
The input parameters are chosen as $m_\mu=105.66\mathrm{MeV}$, $m_b=4.2\mr{GeV}$, and $m_t=172.5\mr{GeV}$ \cite{ParticleDataGroup:2020ssz}. The VLQs can be mainly constrained by the direct search \cite{CMS:2018wpl, CMS:2019eqb, ATLAS:2018ziw, ATLAS:2018dyh} and electro-weak precision observables (EWPO) \cite{Aguilar-Saavedra:2013qpa, Chen:2017hak}. To satisfy these bounds, we choose VLQ mass to be $\mc{O}(\mr{TeV})$ and the input $t-T$ mixing angle to be $\mc{O}(0.1)$. According to the direct search results, the LQ mass is typically required to be above $\mr{TeV}$ \cite{CMS:2018oaj, CMS:2020wzx, ATLAS:2020xov, ATLAS:2021oiz}.

\subsection{Numerical analysis in the one LQ and one VLQ extended models}
\begin{figure}[!htb]
\begin{center}
\includegraphics[scale=0.25]{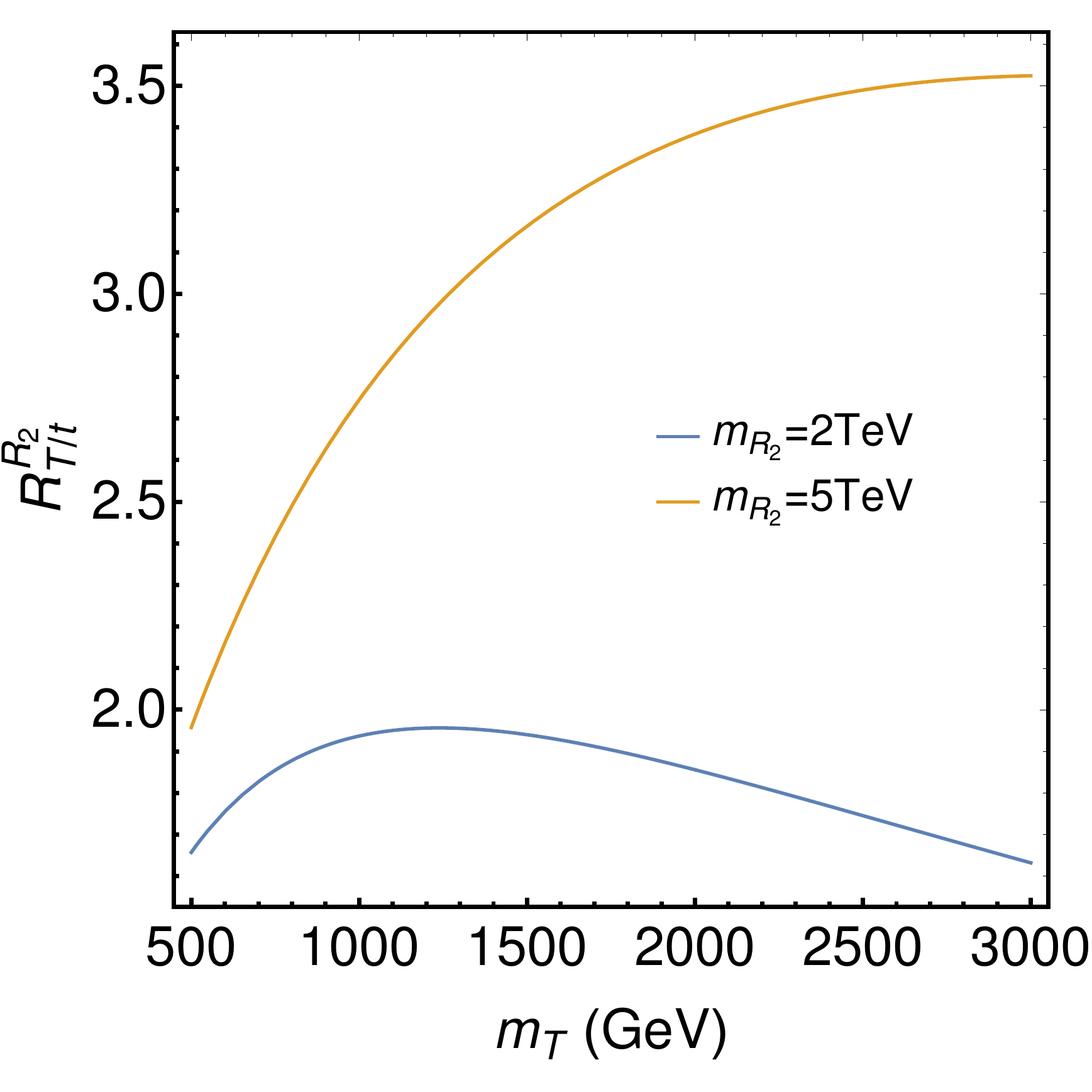}\qquad
\includegraphics[scale=0.25]{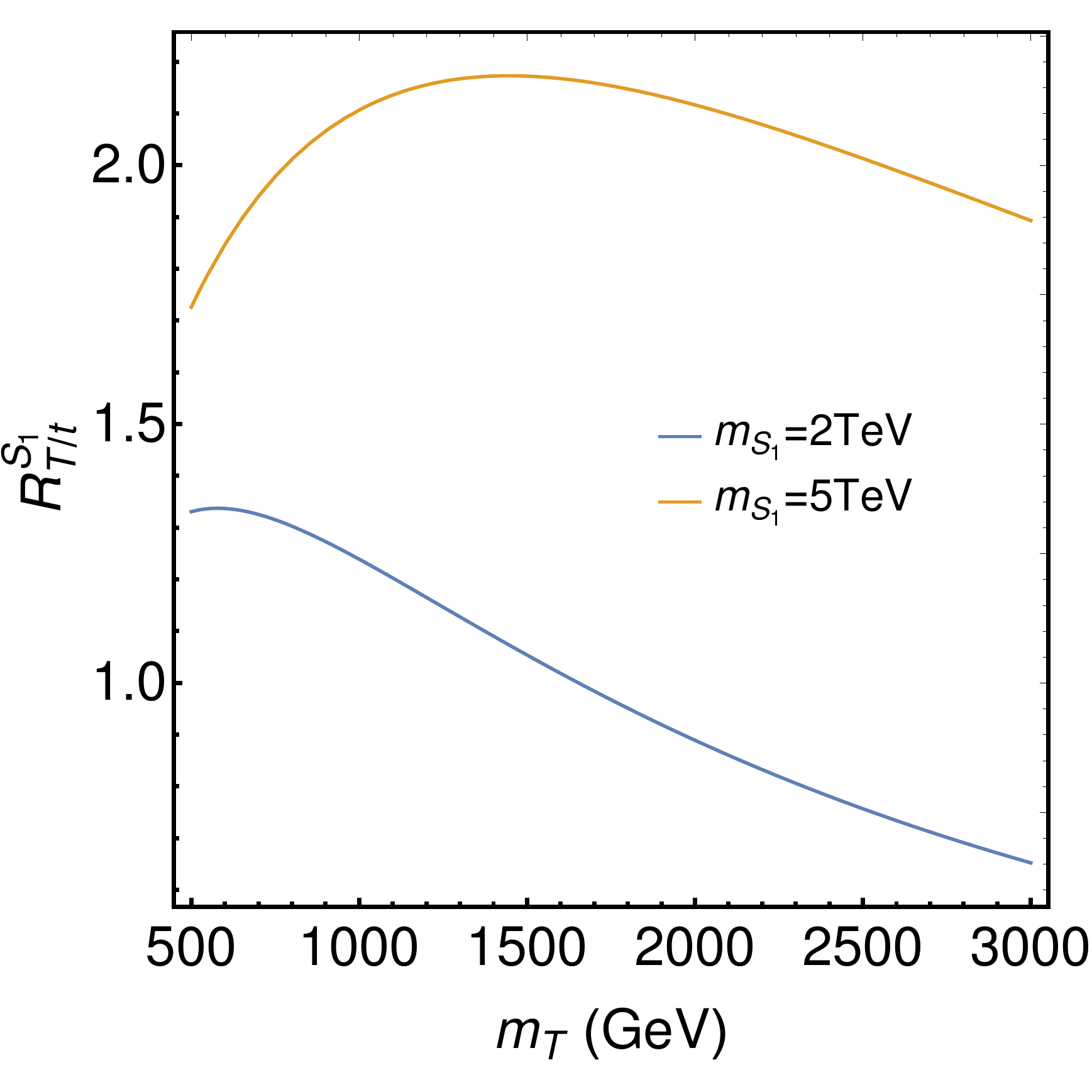}
\caption{The $R_{T/t}^{R_2}$ (left) and $R_{T/t}^{S_1}$ (right) curves as a function of $m_T$ for the LQ masses of 2 TeV (blue) and 5 TeV (yellow).}\label{fig:LQ:R2S1}
\end{center}
\end{figure}

\begin{figure}[!htb]
\begin{center}
\includegraphics[scale=0.25]{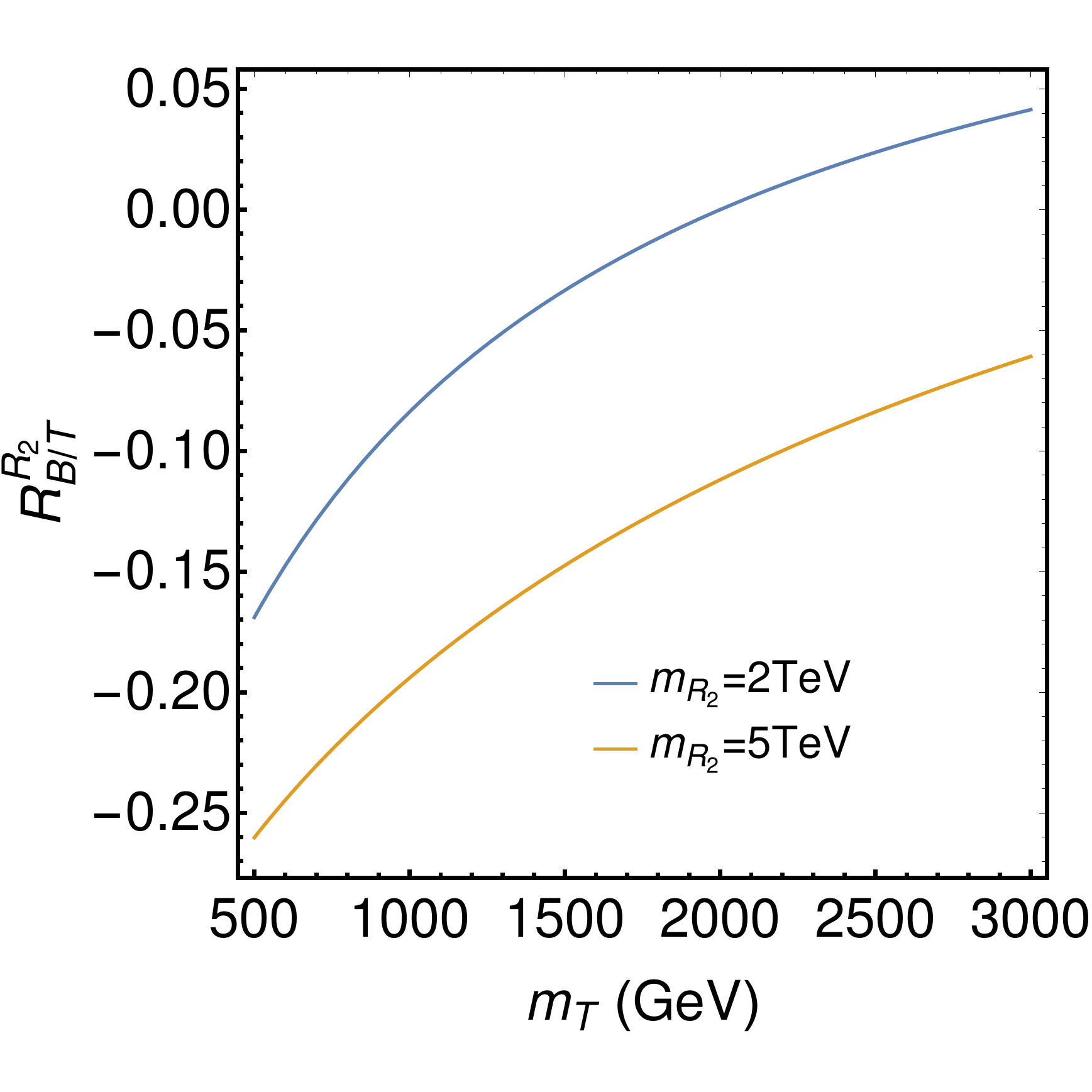}\qquad
\includegraphics[scale=0.25]{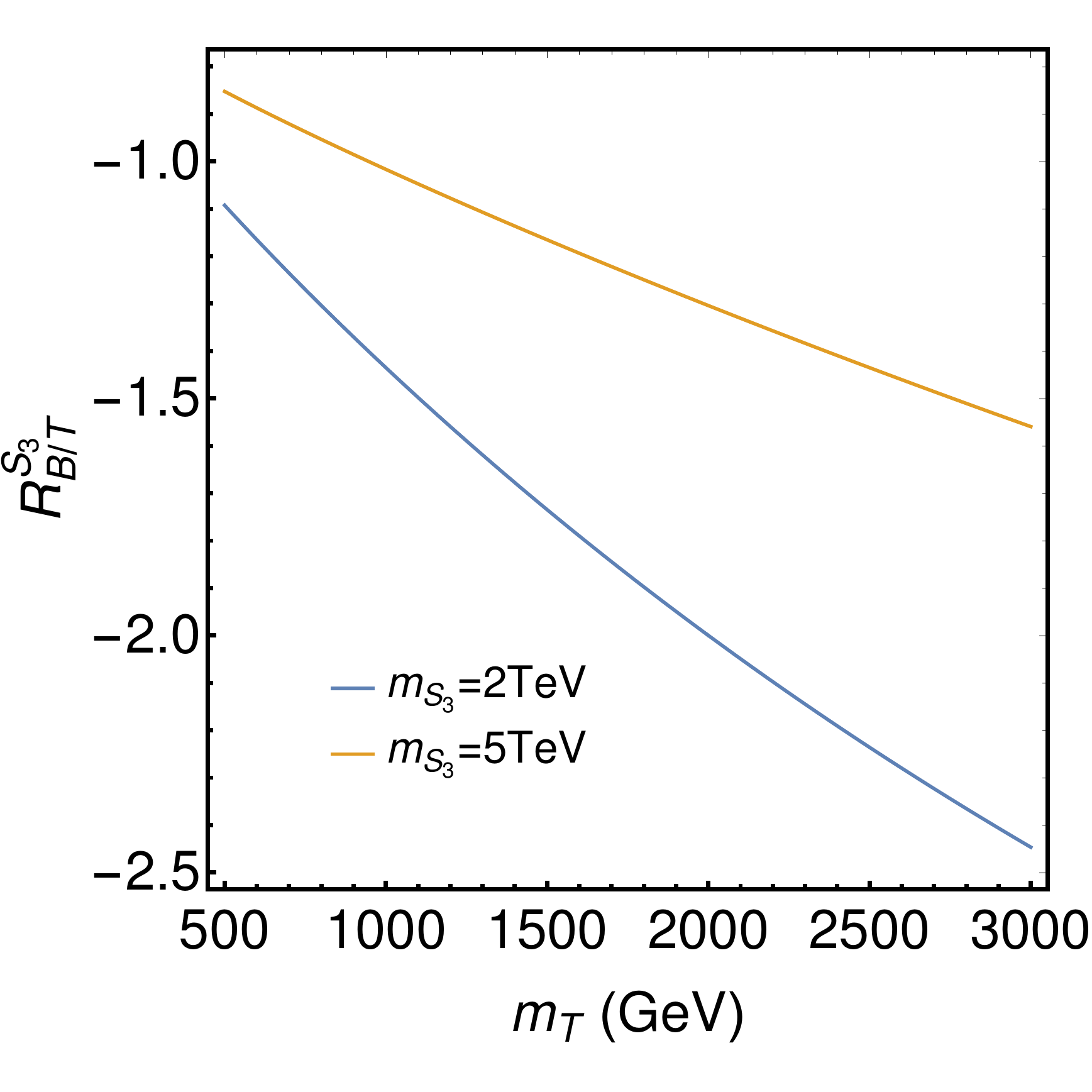}
\caption{The $R_{B/T}^{R_2}$ (left) and $R_{B/T}^{S_3}$ (right) curves as a function of $m_T$ for the LQ masses of 2 TeV (blue) and 5 TeV (yellow).}\label{fig:LQ:R2S3b}
\end{center}
\end{figure}

To estimate the effects of $T$ quark contribution, let us define the following two functions from Eq. \eqref{eqn:LQ:R2S1}:
\begin{align}
&R_{T/t}^{R_2}(m_T,m_{R_2})\equiv\frac{m_Tf_{LR}^{R_2}(m_T^2/m_{R_2}^2)}{m_tf_{LR}^{R_2}(m_t^2/m_{R_2}^2)},\qquad R_{T/t}^{S_1}(m_T,m_{S_1})\equiv\frac{m_Tf_{LR}^{S_1}(m_T^2/m_{S_1}^2)}{m_tf_{LR}^{S_1}(m_t^2/m_{S_1}^2)}.
\end{align}
In Fig. \ref{fig:LQ:R2S1}, we show the plots of $R_{T/t}^{R_2}(m_T,m_{R_2})-m_T$ and $R_{T/t}^{S_1}(m_T,m_{S_1})-m_T$ for fixed LQ masses. According to the plots, we find that the $T$ quark loop integral can be comparable to that of top quark. However, the contributions are suppressed by the mixing angle in most of the models.
In the $R_2/S_3+(X,T,B)_{L,R}$ models, there are also chirally enhanced contributions from $b$ and $B$ quarks. Compared to the $B$ quark contribution, the $b$ quark contribution is always negligible because of the mass and $s_R^b/s_L^b$ suppressions. To compare the effects of $T$ and $B$ quark contributions, we define the following two functions from Eqs. \eqref{eqn:LQ:R2S1} and \eqref{eqn:LQ:R2S3b}:
\begin{align}
&R_{B/T}^{R_2}(m_T,m_{R_2})\equiv\frac{\widetilde{f}_{LR}^{R_2}(m_T^2/m_{R_2}^2)}{f_{LR}^{R_2}(m_T^2/m_{R_2}^2)},\qquad R_{B/T}^{S_3}(m_T,m_{S_3})\equiv\frac{\widetilde{f}_{LR}^{S_3}(m_T^2/m_{S_3}^2)}{f_{LR}^{S_3}(m_T^2/m_{S_3}^2)}.
\end{align}
In Fig. \ref{fig:LQ:R2S3b}, we show the plots of $R_{B/T}^{R_2}(m_T,m_{R_2})-m_T$ and $R_{B/T}^{S_3}(m_T,m_{S_3})-m_T$ for fixed LQ masses. According to the plots, we find that the $B$ quark loop integral can be considerable compared to $T$ quark.

\begin{figure}[!htb]
\begin{center}
\includegraphics[scale=0.27]{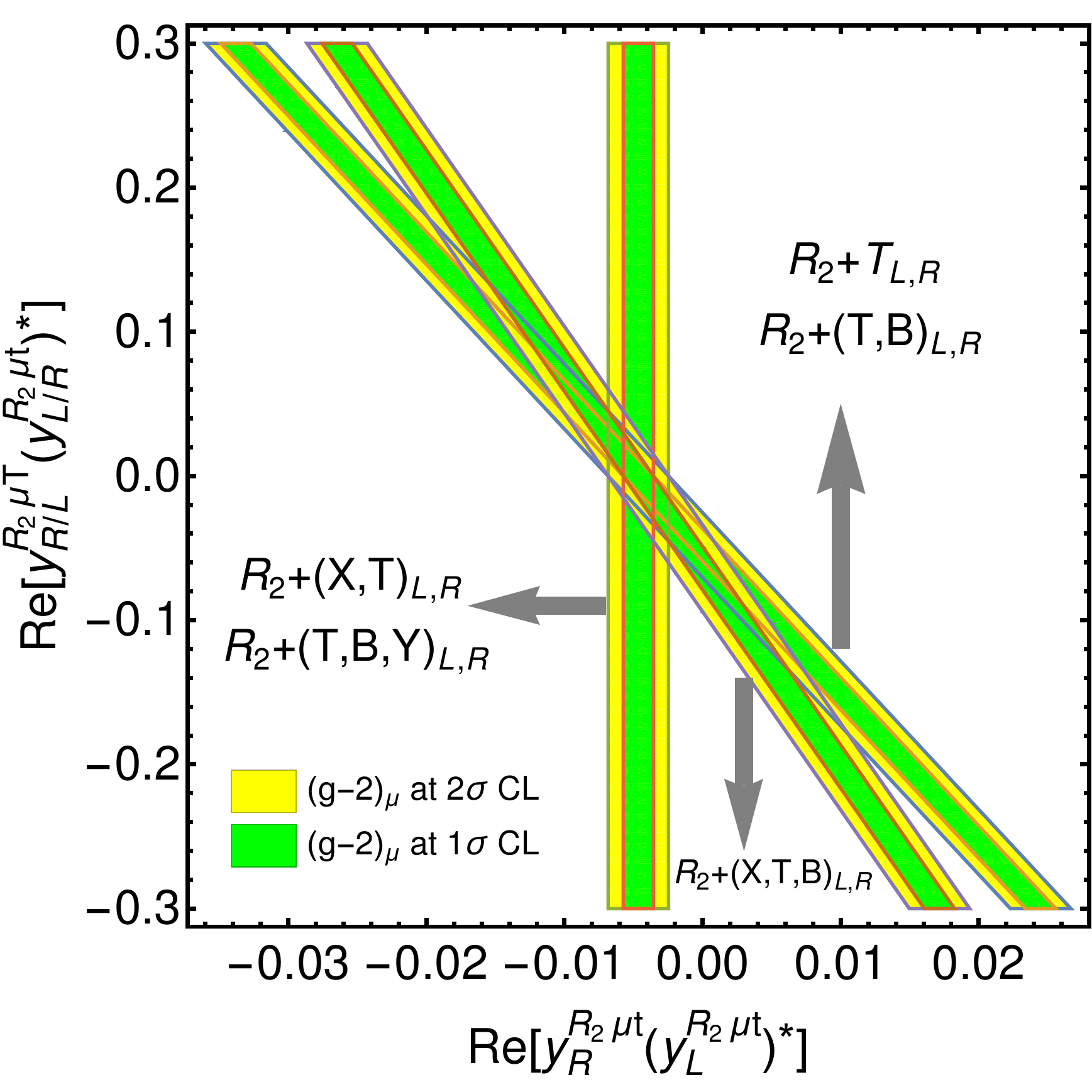}
\includegraphics[scale=0.27]{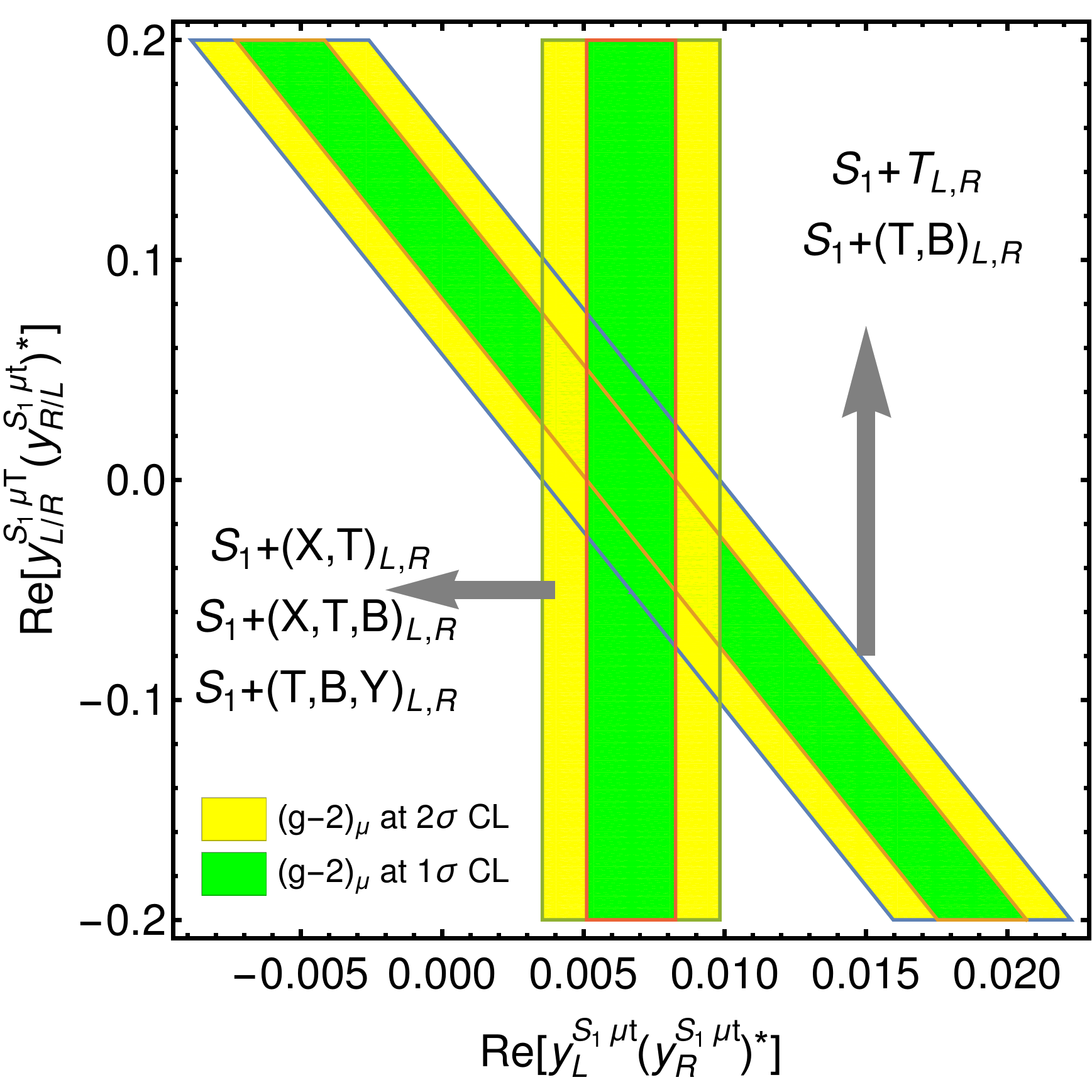}
\includegraphics[scale=0.27]{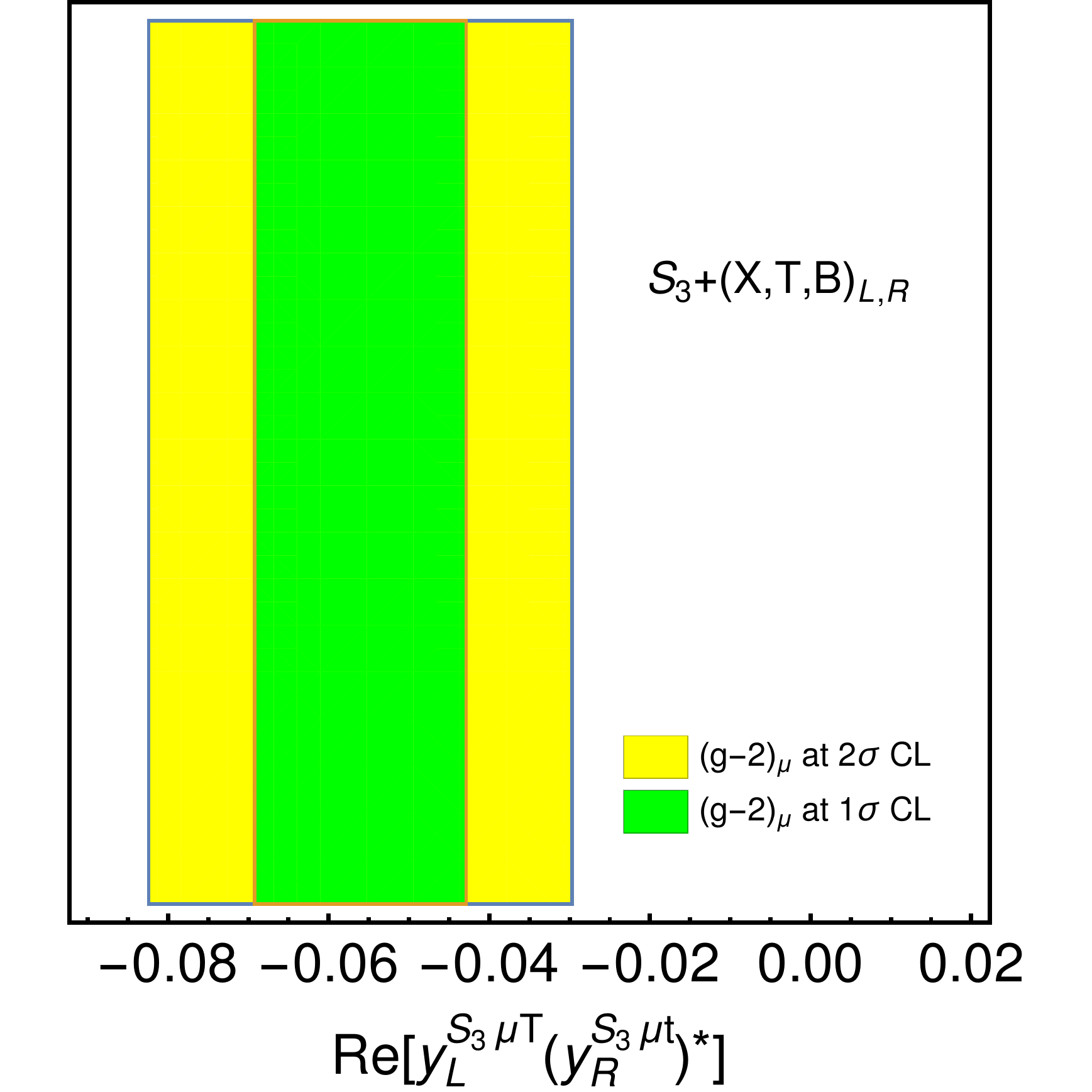}
\caption{The allowed region in the plane of $\mr{Re}[y_{R/L}^{R_2\mu T}(y_{L/R}^{R_2\mu t})^\ast]-\mr{Re}[y_R^{R_2\mu t}(y_L^{R_2\mu t})^\ast]$ (left, $R_2+\mr{VLQ}$ models) and $\mr{Re}[y_{L/R}^{S_1\mu T}(y_{R/L}^{S_1\mu t})^\ast]-\mr{Re}[y_L^{S_1\mu t}(y_R^{S_1\mu t})^\ast]$ (middle, $S_1+\mr{VLQ}$ models). Here, the green and yellow areas are allowed at $1\sigma$ and $2\sigma$ CL, respectively. Note that there are no $y_{L/R}^{R_2\mu T}$ or $y_{L/R}^{S_1\mu T}$ couplings involved in the $R_2+(X,T)_{L,R}/(T,B,Y)_{L,R},S_1+(X,T)_{L,R}/(X,T,B)_{L,R}/(T,B,Y)_{L,R}$ models, but we show them just for comparison. Besides, the subscripts $L,R$ of vertical axis title are reversed between the doublet $(T,B)_{L,R}$ and singlet $T_{L,R}$ cases. The right plot is for the $S_3+(X,T,B)_{L,R}$ model.}\label{fig:LQ+VLQ:g-2num}
\end{center}
\end{figure}

In Tab. \ref{tab:LQ+VLQ:g-2num}, we give the approximate numerical expressions of the $\Delta a_{\mu}$ in different LQ+VLQ models. We adopt the mass parameters to be $m_T=1\mr{TeV}$ and $m_{\mr{LQ}}=2\mr{TeV}$ by default. The input mixing angle in this section is set as $s_L=0.05$ for the singlet and triplet VLQ cases, while it is $s_R=0.05$ for the doublet VLQ cases. Based on the numerical results in Tab. \ref{tab:LQ+VLQ:g-2num}, we show the allowed regions from $(g-2)_{\mu}$ in Fig. \ref{fig:LQ+VLQ:g-2num}. In the $R_2+(X,T)_{L,R}/(T,B,Y)_{L,R}$ models, the $\mr{Re}[y_R^{R_2\mu t}(y_L^{R_2\mu t})^\ast]$ is bounded to be in the ranges $(-5.74,-3.56)\times10^{-3}$ and $(-6.84,-2.46)\times10^{-3}$ at $1\sigma$ and $2\sigma$ CL, respectively. In the $S_1+(X,T)_{L,R}/(X,T,B)_{L,R}/(T,B,Y)_{L,R}$ models, the $\mr{Re}[y_L^{S_1\mu t}(y_R^{S_1\mu t})^\ast]$ is bounded to be in the ranges $(5.11,8.25)\times10^{-3}$ and $(3.54,9.82)\times10^{-3}$ at $1\sigma$ and $2\sigma$ CL, respectively. In the $R_2+T_{L,R}/(T,B)_{L,R}$ models, the $\mr{Re}[y_{R/L}^{R_2\mu T}(y_{L/R}^{R_2\mu t})^\ast]$ can be constrained in the ranges $(-0.059,-0.037)$ and $(-0.07,-0.025)$ with $\mr{Re}[y_R^{R_2\mu t}(y_L^{R_2\mu t})^\ast]$ turned off at $1\sigma$ and $2\sigma$ CL, respectively. In the $R_2+(X,T,B)_{L,R}$ model, the $\mr{Re}[y_R^{R_2\mu T}(y_L^{R_2\mu t})^\ast]$ can be constrained in the ranges $(-0.079, -0.049)$ and $(-0.094, -0.034)$ with $\mr{Re}[y_R^{R_2\mu t}(y_L^{R_2\mu t})^\ast]$ turned off at $1\sigma$ and $2\sigma$ CL, respectively. In the $S_1+T_{L,R}/(T,B)_{L,R}$ models, the $\mr{Re}[y_{L/R}^{S_1\mu T}(y_{R/L}^{S_1\mu t})^\ast]$ can be constrained in the ranges $(0.082,0.133)$ and $(0.057,0.158)$ with $\mr{Re}[y_L^{S_1\mu t}(y_R^{S_1\mu t})^\ast]$ turned off at $1\sigma$ and $2\sigma$ CL, respectively. In the $S_3+(X,T,B)_{L,R}$ model, the $\mr{Re}[y_L^{S_3\mu T}(y_R^{S_3\mu t})^\ast]$ is bounded to be in the ranges $(-0.069,-0.043)$ and $(-0.082,-0.03)$ at $1\sigma$ and $2\sigma$ CL, respectively. Although we only keep the chirally enhanced contributions in Tab. \ref{tab:LQ+VLQ:g-2num}, they are good approximations. According to the numerical estimation, the dropped contributions are always $m_{\mu}/m_t(m_T)\le10^{-3}$ order suppressed.
\begin{table}[!htb]
\begin{center}
%\resizebox{168mm}{18mm}{
\begin{tabular}{c|c|c}
\hline
LQ & VLQ & the leading order expressions of $\Delta a_{\mu}\times10^{7}$ \\
\hline
\multirow{5}{*}{$R_2$} & $T_{L,R}$ & $-0.5238\mr{Re}[y_R^{R_2\mu T}(y_L^{R_2\mu t})^\ast]-5.393\mr{Re}[y_R^{R_2\mu t}(y_L^{R_2\mu t})^\ast]$ \\
\cline{2-3}
& $(X,T)_{L,R}$ & $-5.397\mr{Re}[y_R^{R_2\mu t}(y_L^{R_2\mu t})^\ast]$ \\
\cline{2-3}
& $(T,B)_{L,R}$ & $-0.5238\mr{Re}[y_L^{R_2\mu T}(y_R^{R_2\mu t})^\ast]-5.393\mr{Re}[y_R^{R_2\mu t}(y_L^{R_2\mu t})^\ast]$ \\
\cline{2-3}
& $(X,T,B)_{L,R}$ & $-0.3923\mr{Re}[y_R^{R_2\mu T}(y_L^{R_2\mu t})^\ast]-5.397\mr{Re}[y_R^{R_2\mu t}(y_L^{R_2\mu t})^\ast]$ \\
\cline{2-3}
& $(T,B,Y)_{L,R}$ & $-5.397\mr{Re}[y_R^{R_2\mu t}(y_L^{R_2\mu t})^\ast]$ \\
\hline\hline
\multirow{5}{*}{$S_1$} & $T_{L,R}$ & $0.2331\mr{Re}[y_L^{S_1\mu T}(y_R^{S_1\mu t})^\ast]+3.754\mr{Re}[y_L^{S_1\mu t}(y_R^{S_1\mu t})^\ast]$ \\
\cline{2-3}
& $(X,T)_{L,R}$ & $3.756\mr{Re}[y_L^{S_1\mu t}(y_R^{S_1\mu t})^\ast]$ \\
\cline{2-3}
& $(T,B)_{L,R}$ & $0.2331\mr{Re}[y_R^{S_1\mu T}(y_L^{S_1\mu t})^\ast]+3.754\mr{Re}[y_L^{S_1\mu t}(y_R^{S_1\mu t})^\ast]$ \\
\cline{2-3}
& $(X,T,B)_{L,R}$ & $3.756\mr{Re}[y_L^{S_1\mu t}(y_R^{S_1\mu t})^\ast]$ \\
\cline{2-3}
& $(T,B,Y)_{L,R}$ & $3.756\mr{Re}[y_L^{S_1\mu t}(y_R^{S_1\mu t})^\ast]$ \\
\hline
$S_3$ & $(X,T,B)_{L,R}$ & $-0.4478\mr{Re}[y_L^{S_3\mu T}(y_R^{S_3\mu t})^\ast]$ \\
\hline
\end{tabular}%}
\caption{Leading order numerical expressions of the $\Delta a_{\mu}$ in different LQ+VLQ models. Here, we have chosen the parameters to be $m_T=1\mr{TeV}$ and $m_{\mr{LQ}}=2\mr{TeV}$. The input mixing angle is set as $s_L=0.05$ (singlet and triplet VLQ) and $s_R=0.05$ (doublet VLQ).} \label{tab:LQ+VLQ:g-2num}
\end{center}
\end{table}

\begin{figure}[!htb]
\begin{center}
\includegraphics[scale=0.28]{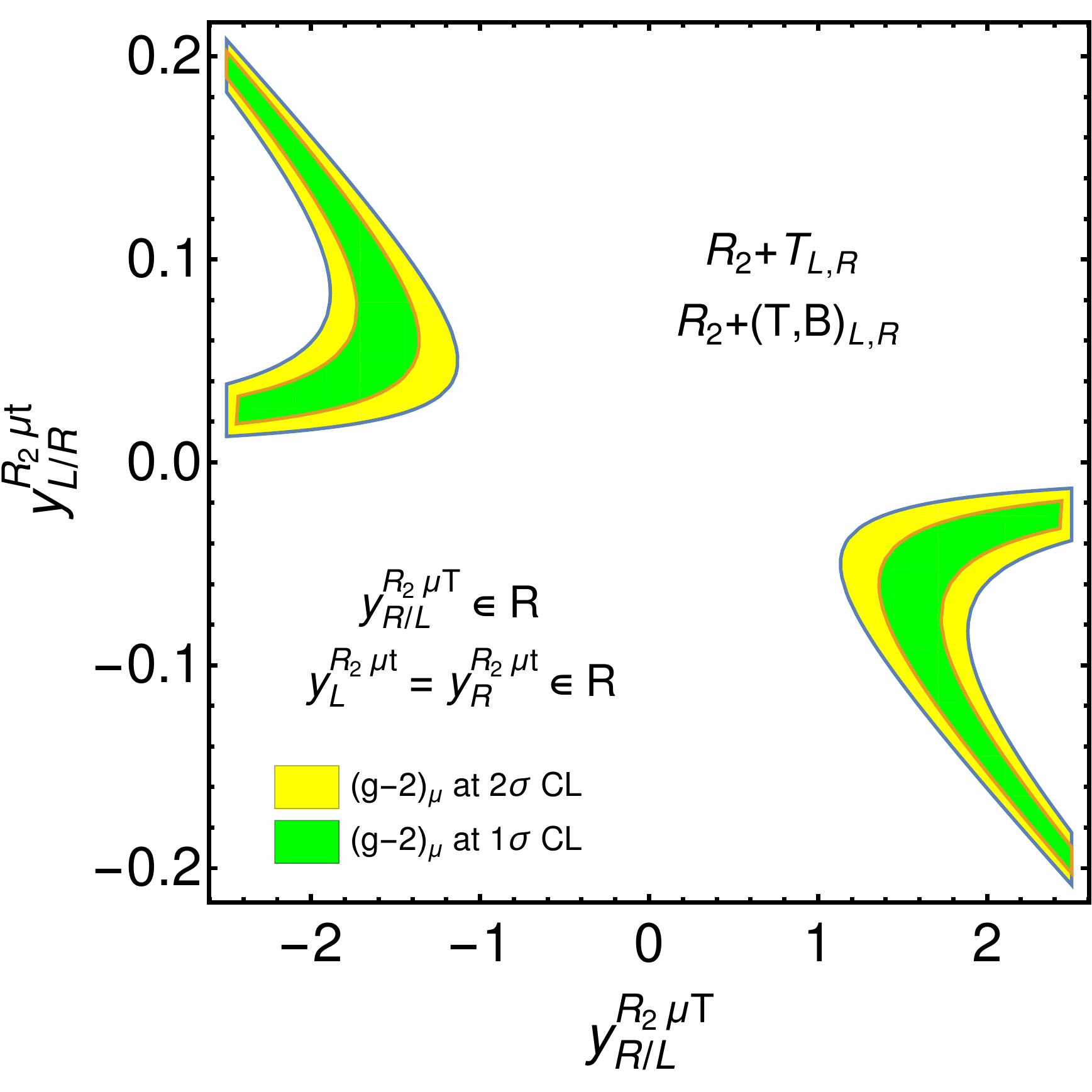}
\includegraphics[scale=0.28]{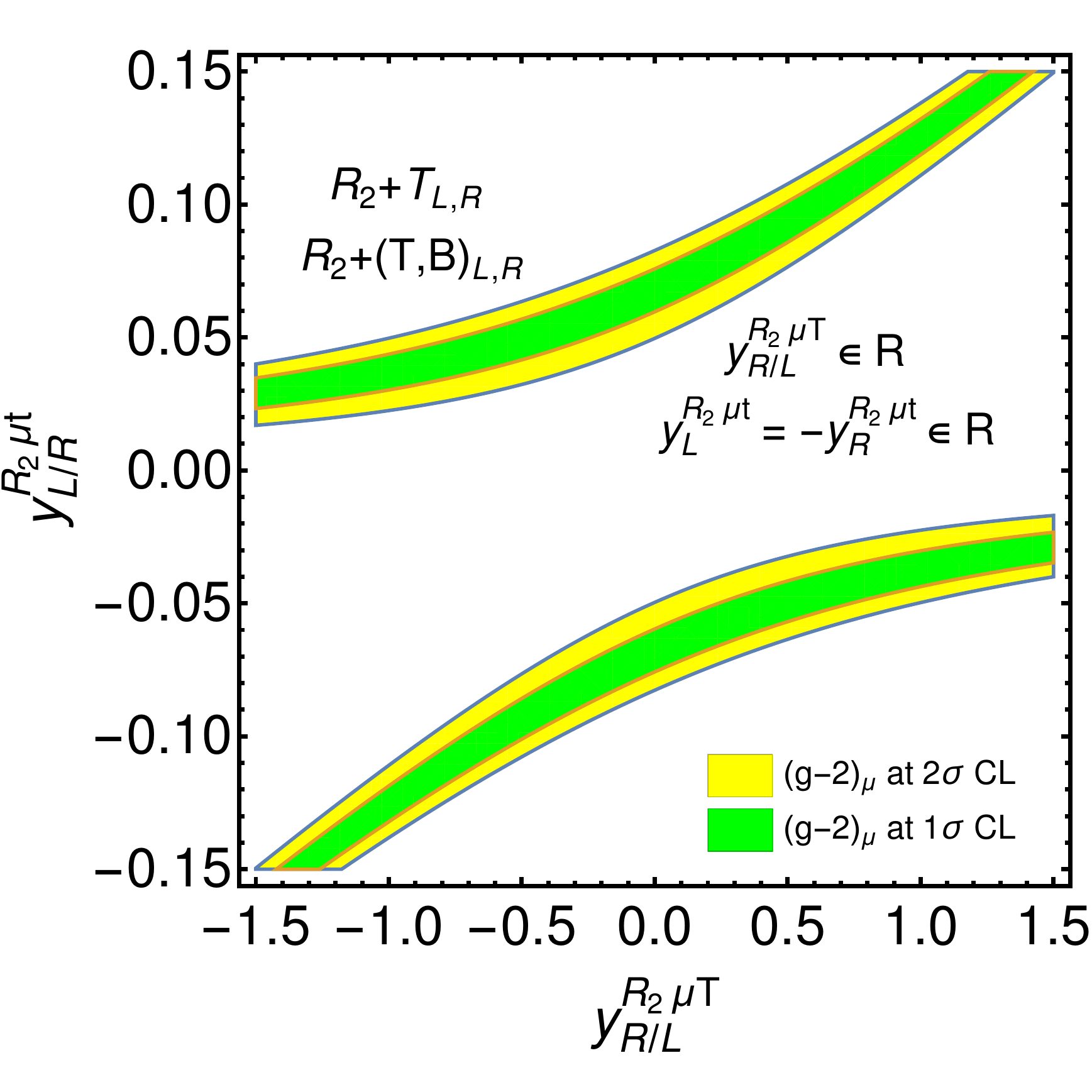}
\includegraphics[scale=0.28]{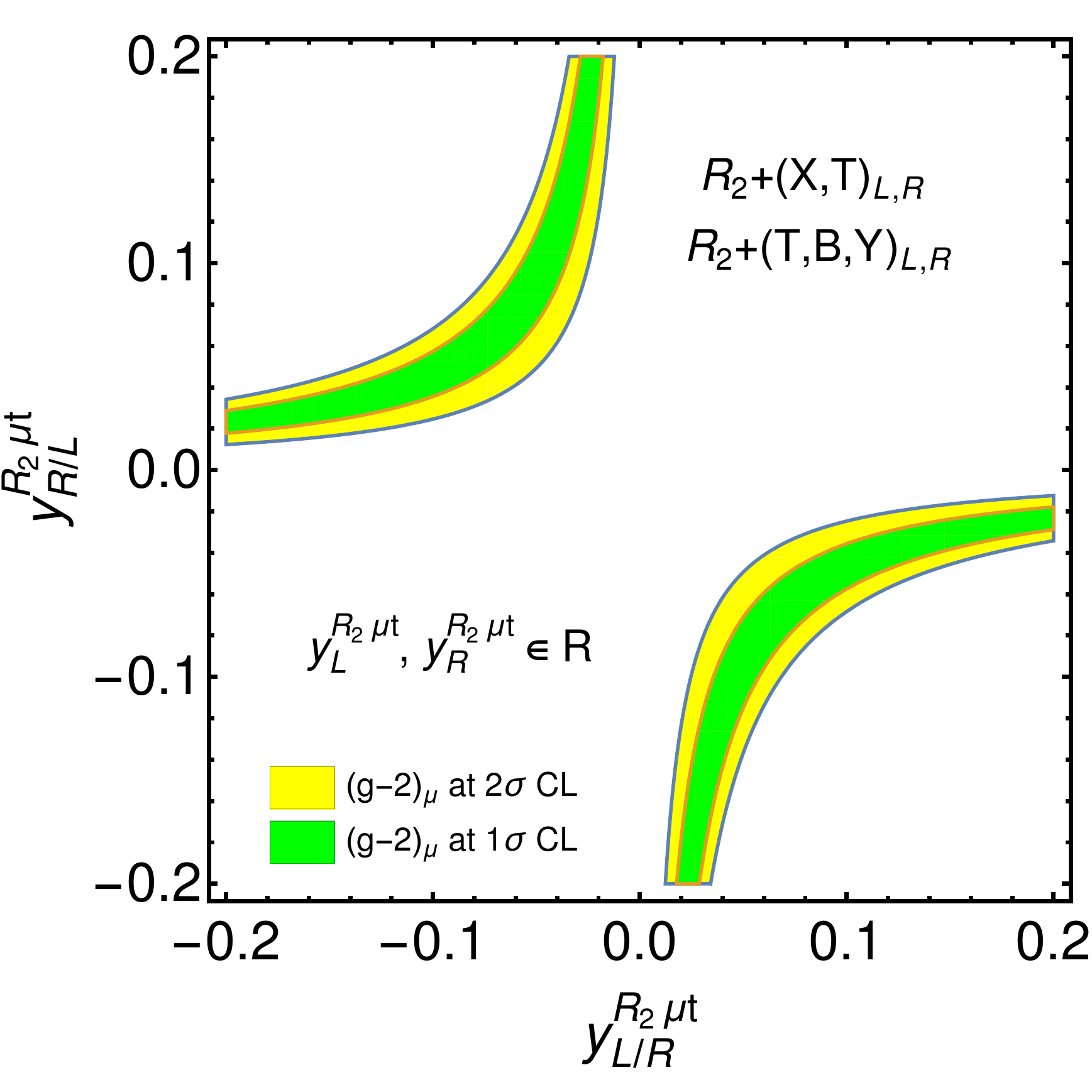}\\
\includegraphics[scale=0.28]{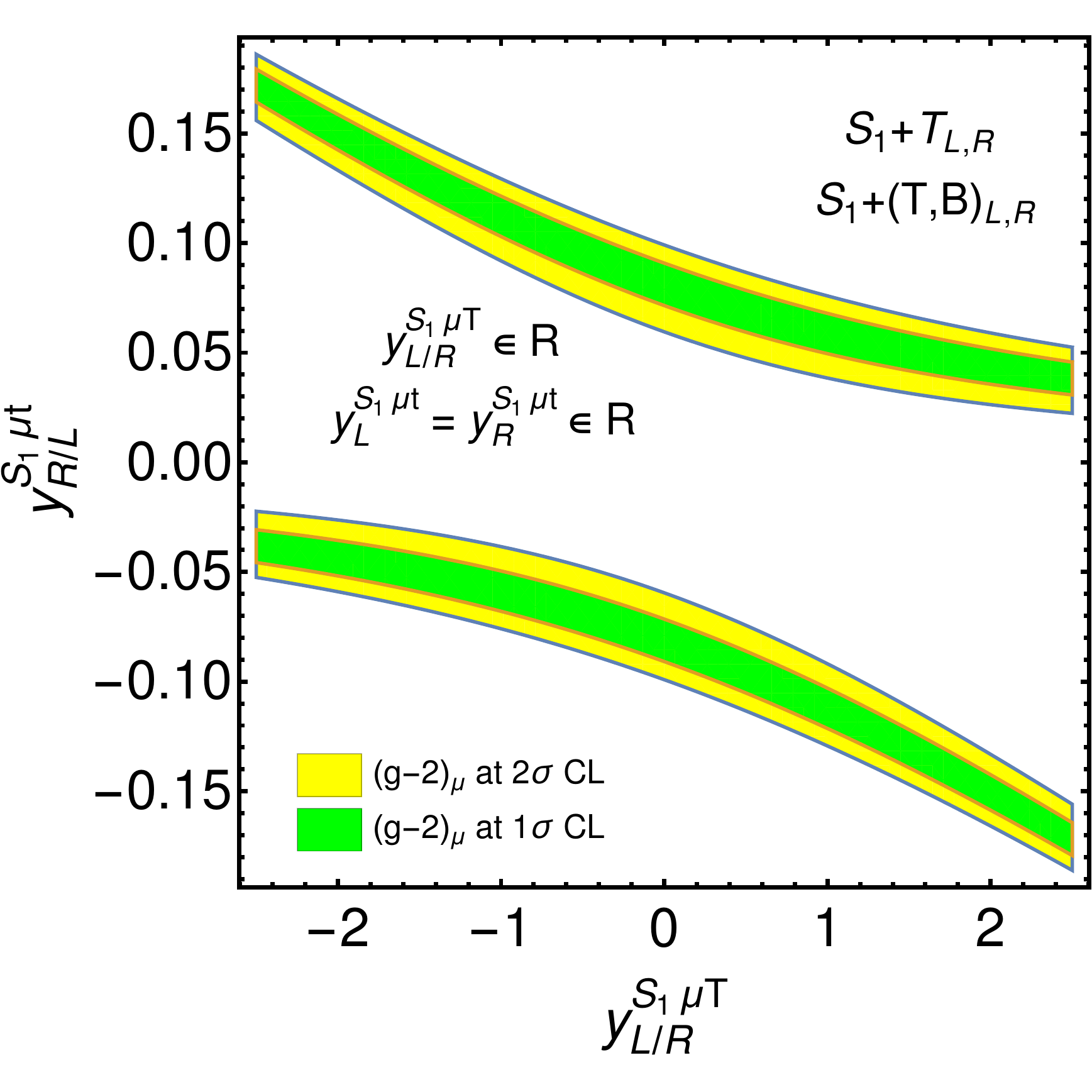}
\includegraphics[scale=0.28]{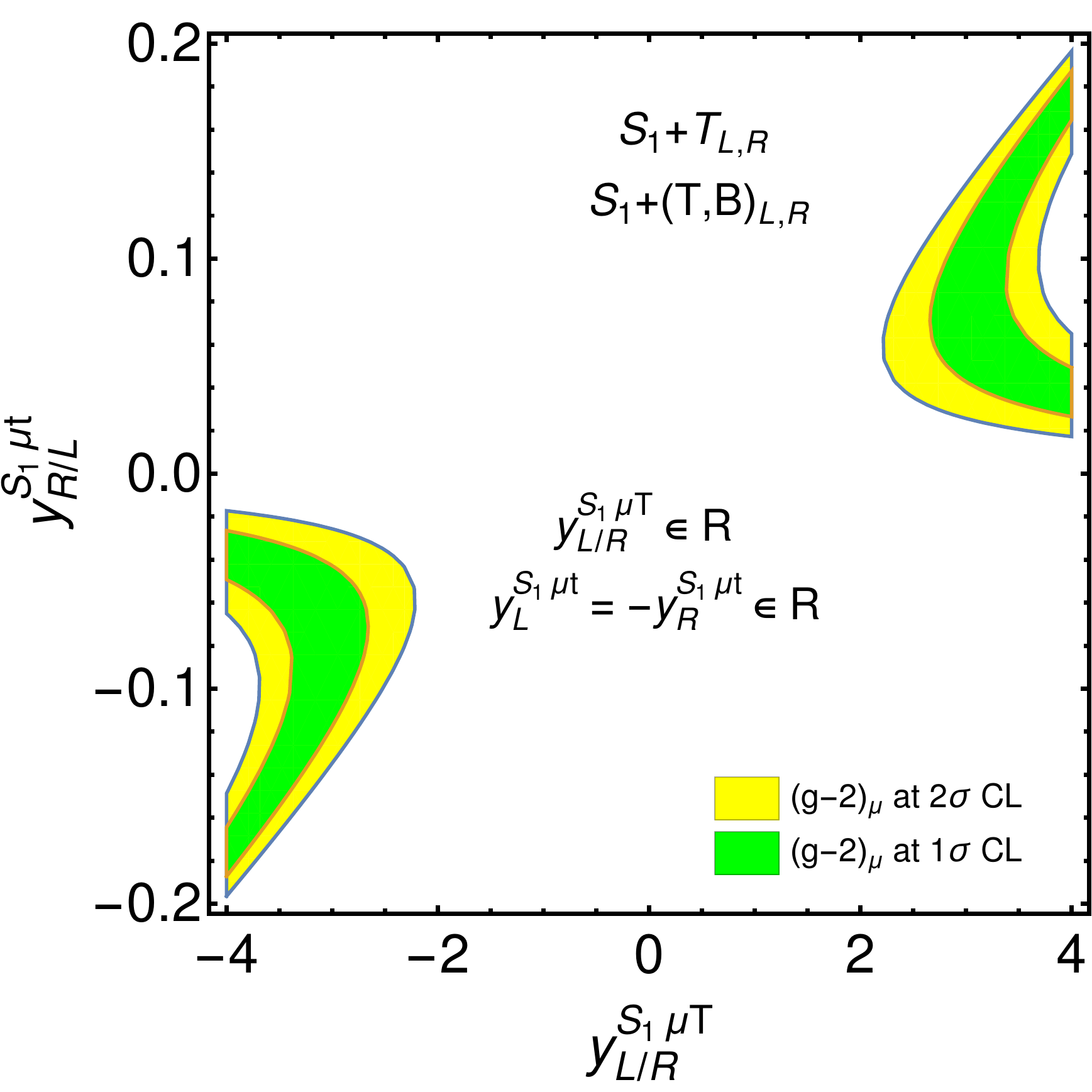}
\includegraphics[scale=0.28]{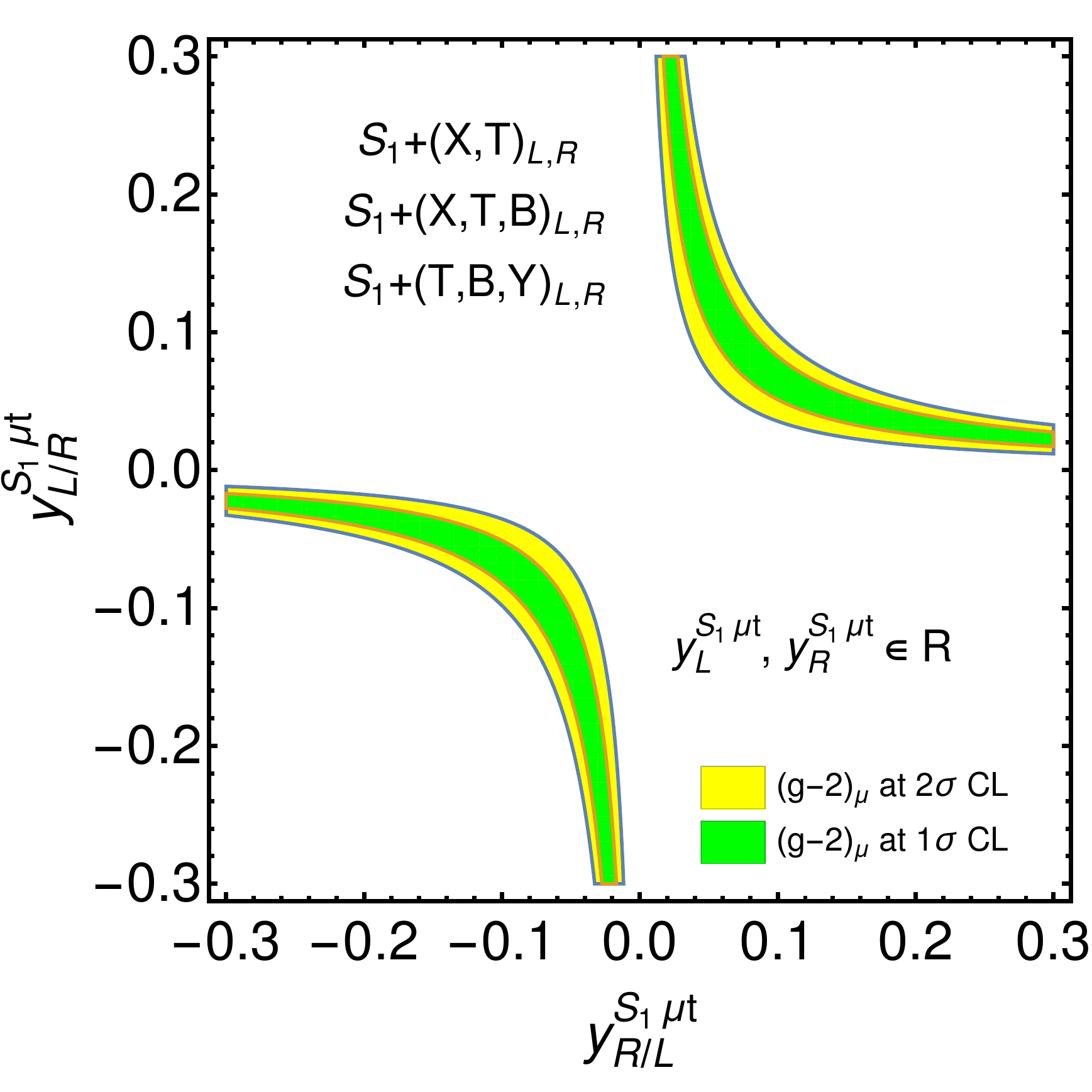}\\
\includegraphics[scale=0.28]{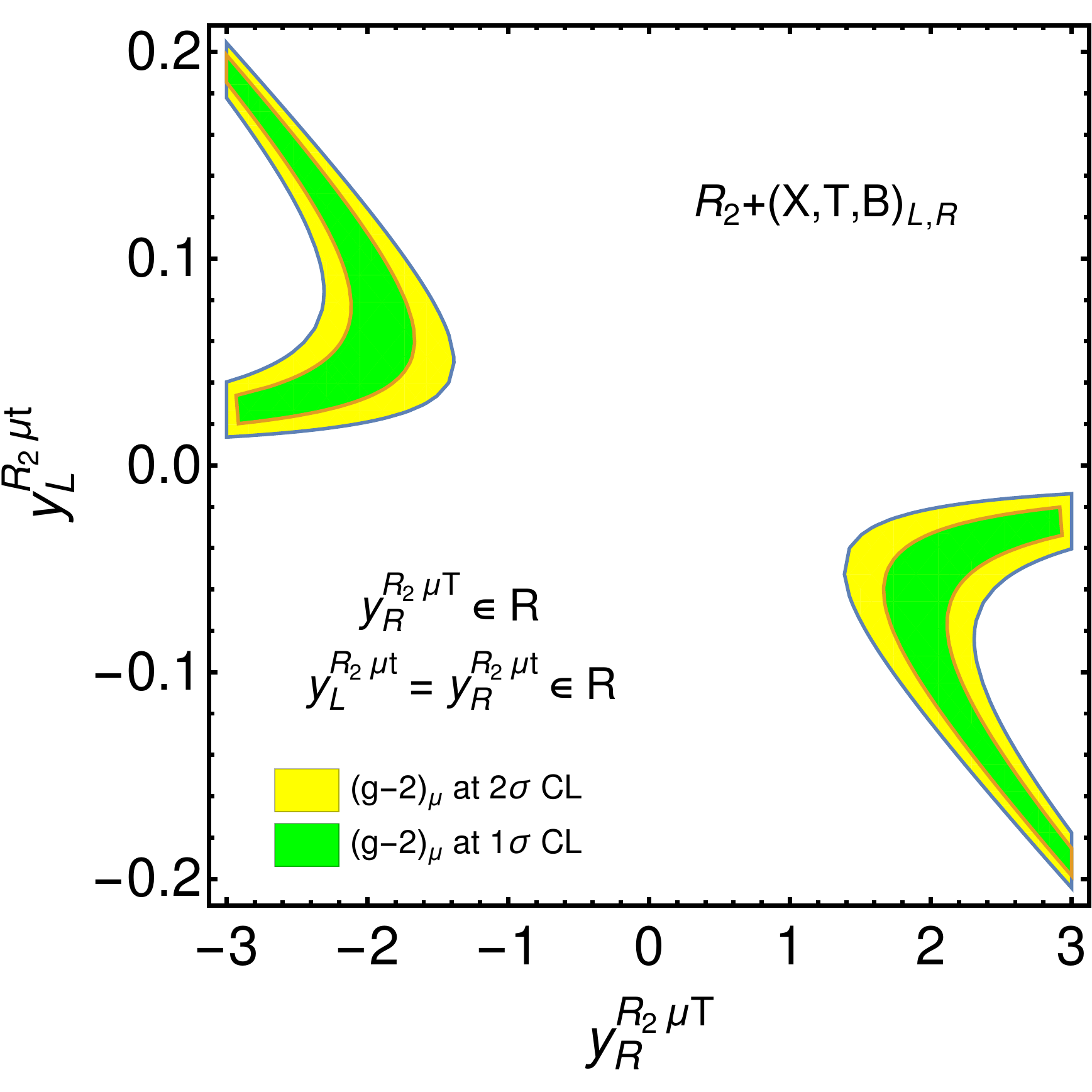}
\includegraphics[scale=0.28]{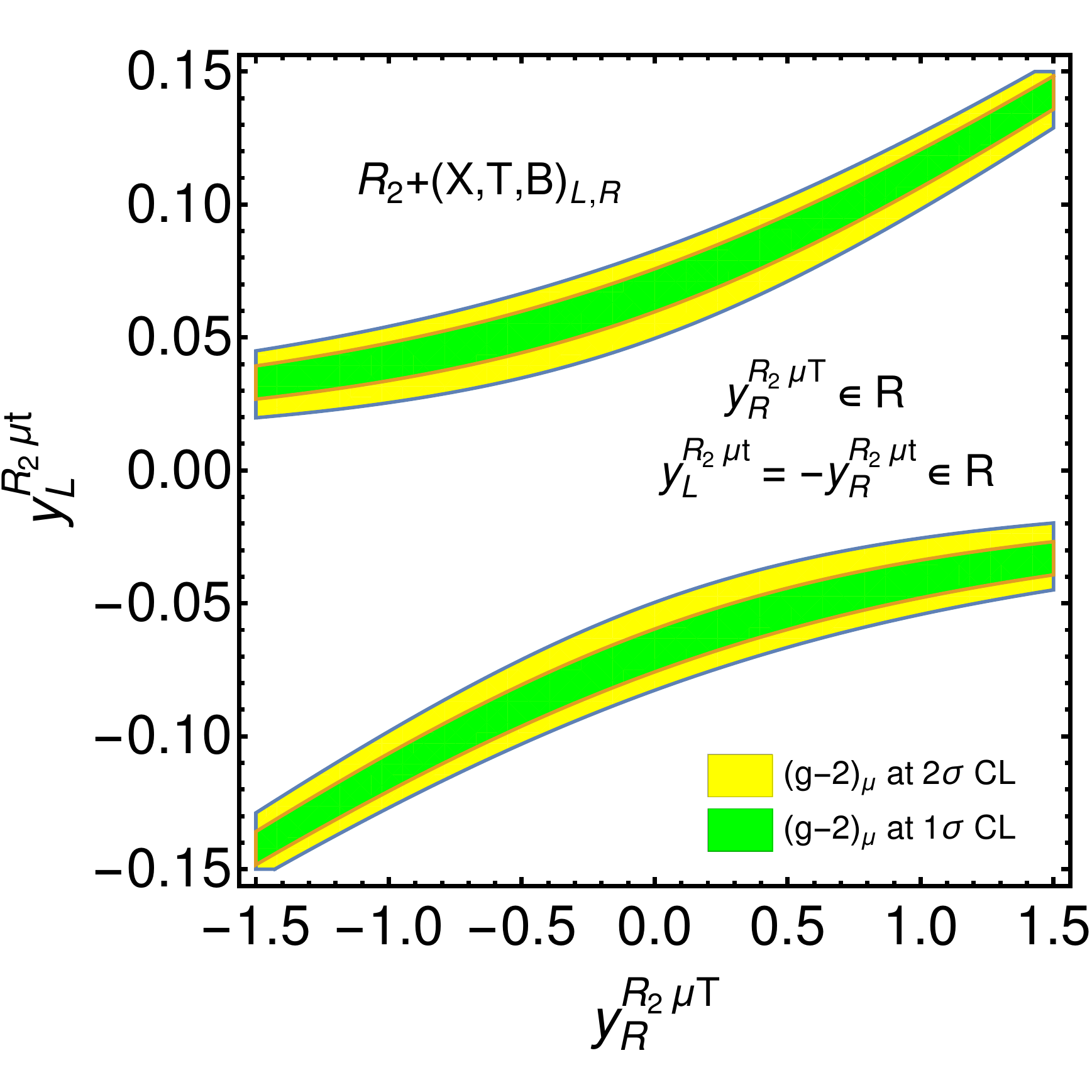}
\includegraphics[scale=0.28]{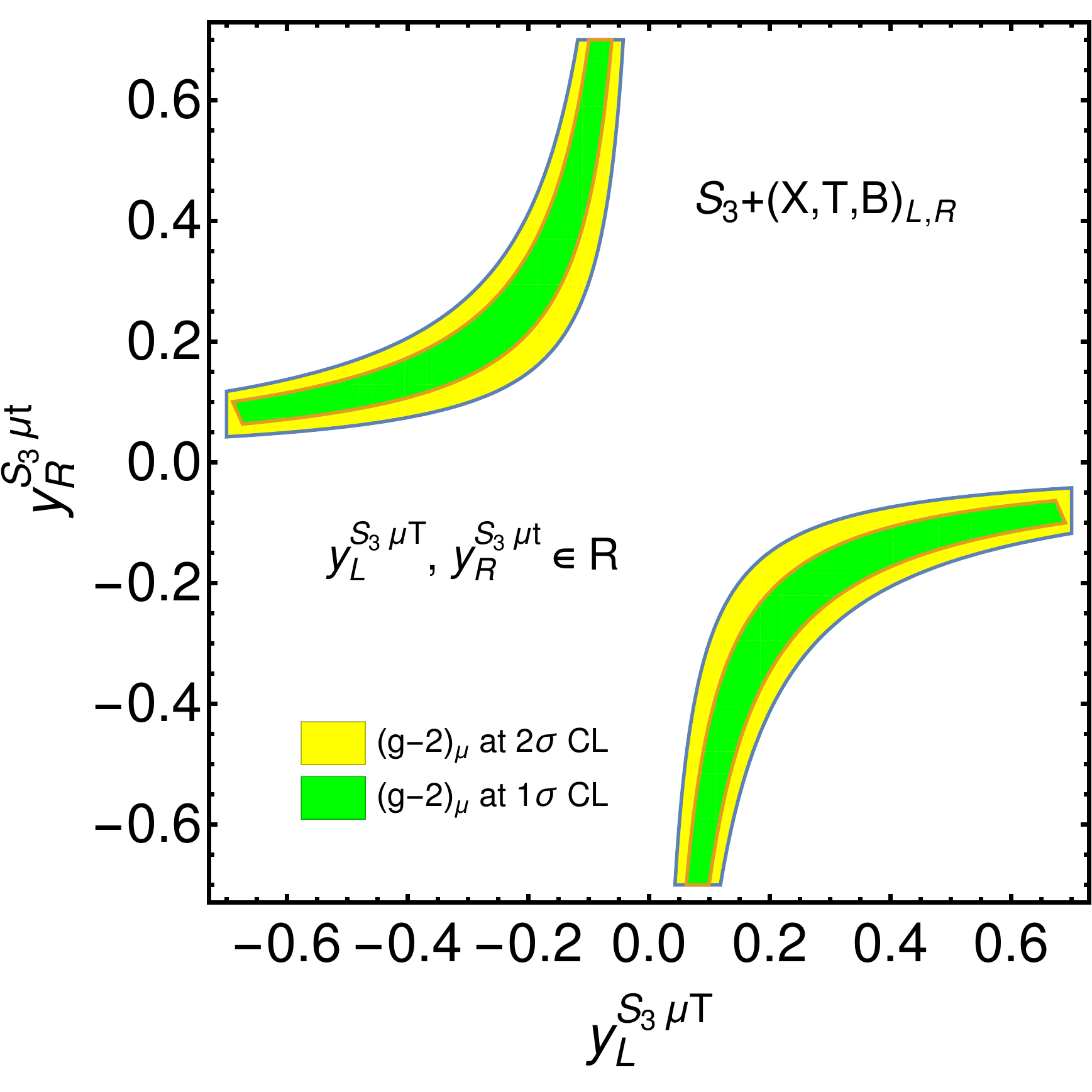}
\caption{The regions allowed by $(g-2)_{\mu}$ at $1\sigma$ (green area) and $2\sigma$ (yellow area) CL, respectively. For the $R_2+T_{L,R}/(T,B)_{L,R}$ models, we consider $y_{L}^{R_2\mu t}=y_{R}^{R_2\mu t}$ (upper left) and $y_{L}^{R_2\mu t}=-y_{R}^{R_2\mu t}$ (upper middle) cases in the plane of $y_{R/L}^{R_2\mu T}-y_{L/R}^{R_2\mu t}$. For the $R_2+(X,T)_{L,R}/(T,B,Y)_{L,R}$ models, we show the regions in the plane of $y_{L/R}^{R_2\mu t}$ and $y_{R/L}^{R_2\mu t}$ (upper right). For the $S_1+T_{L,R}/(T,B)_{L,R}$ models, we consider $y_{L}^{S_1\mu t}=y_{R}^{S_1\mu t}$ (central left) and $y_{L}^{S_1\mu t}=-y_{R}^{S_1\mu t}$ (central middle) cases in the plane of $y_{L/R}^{S_1\mu T}-y_{R/L}^{S_1\mu t}$. For the $S_1+(X,T)_{L,R}/(X,T,B)_{L,R}/(T,B,Y)_{L,R}$ models, we show the regions in the plane of $y_{R/L}^{S_1\mu t}$ and $y_{L/R}^{S_1\mu t}$ (central right). Here, the subscripts in front of the symbol "/" are for the singlet and triplet VLQs, and the subscripts after the symbol "/" are for the doublet VLQs. The three lower plots are for the $R_2+(X,T,B)_{L,R}$ (lower left with $y_{L}^{R_2\mu t}=y_{R}^{R_2\mu t}$ and lower middle with $y_{L}^{R_2\mu t}=-y_{R}^{R_2\mu t}$) and $S_3+(X,T,B)_{L,R}$ (lower right) models.}\label{fig:LQ+VLQ:g-2contour}
\end{center}
\end{figure}

In Fig. \ref{fig:LQ+VLQ:g-2contour}, we show the allowed regions from $(g-2)_{\mu}$ in the plane of $y_L$ and $y_R$. Here, all the LQ couplings are set as real for convenience. For the $R_2+T_{L,R}/(T,B)_{L,R}/(X,T,B)_{L,R}$ models, there are three LQ Yukawa parameters $y_{R/L}^{R_2\mu T},y_L^{R_2\mu t},y_R^{R_2\mu t}$. Then, we consider two cases $y_{L}^{R_2\mu t}=y_{R}^{R_2\mu t}$ and $y_{L}^{R_2\mu t}=-y_{R}^{R_2\mu t}$ to eliminate one parameter. For the $R_2+(X,T)_{L,R}/(T,B,Y)_{L,R}$ models, there are two LQ Yukawa parameters $y_L^{R_2\mu t}$ and $y_R^{R_2\mu t}$. For the $S_1+T_{L,R}/(T,B)_{L,R}$ models, there are three LQ Yukawa parameters $y_{L/R}^{S_1\mu T},y_L^{S_1\mu t},y_R^{S_1\mu t}$. Then, we also consider two cases $y_{L}^{S_1\mu t}=y_{R}^{S_1\mu t}$ and $y_{L}^{S_1\mu t}=-y_{R}^{S_1\mu t}$. For the $S_1+(X,T)_{L,R}/(X,T,B)_{L,R}/(T,B,Y)_{L,R}$ models, there are two LQ Yukawa parameters $y_L^{S_1\mu t}$ and $y_R^{S_1\mu t}$. For the $S_3+(X,T,B)_{L,R}$ model, there are two LQ Yukawa parameters $y_L^{S_3\mu T}$ and $y_R^{S_3\mu t}$. In these plots, we include the full contributions. Obviously, the $y_{L}^{R_2\mu t}=-y_{R}^{R_2\mu t}$ case is favored considering the perturbative unitarity for the $R_2+T_{L,R}/(T,B)_{L,R}/(X,T,B)_{L,R}$ models. For the $S_1+T_{L,R}/(T,B)_{L,R}$ models, the $y_{L}^{S_1\mu t}=y_{R}^{S_1\mu t}$ case is favored considering the perturbative unitarity.

\subsection{Numerical analysis in the one LQ and two VLQs extended models}
Because we have turned off the mixing with top quark in the two VLQ extended models, the constraints from EWPOs can be loose. The free parameters are $m_T,m_{T'},m_{\mr{LQ}},s_L^T,s_R^T,y_{L/R}^{\mr{LQ}\mu T_1},y_{L/R}^{\mr{LQ}\mu T_2}$ with LQ to be $R_2$ or $S_1$. Here, the $s_L^T,s_R^T$ are independent parameters. Then, we consider the two scenarios $s_L^T=0.05,s_R^T=0.1$ and $-s_L^T=s_R^T=0.1$. The VLQ and LQ masses are chosen as $m_T=m_{T'}=1\mr{TeV}$ and $m_{\mr{LQ}}=2\mr{TeV}$ in this section. In the absence of $\mr{LQ}\mu t$ couplings, the muon anomaly can also be explained by the VLQ contributions. In Tab. \ref{tab:LQ+TTB:g-2num}, we give the $(g-2)_{\mu}$ constraints on $\mr{Re}[y_R^{R_2\mu T_1}(y_L^{R_2\mu T_2})^\ast]$ and $\mr{Re}[y_L^{S_1\mu T_1}(y_R^{S_1\mu T_2})^\ast]$ at $1\sigma$ and $2\sigma$ CL, respectively. Assuming $m_T\approx m_{T'}$ and considering $s_{L,R}^T\ll1$ in Eq. \eqref{eqn:LQ+TTB:g-2appR2} and  Eq. \eqref{eqn:LQ+TTB:g-2appS1}, the contributions to $(g-2)_{\mu}$ can be approximated as
\begin{align}
&\Delta a_{\mu}^{R_2+T_{1L,R}+(T_2,B_2)_{L,R}}\approx\frac{m_{\mu}m_T}{4\pi^2m_{R_2}^2}f_{LR}^{R_2}(m_T^2/m_{R_2}^2)\mr{Re}[y_R^{R_2\mu T_1}(y_L^{R_2\mu T_2})^\ast](s_R^T-s_L^T),\nonumber\\
&\Delta a_{\mu}^{S_1+T_{1L,R}+(T_2,B_2)_{L,R}}\approx\frac{m_{\mu}m_T}{4\pi^2m_{S_1}^2}f_{LR}^{S_1}(m_T^2/m_{S_1}^2)\mr{Re}[y_L^{S_1\mu T_1}(y_R^{S_1\mu T_2})^\ast](s_R^T-s_L^T),
\end{align}
which means the contributions are determined by the difference of the mixing angles.

\begin{table}[!h]
%\resizebox{168mm}{18mm}{
\begin{tabular}{c|c|c|c|c|c}
\hline
\multirow{2}{*}{$(m_T,m_{T'},m_{\mr{LQ}})/\mr{TeV}$} & \multirow{2}{*}{$(s_L^T,s_R^T)$} & \multicolumn{2}{c|}{$\mr{Re}[y_R^{R_2\mu T_1}(y_L^{R_2\mu T_2})^\ast]$} & \multicolumn{2}{c}{$\mr{Re}[y_L^{S_1\mu T_1}(y_R^{S_1\mu T_2})^\ast]$}\\
\cline{3-6}
& & $1\sigma$ & $2\sigma$ & $1\sigma$ & $2\sigma$ \\
\hline
$(1,1,2)$ & $(0.05,0.1)$ & $(-0.059,-0.037)$ & $(-0.07,-0.025)$ & $(0.082,0.133)$ & $(0.057,0.158)$ \\
\hline
$(1,1,2)$ & $(-0.1,0.1)$ & $(-0.015,-0.009)$ & $(-0.018,-0.006)$ & $(0.021,0.033)$ & $(0.014,0.04)$ \\
\hline
\end{tabular}%}
\caption{The constraints on $\mr{Re}[y_R^{R_2\mu T_1}(y_L^{R_2\mu T_2})^\ast]$ and $\mr{Re}[y_L^{S_1\mu T_1}(y_R^{S_1\mu T_2})^\ast]$ at $1\sigma$ and $2\sigma$ CL, respectively. Here, we consider the two scenarios $s_L^T=0.05,s_R^T=0.1$ and $-s_L^T=s_R^T=0.1$ in the $\mr{LQ}+T_{L,R}+(T,B)_{L,R}$ model. The VLQ and LQ masses are chosen as $m_T=m_{T'}=1\mr{TeV}$ and $m_{\mr{LQ}}=2\mr{TeV}$.} \label{tab:LQ+TTB:g-2num}
\end{table}

\begin{figure}[!htb]
\begin{center}
\includegraphics[scale=0.29]{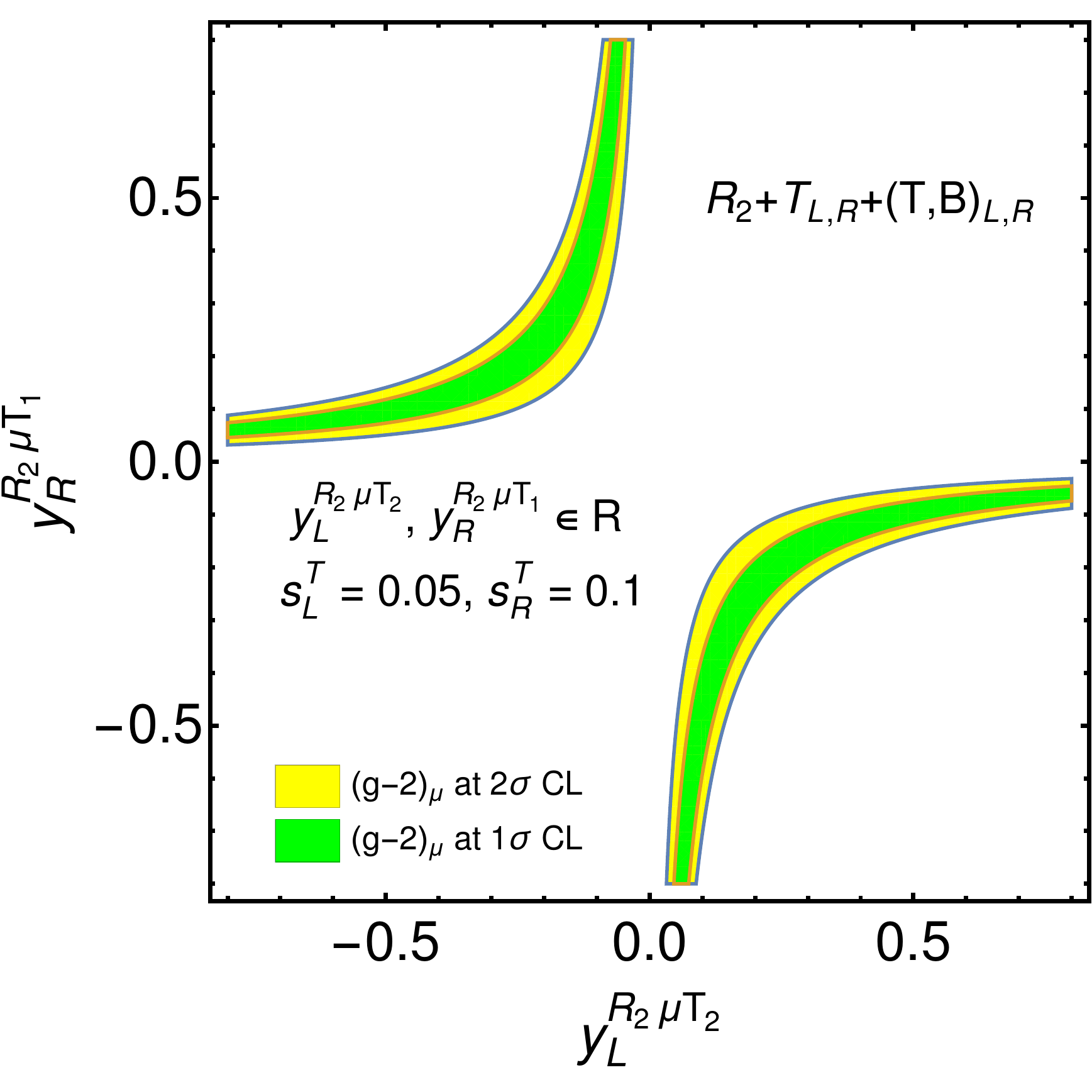}\qquad\qquad
\includegraphics[scale=0.29]{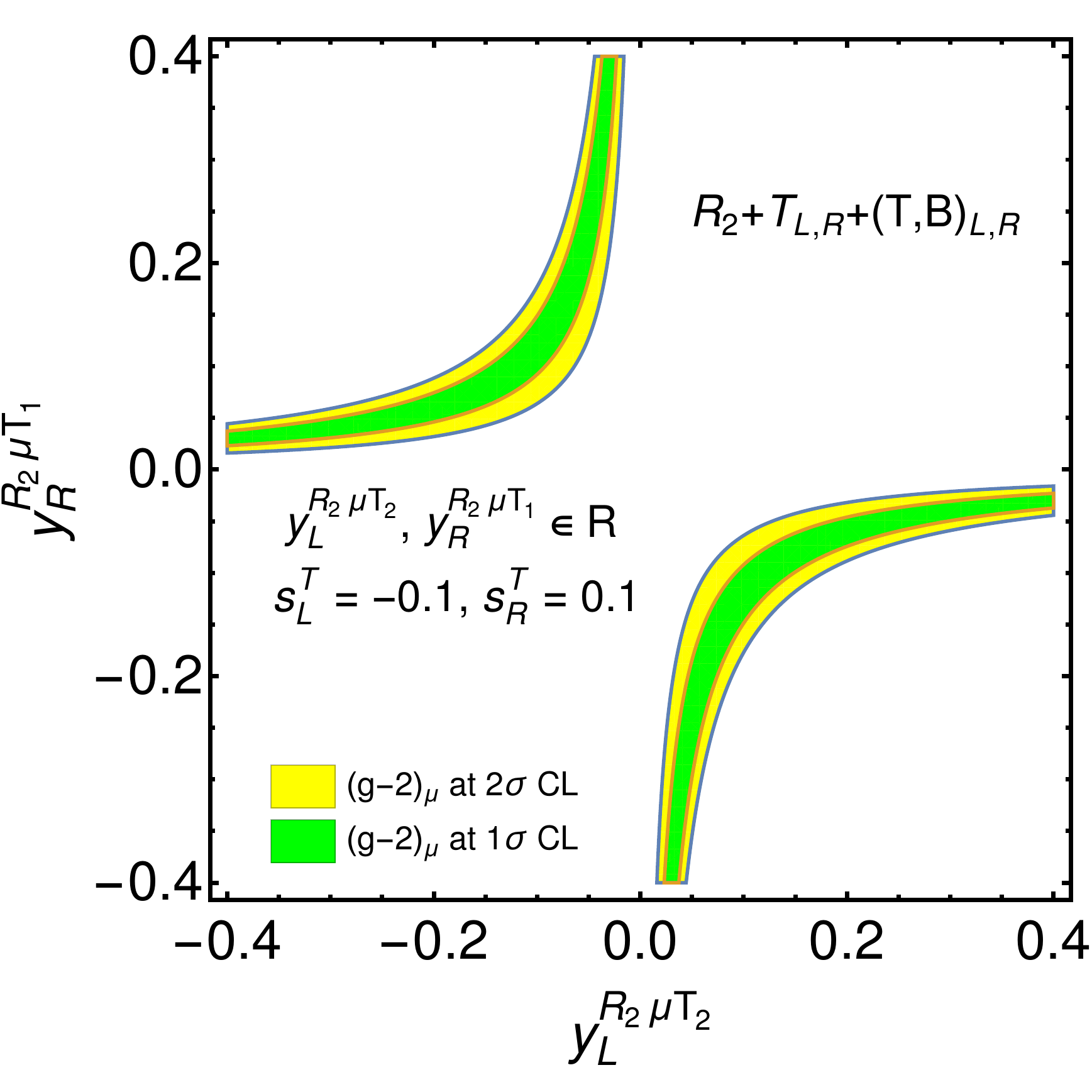}\\
\includegraphics[scale=0.29]{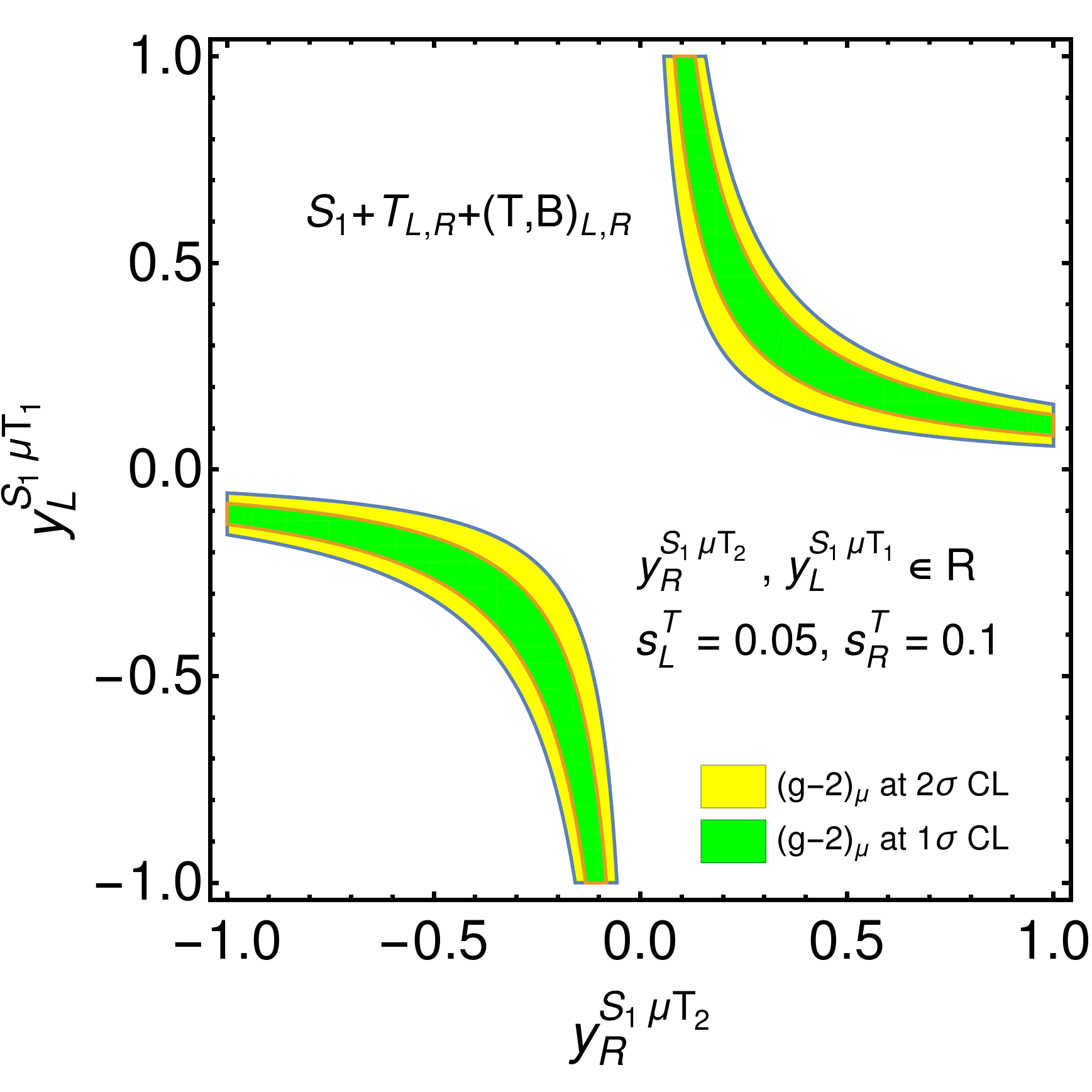}\qquad\qquad
\includegraphics[scale=0.29]{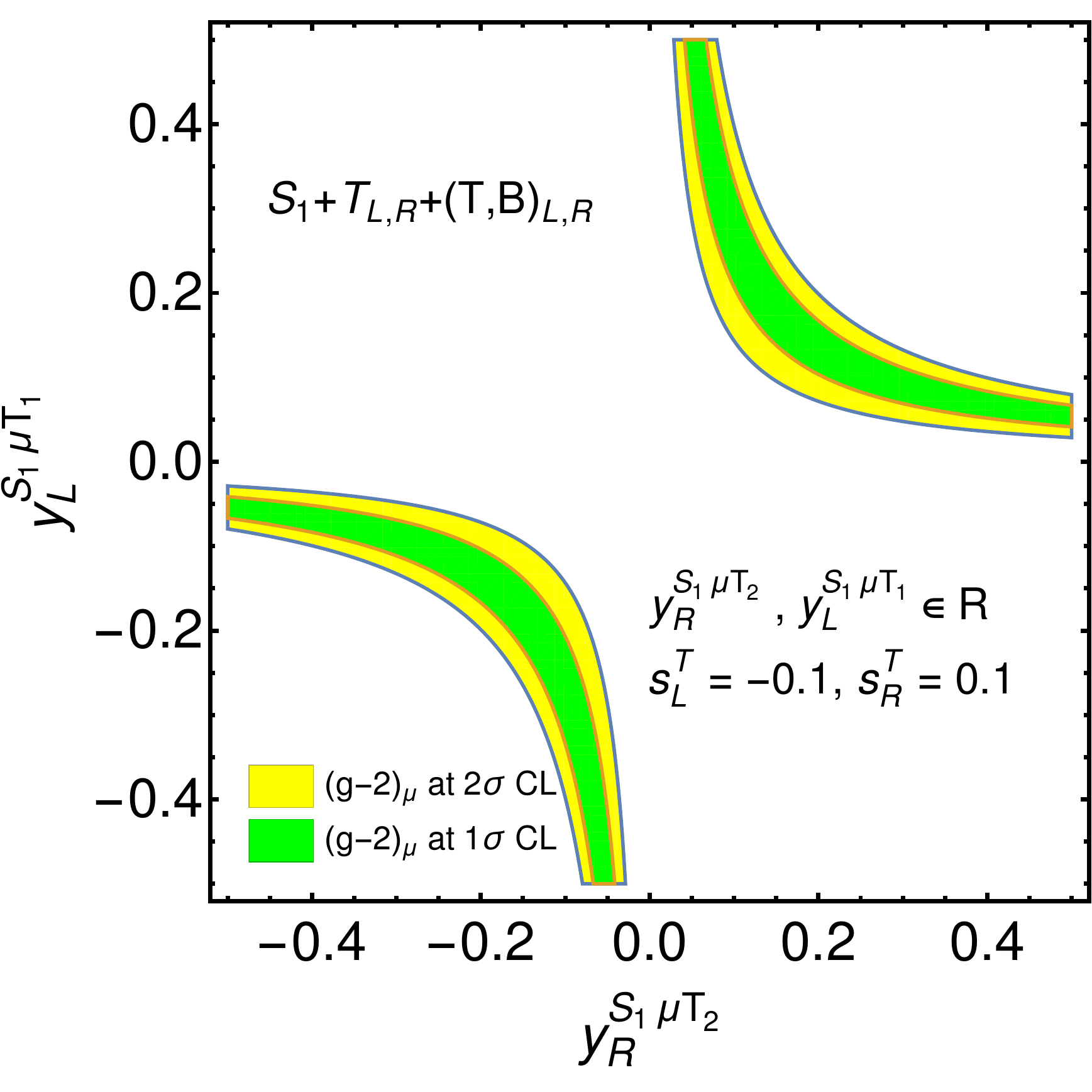}
\caption{For the $R_2+T_{L,R}+(T,B)_{L,R}$ model, we consider the two scenarios with $s_L^T=0.05,s_R^T=0.1$ (upper left) and $-s_L^T=s_R^T=0.1$ (upper right) in the plane of $y_L^{R_2\mu T_2}-y_R^{R_2\mu T_1}$. For the $S_1+T_{L,R}+(T,B)_{L,R}$ model, we also consider the two scenarios with $s_L^T=0.05,s_R^T=0.1$ (lower left) and $-s_L^T=s_R^T=0.1$ (lower right) in the plane of $y_R^{S_1\mu T_2}-y_L^{S_1\mu T_1}$.}\label{fig:LQ+TTB:g-2contour}
\end{center}
\end{figure}

In Fig. \ref{fig:LQ+TTB:g-2contour}, we show the allowed regions from $(g-2)_{\mu}$. Again, all the LQ Yukawa couplings are set as real for simplicity. For the $\mr{LQ}+T_{L,R}+(T,B)_{L,R}$ models, we consider two scenarios with different mixing angles. In these plots, we also include the full contributions rather than those only with chiral enhancements.

%%%%%%%%%%%%%%%%%%%%%%%%%%%%%%%%%%%%%%%%%%%%%%%%%%%%%%%%%%%%%%%%%%%%%
\section{LQ Phenomenology at hadron colliders}

\subsection{LQ decays}
The traditional LQ decay channels are SM quark and lepton, while they can possess the exotic decay channels in our models. Then, we can propose the new LQ search channels. The general LQ decay formulae are given in App. \ref{app:decay:Sf1f2}.   

\begin{figure}[!htb]
\begin{center}
\includegraphics[scale=0.29]{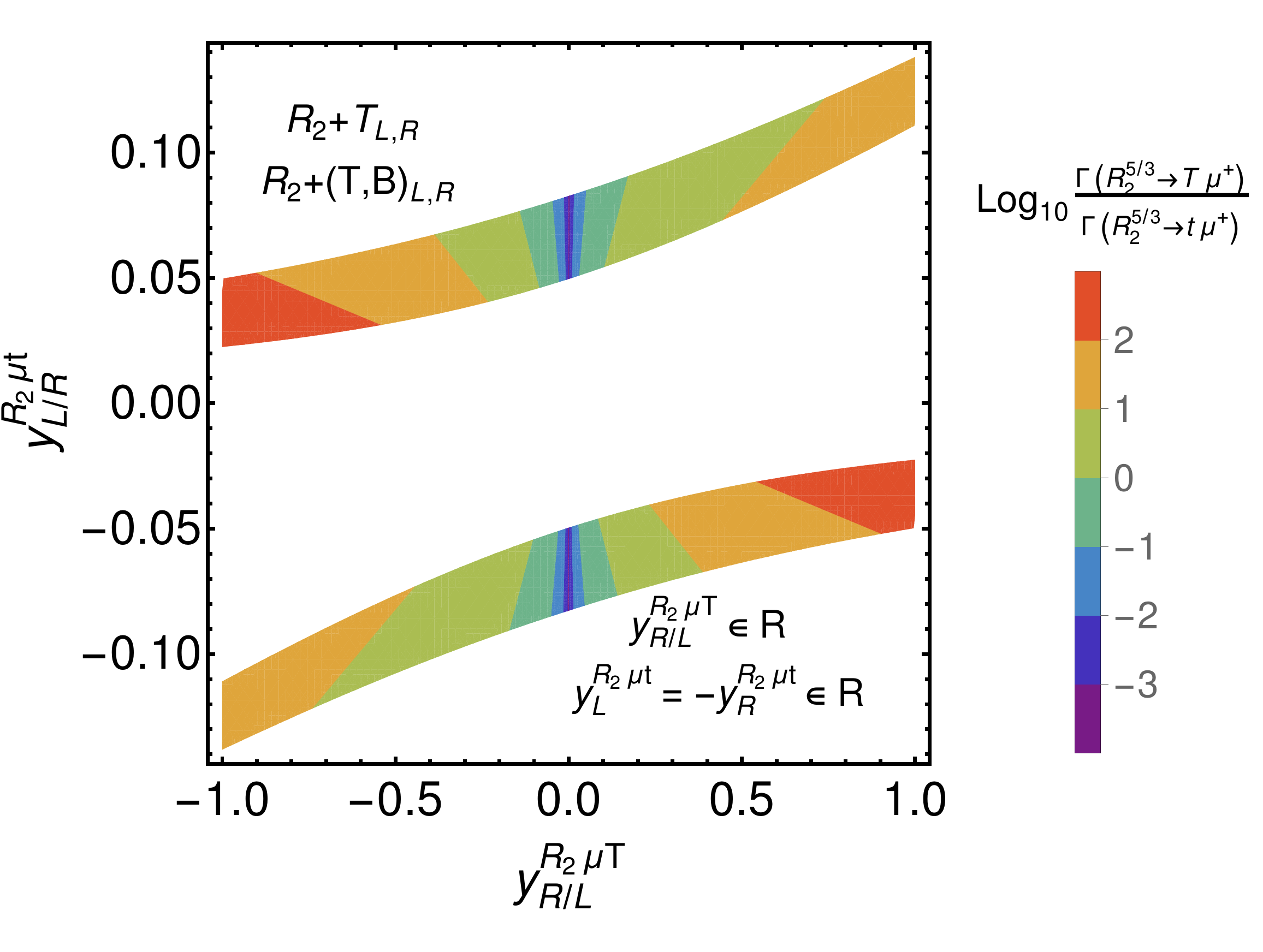}\qquad\qquad
\includegraphics[scale=0.29]{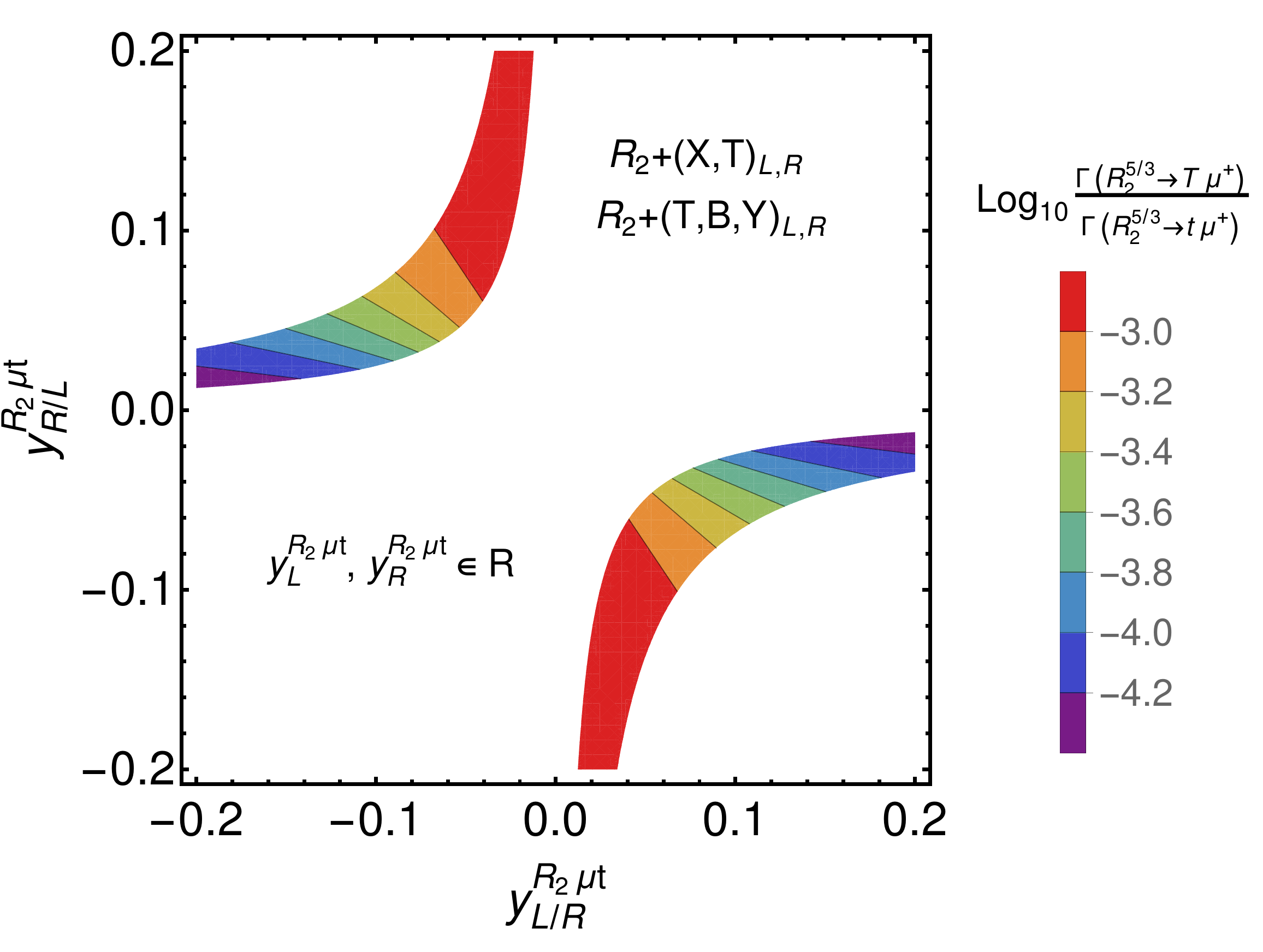}\\
\includegraphics[scale=0.29]{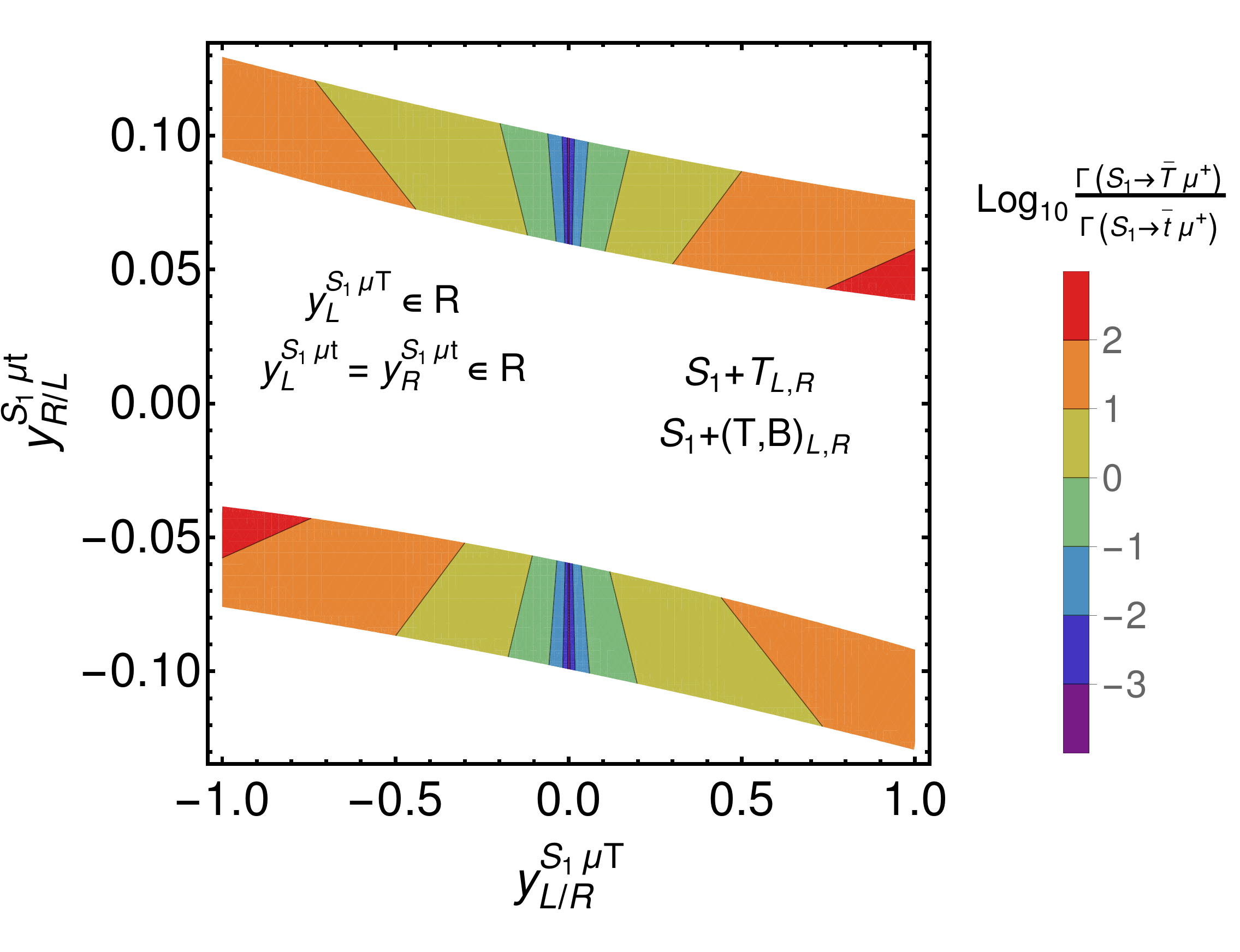}\qquad\qquad
\includegraphics[scale=0.29]{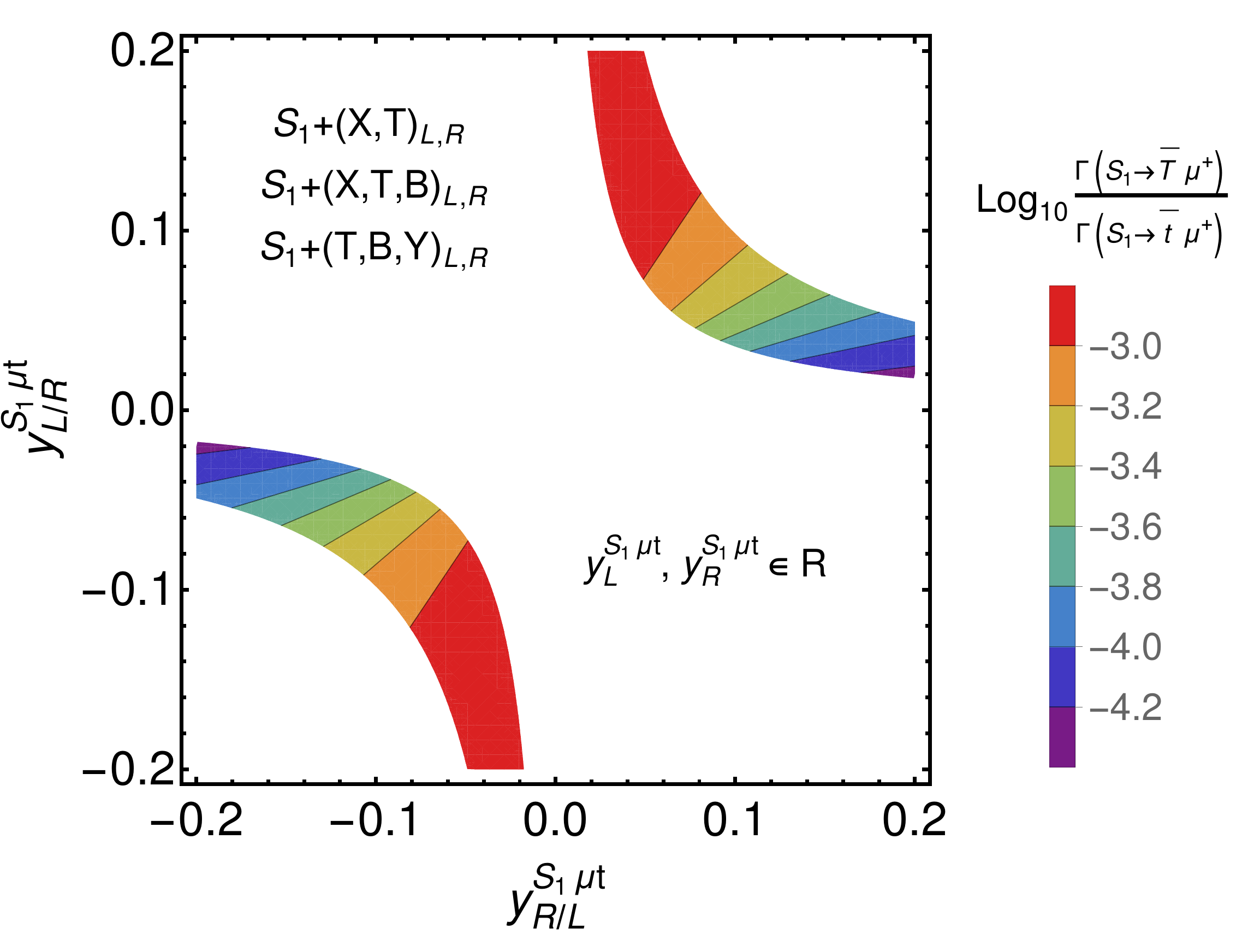}\\
\includegraphics[scale=0.29]{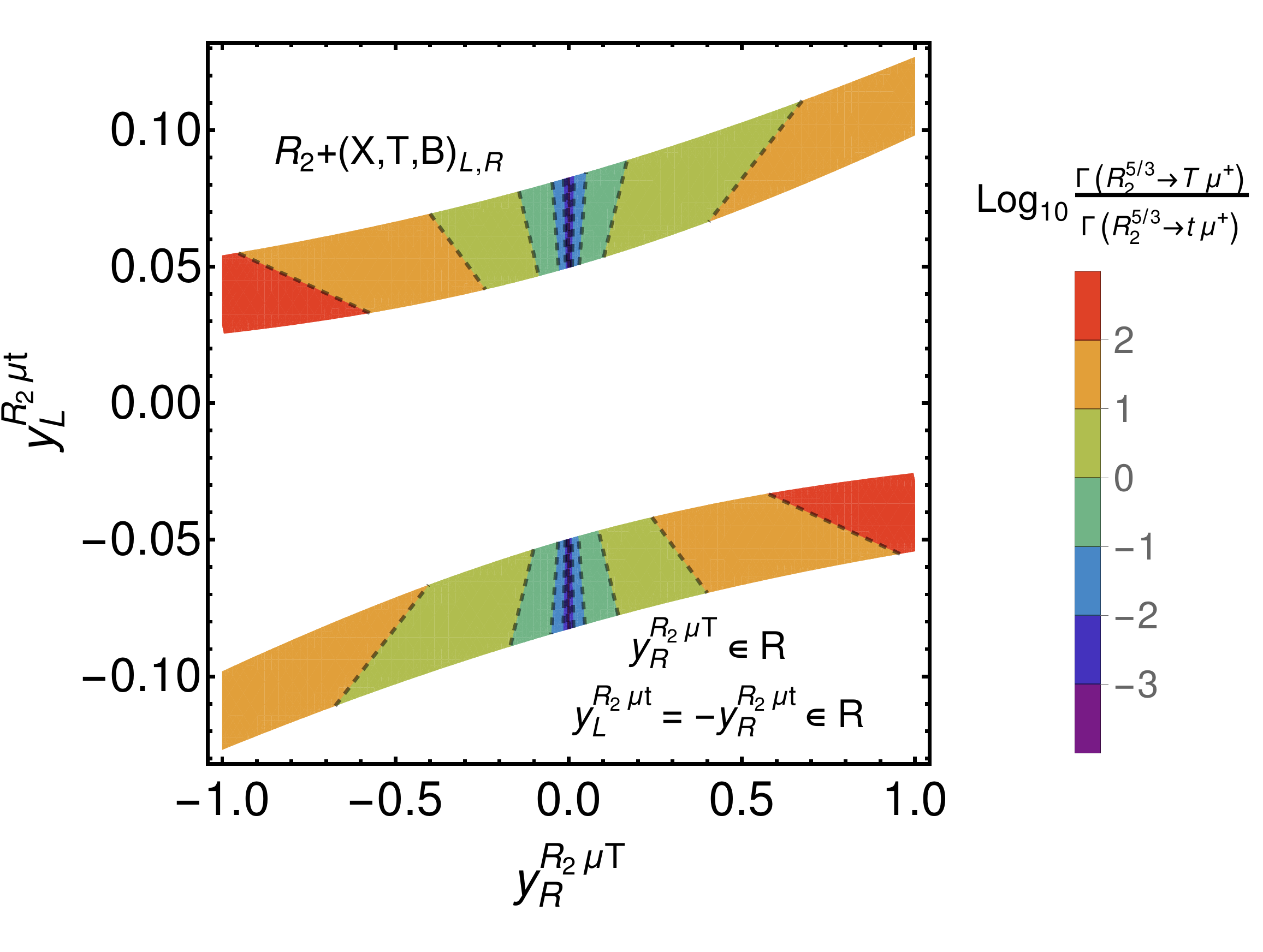}\qquad\qquad
\includegraphics[scale=0.29]{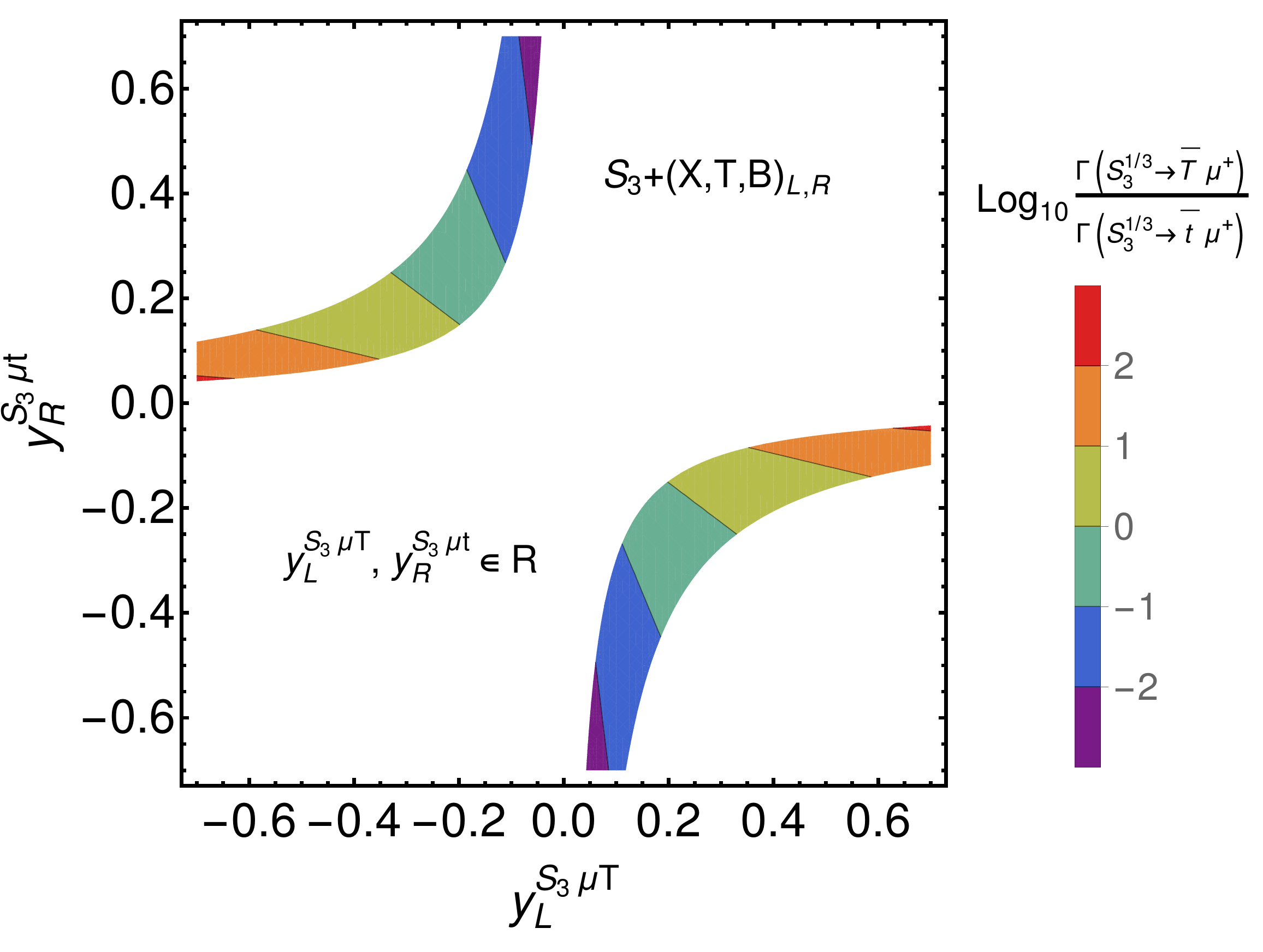}
\caption{The contour plots of $\log_{10}[\Gamma(\mr{LQ}\rightarrow T\mu)/\Gamma(\mr{LQ}\rightarrow t\mu)]$, where the colored regions are allowed by the $(g-2)_{\mu}$ at $2\sigma$ CL. Here, we consider the $R_2+T_{L,R}/(T,B)_{L,R}$ models with $y_{L}^{R_2\mu t}=-y_{R}^{R_2\mu t}$ (upper left), $R_2+(X,T)_{L,R}/(T,B,Y)_{L,R}$ models (upper right), $S_1+T_{L,R}/(T,B)_{L,R}$ models with $y_{L}^{S_1\mu t}=y_{R}^{S_1\mu t}$ (central left), $S_1+(X,T)_{L,R}/(X,T,B)_{L,R}/(T,B,Y)_{L,R}$ models (central right), $R_2+(X,T,B)_{L,R}$ model with $y_{L}^{R_2\mu t}=-y_{R}^{R_2\mu t}$ (lower left), and $S_3+(X,T,B)_{L,R}$ model (lower right).}\label{fig:LQ+VLQ:LQdecay}
\end{center}
\end{figure}

In the LQ+VLQ models, the final states can be $t\mu$ and $T\mu$ \footnote{For the $R_2^{2/3}$, its decay channels are $R_2^{2/3}\rightarrow b\mu^+$ in the $R_2+T_{L,R}/(X,T)_{L,R}$ models and $R_2^{2/3}\rightarrow b\mu^+/B\mu^+$ in the $R_2+(T,B)_{L,R}/(X,T,B)_{L,R}/(T,B,Y)_{L,R}$ models. In the $S_3+(X,T,B)_{L,R}$ model, the LQ decay channels can be $S_3^{4/3}\rightarrow \bar{b}\mu^+/\bar{B}\mu^+$ and $S_3^{-2/3}\rightarrow \bar{X}\mu^+$. While, we will not discuss them here.}. It is reasonable to ignore the masses of top quark and muon compared to VLQ and LQ masses. Assuming $y_{L,R}^{\mr{LQ}\mu T}$ and $y_{L,R}^{\mr{LQ}\mu t}$ to be the same order, we list the approximate formulae of width ratio $\Gamma(\mr{LQ}\rightarrow T\mu)/\Gamma(\mr{LQ}\rightarrow t\mu)$ in Tab. \ref{tab:LQ+VLQ:widthapp}. From the observation, we find that the $t\mu$ decay is dominated in the $R_2+(X,T)_{L,R}/(T,B,Y)_{L,R},S_1+(X,T)_{L,R}/(X,T,B)_{L,R}/(T,B,Y)_{L,R}$ models because of the mixing angle suppression. While, both $T\mu$ and $t\mu$ decay channels are important in the $R_2+T_{L,R}/(T,B)_{L,R}/(X,T,B)_{L,R},S_1+T_{L,R}/(T,B)_{L,R},S_3+(X,T,B)_{L,R}$ models, which can lead to interesting collider signals.
\begin{table}[!htb]
\begin{center}
\begin{tabular}{c|c|c|c}
\hline
LQ & VLQ & the approximate expressions of $\frac{\Gamma(\mr{LQ}\rightarrow T\mu)}{\Gamma(\mr{LQ}\rightarrow t\mu)}$ & suppress or not\\
\hline
\multirow{5}{*}{$R_2$} & $T_{L,R}$ & $(1-\frac{m_T^2}{m_{R_2}^2})^2|y_R^{R_2\mu T}|^2/(|y_L^{R_2\mu t}|^2+|y_R^{R_2\mu t}|^2)$ & No\\
\cline{2-4}
& $(X,T)_{L,R}$ & $(1-\frac{m_T^2}{m_{R_2}^2})^2|y_R^{R_2\mu t}|^2s_R^2/(|y_L^{R_2\mu t}|^2+|y_R^{R_2\mu t}|^2)$ & $s_R^2$\\
\cline{2-4}
& $(T,B)_{L,R}$ & $(1-\frac{m_T^2}{m_{R_2}^2})^2|y_L^{R_2\mu T}|^2/(|y_L^{R_2\mu t}|^2+|y_R^{R_2\mu t}|^2)$ & No\\
\cline{2-4}
& $(X,T,B)_{L,R}$ & $(1-\frac{m_T^2}{m_{R_2}^2})^2|y_R^{R_2\mu T}|^2/(|y_L^{R_2\mu t}|^2+|y_R^{R_2\mu t}|^2)$ & No\\
\cline{2-4}
& $(T,B,Y)_{L,R}$ & $(1-\frac{m_T^2}{m_{R_2}^2})^2|y_L^{R_2\mu t}|^2s_L^2/(|y_L^{R_2\mu t}|^2+|y_R^{R_2\mu t}|^2)$ & $s_L^2$\\
\hline\hline
\multirow{5}{*}{$S_1$} & $T_{L,R}$ & $(1-\frac{m_T^2}{m_{S_1}^2})^2|y_L^{S_1\mu T}|^2/(|y_L^{S_1\mu t}|^2+|y_R^{S_1\mu t}|^2)$ & No\\
\cline{2-4}
& $(X,T)_{L,R}$ & $(1-\frac{m_T^2}{m_{S_1}^2})^2|y_L^{S_1\mu t}|^2s_R^2/(|y_L^{S_1\mu t}|^2+|y_R^{S_1\mu t}|^2)$ & $s_R^2$\\
\cline{2-4}
& $(T,B)_{L,R}$ & $(1-\frac{m_T^2}{m_{S_1}^2})^2|y_R^{S_1\mu T}|^2/(|y_L^{S_1\mu t}|^2+|y_R^{S_1\mu t}|^2)$ & No\\
\cline{2-4}
& $(X,T,B)_{L,R}$ & $(1-\frac{m_T^2}{m_{S_1}^2})^2|y_R^{S_1\mu t}|^2s_L^2/(|y_L^{S_1\mu t}|^2+|y_R^{S_1\mu t}|^2)$ & $s_L^2$\\
\cline{2-4}
& $(T,B,Y)_{L,R}$ & $(1-\frac{m_T^2}{m_{S_1}^2})^2|y_R^{S_1\mu t}|^2s_L^2/(|y_L^{S_1\mu t}|^2+|y_R^{S_1\mu t}|^2)$ & $s_L^2$\\
\hline
$S_3$ & $(X,T,B)_{L,R}$ & $(1-\frac{m_T^2}{m_{S_3}^2})^2|y_L^{S_3\mu T}|^2/|y_R^{S_3\mu t}|^2$ & No\\ \hline
\end{tabular}%}
\caption{The approximate formulae of partial decay width ratio in the LQ+VLQ models for different representations (third column). In the fourth column, we show the $\Gamma(\mr{LQ}\rightarrow T\mu)$ partial decay width suppression factor compared to the $\Gamma(\mr{LQ}\rightarrow t\mu)$ ones.} \label{tab:LQ+VLQ:widthapp}
\end{center}
\end{table}
In Fig. \ref{fig:LQ+VLQ:LQdecay}, we show the contour plots of $\Gamma(\mr{LQ}\rightarrow T\mu)/\Gamma(\mr{LQ}\rightarrow t\mu)$ under the consideration of $(g-2)_{\mu}$ constraints. In these plots, we include the full contributions rather than those only with chiral enhancements. For the $R_2+T_{L,R}/(T,B)_{L,R}/(X,T,B)_{L,R}$ models, we consider the $y_{L}^{R_2\mu t}=-y_{R}^{R_2\mu t}$ case. For the $S_1+T_{L,R}/(T,B)_{L,R}$ models, we consider the $y_{L}^{S_1\mu t}=y_{R}^{S_1\mu t}$ case. Then, we find the $T\mu$ partial decay width can be comparable to and even larger than the $t\mu$ for some parameter space in these models.

\begin{figure}[!htb]
\begin{center}
\includegraphics[scale=0.29]{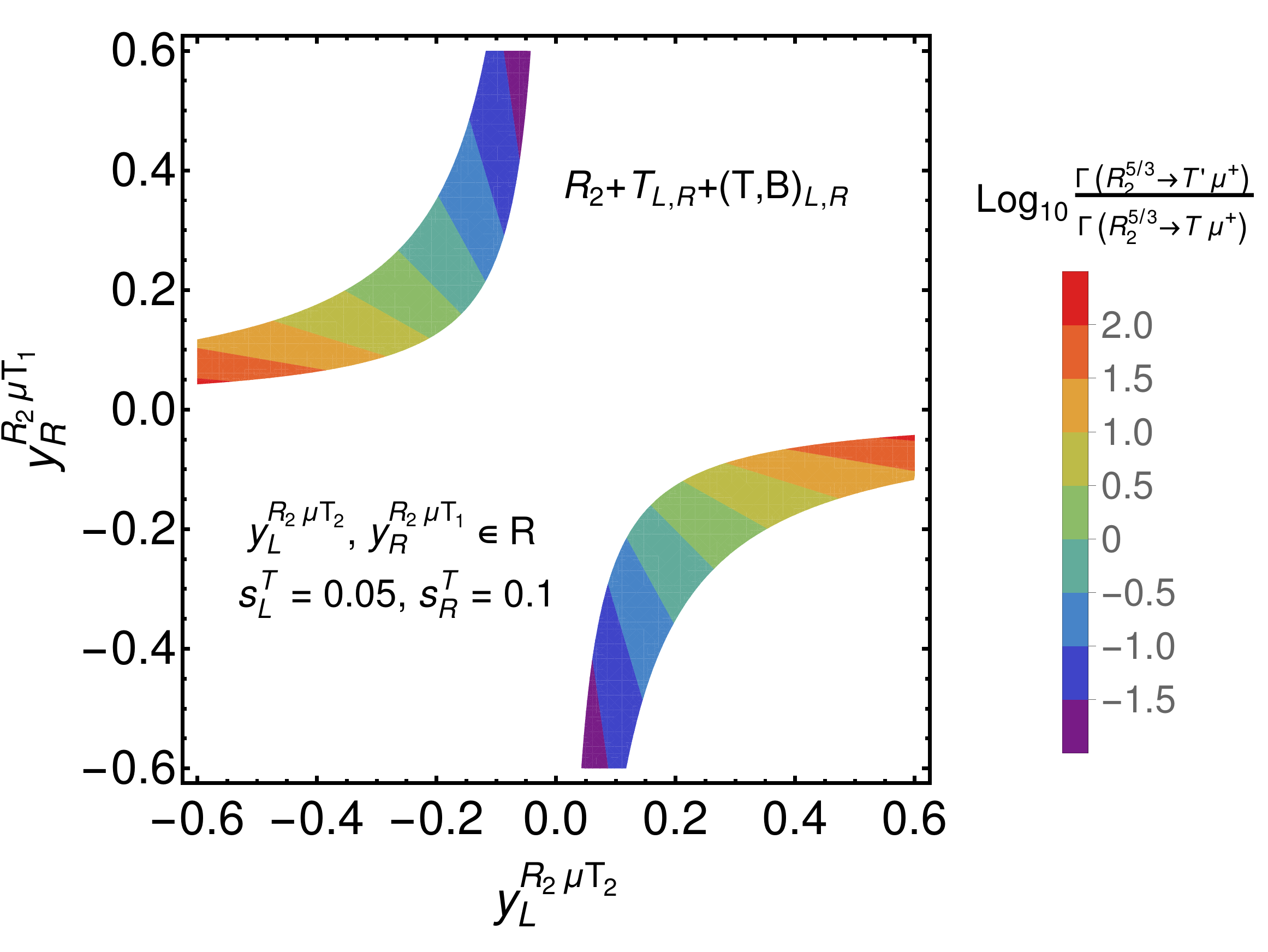}\qquad\qquad
\includegraphics[scale=0.29]{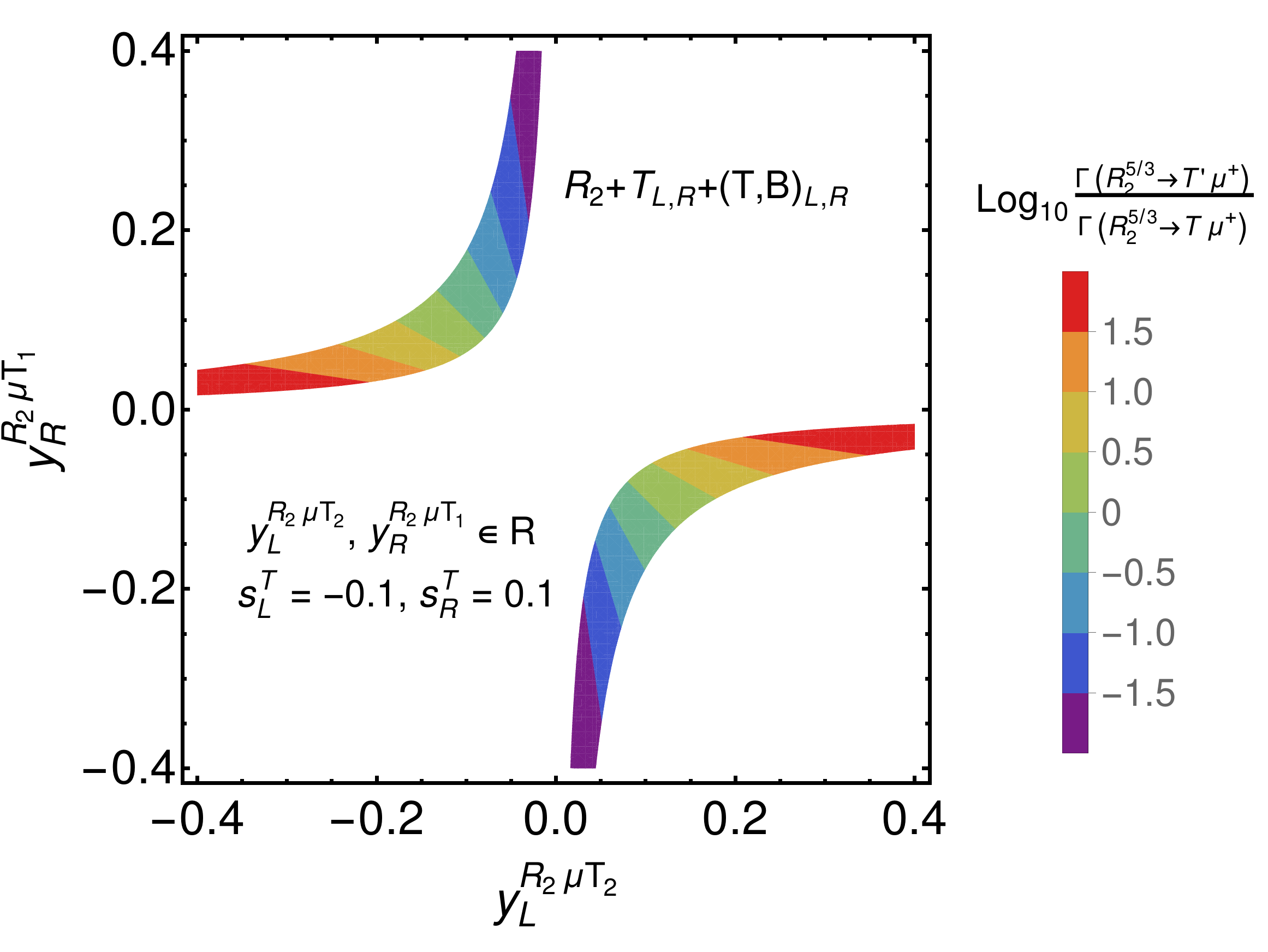}\\
\includegraphics[scale=0.29]{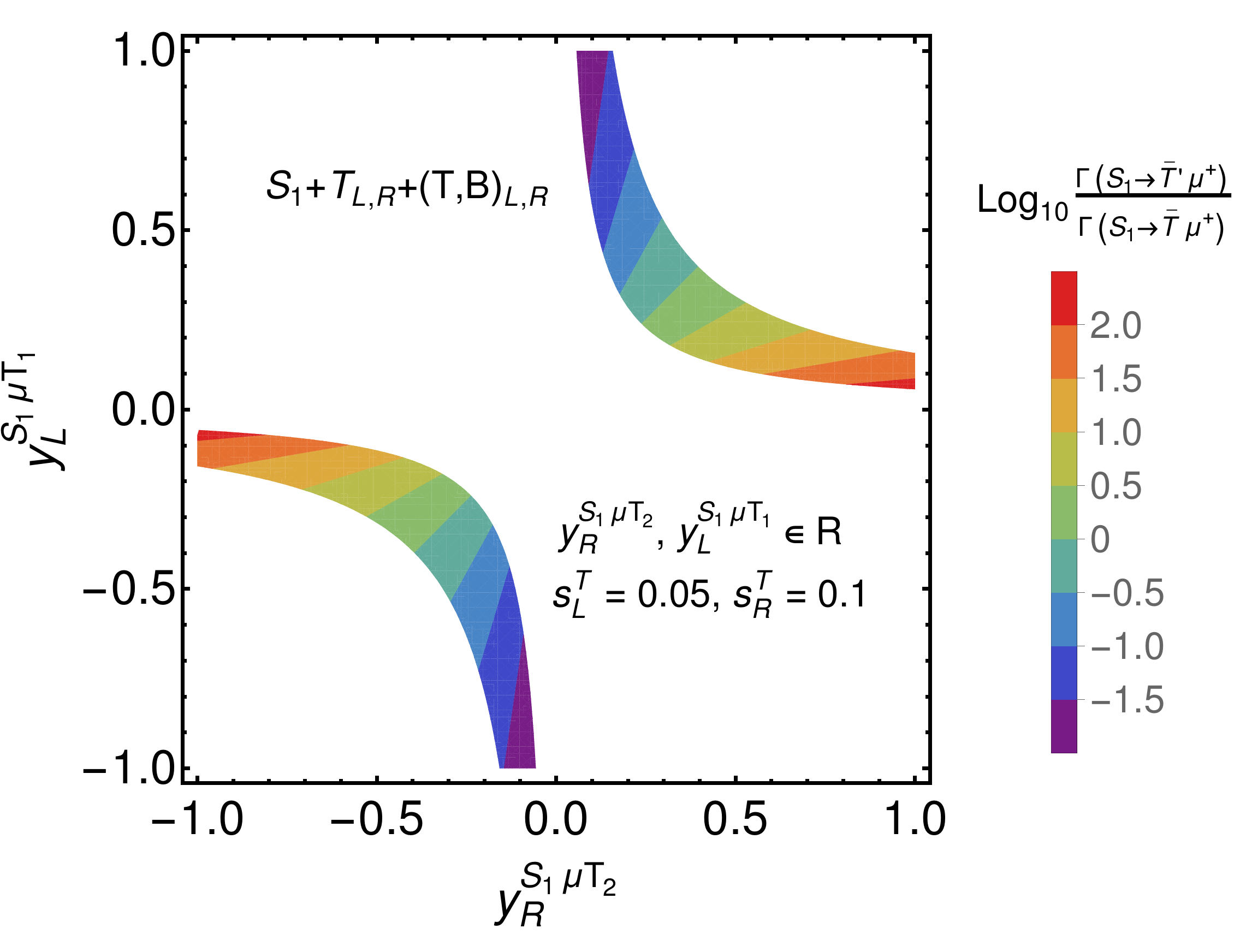}\qquad\qquad
\includegraphics[scale=0.29]{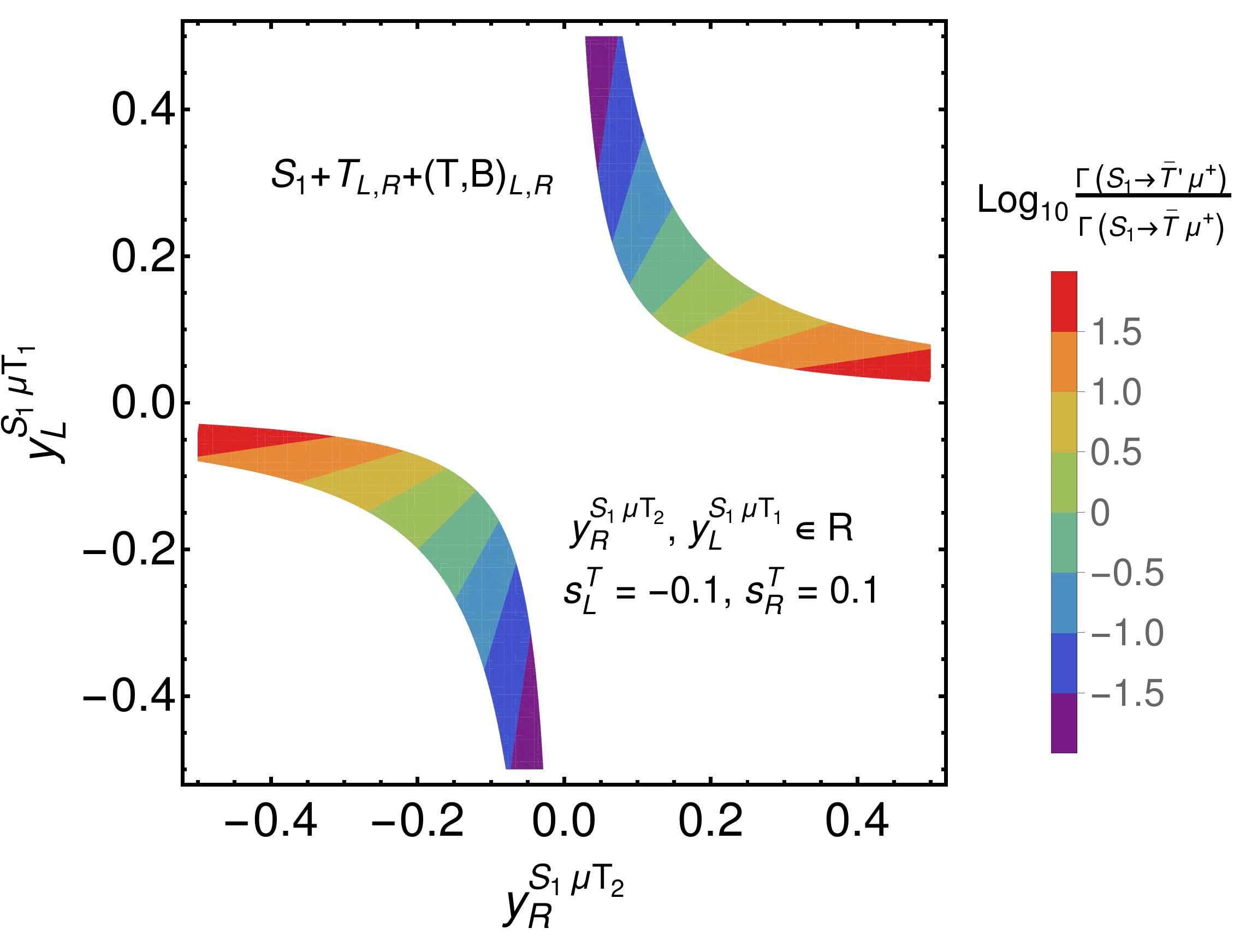}
\caption{The contour plots of $\log_{10}[\Gamma(\mr{LQ}\rightarrow T'\mu)/\Gamma(\mr{LQ}\rightarrow T\mu)]$, where the colored regions are allowed by the $(g-2)_{\mu}$ at $2\sigma$ CL. For the $R_2+T_{L,R}+(T,B)_{L,R}$ model, we consider the $s_L^T=0.05,s_R^T=0.1$ (upper left) and $-s_L^T=s_R^T=0.1$ (upper right) scenarios in the plane of $y_L^{R_2\mu T_2}-y_R^{R_2\mu T_1}$. For the $S_1+T_{L,R}+(T,B)_{L,R}$ model, we also consider the $s_L^T=0.05,s_R^T=0.1$ (lower left) and $-s_L^T=s_R^T=0.1$ (lower right) scenarios in the plane of $y_R^{S_1\mu T_2}-y_L^{S_1\mu T_1}$.}\label{fig:LQ+TTB:LQdecay}
\end{center}
\end{figure}

In the $\mr{LQ}+T_{L,R}+(T,B)_{L,R}$ model, we have turned off the $\mr{LQ}\mu t$ couplings. Thus, the main decay channels are $\mr{LQ}\rightarrow T\mu,T'\mu$ \footnote{For the $R_2^{2/3}$, its decay channel is $R_2^{2/3}\rightarrow B_2\mu^+$ in the $R_2+T_{L,R}+(T,B)_{L,R}$ model. While, we will also not discuss it here.}. According to the Eq. \eqref{eqn:R2+TTB:Lagmass}, Eq. \eqref{eqn:S1+TTB:Lagmass} and Eq. \eqref{eqn:decay:Sf1f2}, we have the following results:
\begin{align}
&\frac{\Gamma(R_2^{5/3}\rightarrow T'\mu^+)}{\Gamma(R_2^{5/3}\rightarrow T\mu^+)}\approx\frac{(m_{R_2}^2-m_{T'}^2)^2}{(m_{R_2}^2-m_T^2)^2}\cdot\frac{|y_L^{R_2\mu T_2}|^2}{|y_R^{R_2\mu T_1}|^2},\nonumber\\
&\frac{\Gamma(S_1\rightarrow \overline{T'}\mu^+)}{\Gamma(S_1\rightarrow \overline{T}\mu^+)}\approx\frac{(m_{S_1}^2-m_{T'}^2)^2}{(m_{S_1}^2-m_T^2)^2}\cdot\frac{|y_R^{S_1\mu T_2}|^2}{|y_L^{S_1\mu T_1}|^2}.
\end{align}
Again, we consider $y_{L/R}^{\mr{LQ}\mu T_1}$ and $y_{L/R}^{\mr{LQ}\mu T_2}$ to be the same order and $s_{L,R}^T\ll1$ in the above results. In Fig. \ref{fig:LQ+TTB:LQdecay}, we show the contour plots of $\Gamma(\mr{LQ}\rightarrow T'\mu)/\Gamma(\mr{LQ}\rightarrow T\mu)$. For the $R_2+T_{L,R}+(T,B)_{L,R}$ and $S_1+T_{L,R}+(T,B)_{L,R}$ models, we consider two scenarios $s_L^T=0.05,s_R^T=0.1$ and $-s_L^T=s_R^T=0.1$. In these plots, we adopt the full expressions of $(g-2)_{\mu}$ and LQ decay width rather than the approximate ones.
\subsection{LQ production at hadron colliders}
\begin{figure}[!htp]
\begin{center}
\includegraphics[scale=0.35]{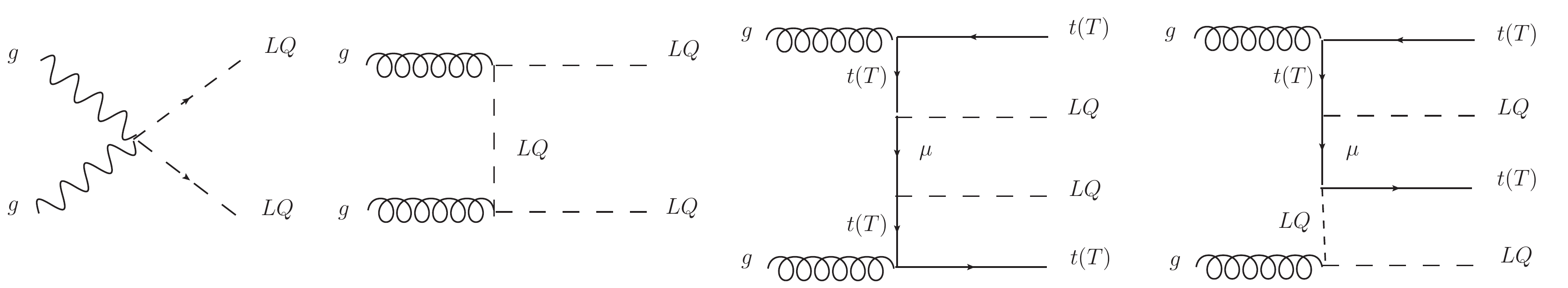}
\caption{The typical Feynman diagrams contributing to the pair production of LQs.}\label{fig:Feyn:pp2LQLQ}
\end{center}
\end{figure}

\begin{figure}[!htp]
\begin{center}
\includegraphics[scale=0.35]{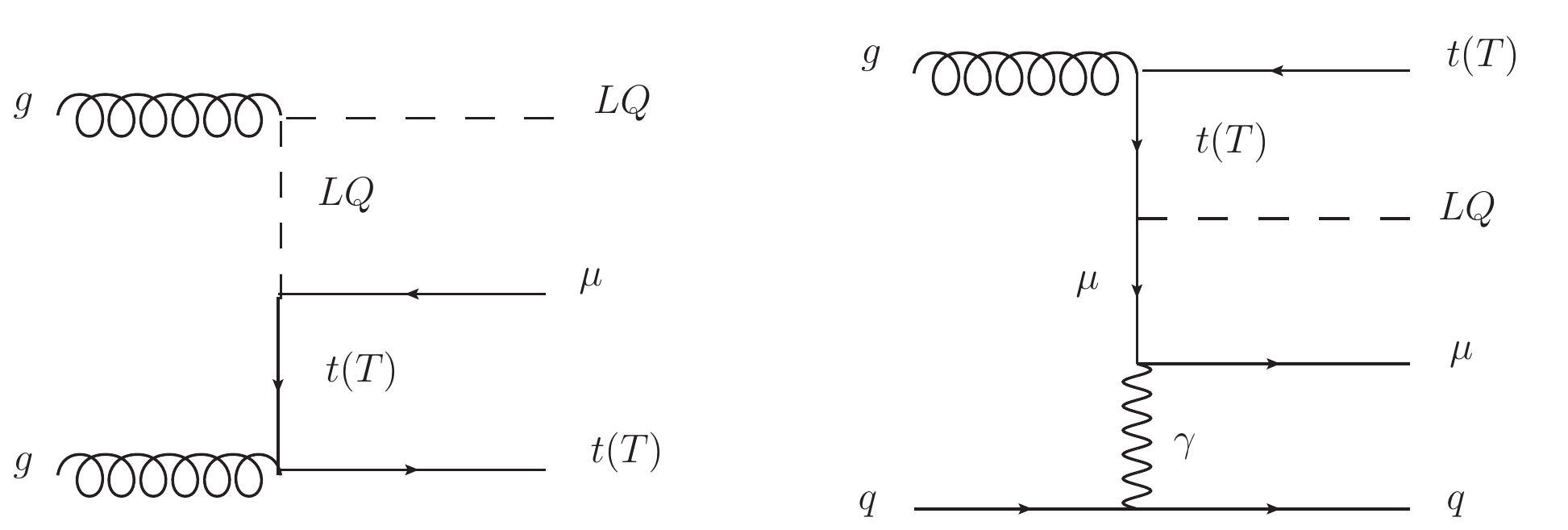}
\caption{The typical Feynman diagrams contributing to the single LQ production.}\label{fig:Feyn:pp2LQ}
\end{center}
\end{figure}

\begin{figure}[!htp]
\begin{center}
\includegraphics[scale=0.35]{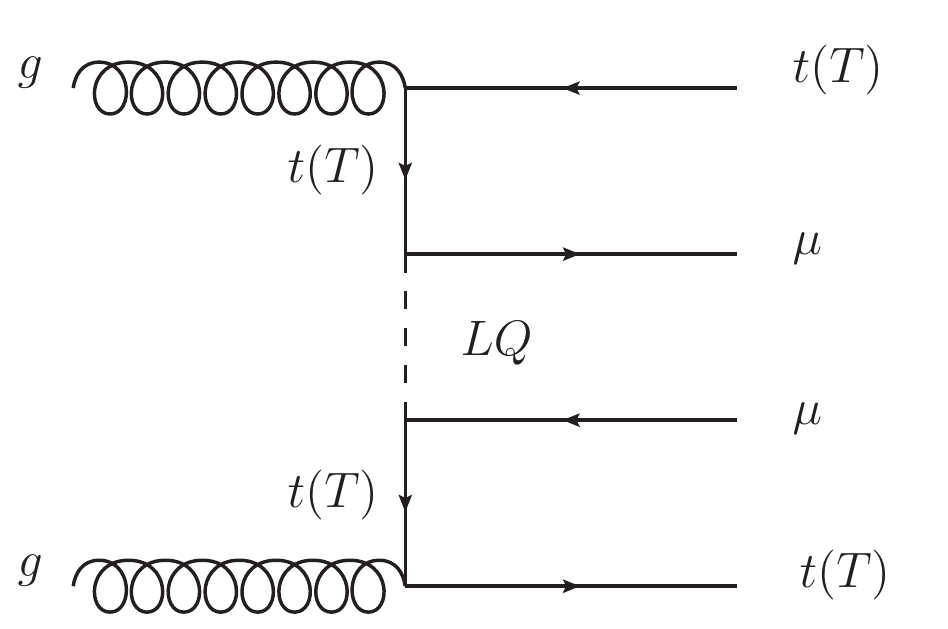}
\caption{The Feynman diagram with the virtual LQ.}\label{fig:Feyn:pp2OSLQ}
\end{center}
\end{figure}

Generally speaking, there are pair and single LQ production channels \cite{Djouadi:1989md, Blumlein:1996qp, Kramer:2004df, Diaz:2017lit, Dorsner:2018ynv, Buonocore:2020erb}. In Fig. \ref{fig:Feyn:pp2LQLQ} and Fig. \ref{fig:Feyn:pp2LQ}, we show the typical Feynman diagrams. Then, the LQ can decay into $t\mu$ or $T(T')\mu$, which has been discussed in the previous parts. Usually speaking, the $T(T')$ quark can further decay into the $bW,tZ,th$ final states. Thus, it will exhibit characteristic multi-top and multi-muon signals. In fact, we can distinguish different models through the LQ production and decay channels. Besides, there are also off-shell LQ channels \cite{Buchmuller:1986zs, Bansal:2018eha}. In Fig. \ref{fig:Feyn:pp2OSLQ}, we show the Feynman diagrams for $t(T)\bar{t}(\bar{T})\mu^+\mu^-$ associated production. In this section, we just show the possible collider signals, and the detailed phenomenology studies are beyond the scope of this work.

%%%%%%%%%%%%%%%%%%%%%%%%%%%%%%%%%%%%%%%%%%%%%%%%%%%%%%%%%%%%%%%%%%%%%
\section{Summary and conclusions}
We consider the models containing the LQ and VLQ simultaneously, which can be the solution to the $(g-2)_{\mu}$ anomaly because of the chiral enhancement. For the one LQ and one VLQ extended models, there are contributions from the $\mr{LQ}\mu t$ and $\mr{LQ}\mu T$ interactions. For the $R_2+(X,T)_{L,R}/(T,B,Y)_{L,R}$ and $S_1+(X,T)_{L,R}/(X,T,B)_{L,R}/(T,B,Y)_{L,R}$ models, the top quark contributions are dominated. For the $R_2+T_{L,R}/(T,B)_{L,R}/(X,T,B)_{L,R}$ and $S_1+T_{L,R}/(T,B)_{L,R}$ models, we need to consider both the top and $T$ quark contributions. In addition to the traditional $R_2$ and $S_1$ choices, we find the new $S_3$ solution to the anomaly in the presence of $(X,T,B)_{L,R}$ triplet, which is dominated by the $T$ and $B$ quark contributions. For the $\mr{LQ}+T_{L,R}+(T,B)_{L,R}$ models, the anomaly can be explained even in the absence of $\mr{LQ}\mu t$ interactions. Based on the constraints from $(g-2)_{\mu}$, we propose new LQ search channels. Besides the traditional $t\mu$ decay channel, the LQ can also decay into $T\mu$ final states. It will lead to the characteristic multi-top and multi-muon signals in these models, which can be tested at hadron colliders.\\

\textit{Note added:} After submitted to the arxiv, we have a discussion with Chen Zhang. Then, we notice there are already some works to explain the flavor anomalies in the LQ and VLQ extended models. In Ref. \cite{Bigaran:2019bqv}, the authors considered the VLQ representation $(3,2,-5/6)$. The VLQs in their model are mainly for the $B$ physics anomalies and neutrino mass, while the contributions to the $(g-2)_{\mu}$ are the same as those in the minimal LQ models. In Ref. \cite{Sahoo:2021vug}, the authors studied the VLQ with $(3,2,1/6)$ representation, where no chiral enhancements are involved because of the $Z_2$ symmetry.
%%%%%%%%%%%%%%%%%%%%%%%%%%%%%%%%%%%%%%%%%%%%%%%%%%%%%%%%%%%%%%%%%%%%%
\begin{acknowledgments}
We would like to thank Rui Zhang and Hiroshi Okada for helpful discussions. We also thank Juan Antonio Aguilar Saavedra for email discussion to clarify the $T$ quark convention difference in the triplet $(T,B,Y)_{L,R}$. This research was supported by an appointment to the Young Scientist Training Program at the APCTP through the Science and Technology Promotion Fund and Lottery Fund of the Korean Government. This was also supported by the Korean Local Governments-Gyeongsangbuk-do Province and Pohang City.
\end{acknowledgments}
%%%%%%%%%%%%%%%%%%%%%%%%%%%%%%%%%%%%%%%%%%%%%%%%%%%%%%%%%%%%%%%%%%%%%
%%%%%%%%%%%%%%%%%%%%%%%%%%%%%%%%%%%%%%%%%%%%%%%%%%%%%%%%%%%%%%%%%%%%%
\appendix
\section{The decay width formula}\label{app:decay:Sf1f2}
Let us consider the interaction $\mc{L}\supset S\overline{f_1}(y_L\omega_-+y_R\omega_+)f_2$, where the masses of $S,f_1,f_2$ are labelled as  $m_S,m_1,m_2$. If $m_S>m_1+m_2$, we can obtain the following decay width formula:
\begin{align}\label{eqn:decay:Sf1f2}
&\Gamma(S\rightarrow \overline{f_2}f_1)=\frac{1}{16\pi m_S}\sqrt{(1-\frac{m_1^2+m_2^2}{m_S^2})^2-\frac{4m_1^2m_2^2}{m_S^4}}~\cdot\nonumber\\
&\Big[(m_S^2-m_1^2-m_2^2)(|y_L|^2+|y_R|^2)-2m_1m_2\big(y_L(y_R)^{\ast}+y_R(y_L)^{\ast}\big)\Big].
\end{align}
If $m_S,m_1\gg m_2$, it can be approximated as
\begin{align}
&\Gamma(S\rightarrow \overline{f_2}f_1)\approx\frac{m_S}{16\pi}(1-\frac{m_1^2}{m_S^2})^2(|y_L|^2+|y_R|^2).
\end{align}
If $m_S\gg m_1,m_2$, it can be approximated as
\begin{align}
&\Gamma(S\rightarrow \overline{f_2}f_1)\approx\frac{m_S}{16\pi}(|y_L|^2+|y_R|^2).
\end{align}
Different from the $\mr{Re}[y_L(y_R)^{\ast}]$ dominated $(g-2)_{\mu}$ case, the main contributions are $(|y_L|^2+|y_R|^2)$ parts in the decay width formula.
%%%%%%%%%%%%%%%%%%%%%%%%%%%%%%%%%%%%%%%%%%%%%%%%%%%%%%%%%%%%%%%%%%%%%
\end{sloppypar}
\bibliography{muongm2-LQ-VLQ-v3}

\begin{thebibliography}{98}%
\makeatletter
\providecommand \@ifxundefined [1]{%
 \@ifx{#1\undefined}
}%
\providecommand \@ifnum [1]{%
 \ifnum #1\expandafter \@firstoftwo
 \else \expandafter \@secondoftwo
 \fi
}%
\providecommand \@ifx [1]{%
 \ifx #1\expandafter \@firstoftwo
 \else \expandafter \@secondoftwo
 \fi
}%
\providecommand \natexlab [1]{#1}%
\providecommand \enquote  [1]{``#1''}%
\providecommand \bibnamefont  [1]{#1}%
\providecommand \bibfnamefont [1]{#1}%
\providecommand \citenamefont [1]{#1}%
\providecommand \href@noop [0]{\@secondoftwo}%
\providecommand \href [0]{\begingroup \@sanitize@url \@href}%
\providecommand \@href[1]{\@@startlink{#1}\@@href}%
\providecommand \@@href[1]{\endgroup#1\@@endlink}%
\providecommand \@sanitize@url [0]{\catcode `\\12\catcode `\$12\catcode
  `\&12\catcode `\#12\catcode `\^12\catcode `\_12\catcode `\%12\relax}%
\providecommand \@@startlink[1]{}%
\providecommand \@@endlink[0]{}%
\providecommand \url  [0]{\begingroup\@sanitize@url \@url }%
\providecommand \@url [1]{\endgroup\@href {#1}{\urlprefix }}%
\providecommand \urlprefix  [0]{URL }%
\providecommand \Eprint [0]{\href }%
\providecommand \doibase [0]{https://doi.org/}%
\providecommand \selectlanguage [0]{\@gobble}%
\providecommand \bibinfo  [0]{\@secondoftwo}%
\providecommand \bibfield  [0]{\@secondoftwo}%
\providecommand \translation [1]{[#1]}%
\providecommand \BibitemOpen [0]{}%
\providecommand \bibitemStop [0]{}%
\providecommand \bibitemNoStop [0]{.\EOS\space}%
\providecommand \EOS [0]{\spacefactor3000\relax}%
\providecommand \BibitemShut  [1]{\csname bibitem#1\endcsname}%
\let\auto@bib@innerbib\@empty
%</preamble>
\bibitem [{\citenamefont {Jegerlehner}\ and\ \citenamefont
  {Nyffeler}(2009)}]{Jegerlehner:2009ry}%
  \BibitemOpen
  \bibfield  {author} {\bibinfo {author} {\bibfnamefont {F.}~\bibnamefont
  {Jegerlehner}}\ and\ \bibinfo {author} {\bibfnamefont {A.}~\bibnamefont
  {Nyffeler}},\ }\bibfield  {title} {\bibinfo {title} {{The Muon g-2}},\ }\href
  {https://doi.org/10.1016/j.physrep.2009.04.003} {\bibfield  {journal}
  {\bibinfo  {journal} {Phys. Rept.}\ }\textbf {\bibinfo {volume} {477}},\
  \bibinfo {pages} {1} (\bibinfo {year} {2009})},\ \Eprint
  {https://arxiv.org/abs/0902.3360} {arXiv:0902.3360 [hep-ph]} \BibitemShut
  {NoStop}%
%%CITATION = ARXIV:0902.3360;%%
\bibitem [{\citenamefont {Jegerlehner}(2017)}]{Jegerlehner:2017gek}%
  \BibitemOpen
  \bibfield  {author} {\bibinfo {author} {\bibfnamefont {F.}~\bibnamefont
  {Jegerlehner}},\ }\href {https://doi.org/10.1007/978-3-319-63577-4} {\emph
  {\bibinfo {title} {{The Anomalous Magnetic Moment of the Muon}}}},\ Vol.\
  \bibinfo {volume} {274}\ (\bibinfo  {publisher} {Springer},\ \bibinfo
  {address} {Cham},\ \bibinfo {year} {2017})\BibitemShut {NoStop}%
\bibitem [{\citenamefont {Bennett}\ \emph {et~al.}(2006)\citenamefont {Bennett}
  \emph {et~al.}}]{Muong-2:2006rrc}%
  \BibitemOpen
  \bibfield  {author} {\bibinfo {author} {\bibfnamefont {G.~W.}\ \bibnamefont
  {Bennett}} \emph {et~al.} (\bibinfo {collaboration} {Muon g-2}),\ }\bibfield
  {title} {\bibinfo {title} {{Final Report of the Muon E821 Anomalous Magnetic
  Moment Measurement at BNL}},\ }\href
  {https://doi.org/10.1103/PhysRevD.73.072003} {\bibfield  {journal} {\bibinfo
  {journal} {Phys. Rev. D}\ }\textbf {\bibinfo {volume} {73}},\ \bibinfo
  {pages} {072003} (\bibinfo {year} {2006})},\ \Eprint
  {https://arxiv.org/abs/hep-ex/0602035} {arXiv:hep-ex/0602035} \BibitemShut
  {NoStop}%
\bibitem [{\citenamefont {Abi}\ \emph {et~al.}(2021)\citenamefont {Abi} \emph
  {et~al.}}]{Muong-2:2021ojo}%
  \BibitemOpen
  \bibfield  {author} {\bibinfo {author} {\bibfnamefont {B.}~\bibnamefont
  {Abi}} \emph {et~al.} (\bibinfo {collaboration} {Muon g-2}),\ }\bibfield
  {title} {\bibinfo {title} {{Measurement of the Positive Muon Anomalous
  Magnetic Moment to 0.46 ppm}},\ }\href
  {https://doi.org/10.1103/PhysRevLett.126.141801} {\bibfield  {journal}
  {\bibinfo  {journal} {Phys. Rev. Lett.}\ }\textbf {\bibinfo {volume} {126}},\
  \bibinfo {pages} {141801} (\bibinfo {year} {2021})},\ \Eprint
  {https://arxiv.org/abs/2104.03281} {arXiv:2104.03281 [hep-ex]} \BibitemShut
  {NoStop}%
\bibitem [{\citenamefont {Aoyama}\ \emph {et~al.}(2012)\citenamefont {Aoyama},
  \citenamefont {Hayakawa}, \citenamefont {Kinoshita},\ and\ \citenamefont
  {Nio}}]{Aoyama:2012wk}%
  \BibitemOpen
  \bibfield  {author} {\bibinfo {author} {\bibfnamefont {T.}~\bibnamefont
  {Aoyama}}, \bibinfo {author} {\bibfnamefont {M.}~\bibnamefont {Hayakawa}},
  \bibinfo {author} {\bibfnamefont {T.}~\bibnamefont {Kinoshita}},\ and\
  \bibinfo {author} {\bibfnamefont {M.}~\bibnamefont {Nio}},\ }\bibfield
  {title} {\bibinfo {title} {{Complete Tenth-Order QED Contribution to the Muon
  g-2}},\ }\href {https://doi.org/10.1103/PhysRevLett.109.111808} {\bibfield
  {journal} {\bibinfo  {journal} {Phys. Rev. Lett.}\ }\textbf {\bibinfo
  {volume} {109}},\ \bibinfo {pages} {111808} (\bibinfo {year} {2012})},\
  \Eprint {https://arxiv.org/abs/1205.5370} {arXiv:1205.5370 [hep-ph]}
  \BibitemShut {NoStop}%
\bibitem [{\citenamefont {Aoyama}\ \emph {et~al.}(2019)\citenamefont {Aoyama},
  \citenamefont {Kinoshita},\ and\ \citenamefont {Nio}}]{Aoyama:2019ryr}%
  \BibitemOpen
  \bibfield  {author} {\bibinfo {author} {\bibfnamefont {T.}~\bibnamefont
  {Aoyama}}, \bibinfo {author} {\bibfnamefont {T.}~\bibnamefont {Kinoshita}},\
  and\ \bibinfo {author} {\bibfnamefont {M.}~\bibnamefont {Nio}},\ }\bibfield
  {title} {\bibinfo {title} {{Theory of the Anomalous Magnetic Moment of the
  Electron}},\ }\href {https://doi.org/10.3390/atoms7010028} {\bibfield
  {journal} {\bibinfo  {journal} {Atoms}\ }\textbf {\bibinfo {volume} {7}},\
  \bibinfo {pages} {28} (\bibinfo {year} {2019})}\BibitemShut {NoStop}%
\bibitem [{\citenamefont {Czarnecki}\ \emph {et~al.}(2003)\citenamefont
  {Czarnecki}, \citenamefont {Marciano},\ and\ \citenamefont
  {Vainshtein}}]{Czarnecki:2002nt}%
  \BibitemOpen
  \bibfield  {author} {\bibinfo {author} {\bibfnamefont {A.}~\bibnamefont
  {Czarnecki}}, \bibinfo {author} {\bibfnamefont {W.~J.}\ \bibnamefont
  {Marciano}},\ and\ \bibinfo {author} {\bibfnamefont {A.}~\bibnamefont
  {Vainshtein}},\ }\bibfield  {title} {\bibinfo {title} {{Refinements in
  electroweak contributions to the muon anomalous magnetic moment}},\ }\href
  {https://doi.org/10.1103/PhysRevD.67.073006} {\bibfield  {journal} {\bibinfo
  {journal} {Phys. Rev. D}\ }\textbf {\bibinfo {volume} {67}},\ \bibinfo
  {pages} {073006} (\bibinfo {year} {2003})},\ \bibinfo {note} {[Erratum:
  Phys.Rev.D 73, 119901 (2006)]},\ \Eprint
  {https://arxiv.org/abs/hep-ph/0212229} {arXiv:hep-ph/0212229} \BibitemShut
  {NoStop}%
\bibitem [{\citenamefont {Gnendiger}\ \emph {et~al.}(2013)\citenamefont
  {Gnendiger}, \citenamefont {St\"ockinger},\ and\ \citenamefont
  {St\"ockinger-Kim}}]{Gnendiger:2013pva}%
  \BibitemOpen
  \bibfield  {author} {\bibinfo {author} {\bibfnamefont {C.}~\bibnamefont
  {Gnendiger}}, \bibinfo {author} {\bibfnamefont {D.}~\bibnamefont
  {St\"ockinger}},\ and\ \bibinfo {author} {\bibfnamefont {H.}~\bibnamefont
  {St\"ockinger-Kim}},\ }\bibfield  {title} {\bibinfo {title} {{The electroweak
  contributions to $(g-2)_\mu$ after the Higgs boson mass measurement}},\
  }\href {https://doi.org/10.1103/PhysRevD.88.053005} {\bibfield  {journal}
  {\bibinfo  {journal} {Phys. Rev. D}\ }\textbf {\bibinfo {volume} {88}},\
  \bibinfo {pages} {053005} (\bibinfo {year} {2013})},\ \Eprint
  {https://arxiv.org/abs/1306.5546} {arXiv:1306.5546 [hep-ph]} \BibitemShut
  {NoStop}%
\bibitem [{\citenamefont {Davier}\ \emph {et~al.}(2017)\citenamefont {Davier},
  \citenamefont {Hoecker}, \citenamefont {Malaescu},\ and\ \citenamefont
  {Zhang}}]{Davier:2017zfy}%
  \BibitemOpen
  \bibfield  {author} {\bibinfo {author} {\bibfnamefont {M.}~\bibnamefont
  {Davier}}, \bibinfo {author} {\bibfnamefont {A.}~\bibnamefont {Hoecker}},
  \bibinfo {author} {\bibfnamefont {B.}~\bibnamefont {Malaescu}},\ and\
  \bibinfo {author} {\bibfnamefont {Z.}~\bibnamefont {Zhang}},\ }\bibfield
  {title} {\bibinfo {title} {{Reevaluation of the hadronic vacuum polarisation
  contributions to the Standard Model predictions of the muon $g-2$ and
  ${\alpha (m_Z^2)}$ using newest hadronic cross-section data}},\ }\href
  {https://doi.org/10.1140/epjc/s10052-017-5161-6} {\bibfield  {journal}
  {\bibinfo  {journal} {Eur. Phys. J. C}\ }\textbf {\bibinfo {volume} {77}},\
  \bibinfo {pages} {827} (\bibinfo {year} {2017})},\ \Eprint
  {https://arxiv.org/abs/1706.09436} {arXiv:1706.09436 [hep-ph]} \BibitemShut
  {NoStop}%
\bibitem [{\citenamefont {Keshavarzi}\ \emph {et~al.}(2018)\citenamefont
  {Keshavarzi}, \citenamefont {Nomura},\ and\ \citenamefont
  {Teubner}}]{Keshavarzi:2018mgv}%
  \BibitemOpen
  \bibfield  {author} {\bibinfo {author} {\bibfnamefont {A.}~\bibnamefont
  {Keshavarzi}}, \bibinfo {author} {\bibfnamefont {D.}~\bibnamefont {Nomura}},\
  and\ \bibinfo {author} {\bibfnamefont {T.}~\bibnamefont {Teubner}},\
  }\bibfield  {title} {\bibinfo {title} {{Muon $g-2$ and $\alpha(M_Z^2)$: a new
  data-based analysis}},\ }\href {https://doi.org/10.1103/PhysRevD.97.114025}
  {\bibfield  {journal} {\bibinfo  {journal} {Phys. Rev. D}\ }\textbf {\bibinfo
  {volume} {97}},\ \bibinfo {pages} {114025} (\bibinfo {year} {2018})},\
  \Eprint {https://arxiv.org/abs/1802.02995} {arXiv:1802.02995 [hep-ph]}
  \BibitemShut {NoStop}%
\bibitem [{\citenamefont {Colangelo}\ \emph {et~al.}(2019)\citenamefont
  {Colangelo}, \citenamefont {Hoferichter},\ and\ \citenamefont
  {Stoffer}}]{Colangelo:2018mtw}%
  \BibitemOpen
  \bibfield  {author} {\bibinfo {author} {\bibfnamefont {G.}~\bibnamefont
  {Colangelo}}, \bibinfo {author} {\bibfnamefont {M.}~\bibnamefont
  {Hoferichter}},\ and\ \bibinfo {author} {\bibfnamefont {P.}~\bibnamefont
  {Stoffer}},\ }\bibfield  {title} {\bibinfo {title} {{Two-pion contribution to
  hadronic vacuum polarization}},\ }\href
  {https://doi.org/10.1007/JHEP02(2019)006} {\bibfield  {journal} {\bibinfo
  {journal} {JHEP}\ }\textbf {\bibinfo {volume} {02}},\ \bibinfo {pages}
  {006}},\ \Eprint {https://arxiv.org/abs/1810.00007} {arXiv:1810.00007
  [hep-ph]} \BibitemShut {NoStop}%
\bibitem [{\citenamefont {Hoferichter}\ \emph {et~al.}(2019)\citenamefont
  {Hoferichter}, \citenamefont {Hoid},\ and\ \citenamefont
  {Kubis}}]{Hoferichter:2019mqg}%
  \BibitemOpen
  \bibfield  {author} {\bibinfo {author} {\bibfnamefont {M.}~\bibnamefont
  {Hoferichter}}, \bibinfo {author} {\bibfnamefont {B.-L.}\ \bibnamefont
  {Hoid}},\ and\ \bibinfo {author} {\bibfnamefont {B.}~\bibnamefont {Kubis}},\
  }\bibfield  {title} {\bibinfo {title} {{Three-pion contribution to hadronic
  vacuum polarization}},\ }\href {https://doi.org/10.1007/JHEP08(2019)137}
  {\bibfield  {journal} {\bibinfo  {journal} {JHEP}\ }\textbf {\bibinfo
  {volume} {08}},\ \bibinfo {pages} {137}},\ \Eprint
  {https://arxiv.org/abs/1907.01556} {arXiv:1907.01556 [hep-ph]} \BibitemShut
  {NoStop}%
\bibitem [{\citenamefont {Davier}\ \emph {et~al.}(2020)\citenamefont {Davier},
  \citenamefont {Hoecker}, \citenamefont {Malaescu},\ and\ \citenamefont
  {Zhang}}]{Davier:2019can}%
  \BibitemOpen
  \bibfield  {author} {\bibinfo {author} {\bibfnamefont {M.}~\bibnamefont
  {Davier}}, \bibinfo {author} {\bibfnamefont {A.}~\bibnamefont {Hoecker}},
  \bibinfo {author} {\bibfnamefont {B.}~\bibnamefont {Malaescu}},\ and\
  \bibinfo {author} {\bibfnamefont {Z.}~\bibnamefont {Zhang}},\ }\bibfield
  {title} {\bibinfo {title} {{A new evaluation of the hadronic vacuum
  polarisation contributions to the muon anomalous magnetic moment and to
  $\mathbf{\boldsymbol\alpha(m_Z^2)}$}},\ }\href
  {https://doi.org/10.1140/epjc/s10052-020-7792-2} {\bibfield  {journal}
  {\bibinfo  {journal} {Eur. Phys. J. C}\ }\textbf {\bibinfo {volume} {80}},\
  \bibinfo {pages} {241} (\bibinfo {year} {2020})},\ \bibinfo {note} {[Erratum:
  Eur.Phys.J.C 80, 410 (2020)]},\ \Eprint {https://arxiv.org/abs/1908.00921}
  {arXiv:1908.00921 [hep-ph]} \BibitemShut {NoStop}%
\bibitem [{\citenamefont {Keshavarzi}\ \emph {et~al.}(2020)\citenamefont
  {Keshavarzi}, \citenamefont {Nomura},\ and\ \citenamefont
  {Teubner}}]{Keshavarzi:2019abf}%
  \BibitemOpen
  \bibfield  {author} {\bibinfo {author} {\bibfnamefont {A.}~\bibnamefont
  {Keshavarzi}}, \bibinfo {author} {\bibfnamefont {D.}~\bibnamefont {Nomura}},\
  and\ \bibinfo {author} {\bibfnamefont {T.}~\bibnamefont {Teubner}},\
  }\bibfield  {title} {\bibinfo {title} {{$g-2$ of charged leptons, $\alpha
  (M^2_Z)$ , and the hyperfine splitting of muonium}},\ }\href
  {https://doi.org/10.1103/PhysRevD.101.014029} {\bibfield  {journal} {\bibinfo
   {journal} {Phys. Rev. D}\ }\textbf {\bibinfo {volume} {101}},\ \bibinfo
  {pages} {014029} (\bibinfo {year} {2020})},\ \Eprint
  {https://arxiv.org/abs/1911.00367} {arXiv:1911.00367 [hep-ph]} \BibitemShut
  {NoStop}%
\bibitem [{\citenamefont {Kurz}\ \emph {et~al.}(2014)\citenamefont {Kurz},
  \citenamefont {Liu}, \citenamefont {Marquard},\ and\ \citenamefont
  {Steinhauser}}]{Kurz:2014wya}%
  \BibitemOpen
  \bibfield  {author} {\bibinfo {author} {\bibfnamefont {A.}~\bibnamefont
  {Kurz}}, \bibinfo {author} {\bibfnamefont {T.}~\bibnamefont {Liu}}, \bibinfo
  {author} {\bibfnamefont {P.}~\bibnamefont {Marquard}},\ and\ \bibinfo
  {author} {\bibfnamefont {M.}~\bibnamefont {Steinhauser}},\ }\bibfield
  {title} {\bibinfo {title} {{Hadronic contribution to the muon anomalous
  magnetic moment to next-to-next-to-leading order}},\ }\href
  {https://doi.org/10.1016/j.physletb.2014.05.043} {\bibfield  {journal}
  {\bibinfo  {journal} {Phys. Lett. B}\ }\textbf {\bibinfo {volume} {734}},\
  \bibinfo {pages} {144} (\bibinfo {year} {2014})},\ \Eprint
  {https://arxiv.org/abs/1403.6400} {arXiv:1403.6400 [hep-ph]} \BibitemShut
  {NoStop}%
\bibitem [{\citenamefont {Melnikov}\ and\ \citenamefont
  {Vainshtein}(2004)}]{Melnikov:2003xd}%
  \BibitemOpen
  \bibfield  {author} {\bibinfo {author} {\bibfnamefont {K.}~\bibnamefont
  {Melnikov}}\ and\ \bibinfo {author} {\bibfnamefont {A.}~\bibnamefont
  {Vainshtein}},\ }\bibfield  {title} {\bibinfo {title} {{Hadronic
  light-by-light scattering contribution to the muon anomalous magnetic moment
  revisited}},\ }\href {https://doi.org/10.1103/PhysRevD.70.113006} {\bibfield
  {journal} {\bibinfo  {journal} {Phys. Rev. D}\ }\textbf {\bibinfo {volume}
  {70}},\ \bibinfo {pages} {113006} (\bibinfo {year} {2004})},\ \Eprint
  {https://arxiv.org/abs/hep-ph/0312226} {arXiv:hep-ph/0312226} \BibitemShut
  {NoStop}%
\bibitem [{\citenamefont {Masjuan}\ and\ \citenamefont
  {Sanchez-Puertas}(2017)}]{Masjuan:2017tvw}%
  \BibitemOpen
  \bibfield  {author} {\bibinfo {author} {\bibfnamefont {P.}~\bibnamefont
  {Masjuan}}\ and\ \bibinfo {author} {\bibfnamefont {P.}~\bibnamefont
  {Sanchez-Puertas}},\ }\bibfield  {title} {\bibinfo {title}
  {{Pseudoscalar-pole contribution to the $(g_{\mu}-2)$: a rational
  approach}},\ }\href {https://doi.org/10.1103/PhysRevD.95.054026} {\bibfield
  {journal} {\bibinfo  {journal} {Phys. Rev. D}\ }\textbf {\bibinfo {volume}
  {95}},\ \bibinfo {pages} {054026} (\bibinfo {year} {2017})},\ \Eprint
  {https://arxiv.org/abs/1701.05829} {arXiv:1701.05829 [hep-ph]} \BibitemShut
  {NoStop}%
\bibitem [{\citenamefont {Colangelo}\ \emph {et~al.}(2017)\citenamefont
  {Colangelo}, \citenamefont {Hoferichter}, \citenamefont {Procura},\ and\
  \citenamefont {Stoffer}}]{Colangelo:2017fiz}%
  \BibitemOpen
  \bibfield  {author} {\bibinfo {author} {\bibfnamefont {G.}~\bibnamefont
  {Colangelo}}, \bibinfo {author} {\bibfnamefont {M.}~\bibnamefont
  {Hoferichter}}, \bibinfo {author} {\bibfnamefont {M.}~\bibnamefont
  {Procura}},\ and\ \bibinfo {author} {\bibfnamefont {P.}~\bibnamefont
  {Stoffer}},\ }\bibfield  {title} {\bibinfo {title} {{Dispersion relation for
  hadronic light-by-light scattering: two-pion contributions}},\ }\href
  {https://doi.org/10.1007/JHEP04(2017)161} {\bibfield  {journal} {\bibinfo
  {journal} {JHEP}\ }\textbf {\bibinfo {volume} {04}},\ \bibinfo {pages}
  {161}},\ \Eprint {https://arxiv.org/abs/1702.07347} {arXiv:1702.07347
  [hep-ph]} \BibitemShut {NoStop}%
\bibitem [{\citenamefont {Hoferichter}\ \emph {et~al.}(2018)\citenamefont
  {Hoferichter}, \citenamefont {Hoid}, \citenamefont {Kubis}, \citenamefont
  {Leupold},\ and\ \citenamefont {Schneider}}]{Hoferichter:2018kwz}%
  \BibitemOpen
  \bibfield  {author} {\bibinfo {author} {\bibfnamefont {M.}~\bibnamefont
  {Hoferichter}}, \bibinfo {author} {\bibfnamefont {B.-L.}\ \bibnamefont
  {Hoid}}, \bibinfo {author} {\bibfnamefont {B.}~\bibnamefont {Kubis}},
  \bibinfo {author} {\bibfnamefont {S.}~\bibnamefont {Leupold}},\ and\ \bibinfo
  {author} {\bibfnamefont {S.~P.}\ \bibnamefont {Schneider}},\ }\bibfield
  {title} {\bibinfo {title} {{Dispersion relation for hadronic light-by-light
  scattering: pion pole}},\ }\href {https://doi.org/10.1007/JHEP10(2018)141}
  {\bibfield  {journal} {\bibinfo  {journal} {JHEP}\ }\textbf {\bibinfo
  {volume} {10}},\ \bibinfo {pages} {141}},\ \Eprint
  {https://arxiv.org/abs/1808.04823} {arXiv:1808.04823 [hep-ph]} \BibitemShut
  {NoStop}%
\bibitem [{\citenamefont {G\'erardin}\ \emph {et~al.}(2019)\citenamefont
  {G\'erardin}, \citenamefont {Meyer},\ and\ \citenamefont
  {Nyffeler}}]{Gerardin:2019vio}%
  \BibitemOpen
  \bibfield  {author} {\bibinfo {author} {\bibfnamefont {A.}~\bibnamefont
  {G\'erardin}}, \bibinfo {author} {\bibfnamefont {H.~B.}\ \bibnamefont
  {Meyer}},\ and\ \bibinfo {author} {\bibfnamefont {A.}~\bibnamefont
  {Nyffeler}},\ }\bibfield  {title} {\bibinfo {title} {{Lattice calculation of
  the pion transition form factor with $N_f=2+1$ Wilson quarks}},\ }\href
  {https://doi.org/10.1103/PhysRevD.100.034520} {\bibfield  {journal} {\bibinfo
   {journal} {Phys. Rev. D}\ }\textbf {\bibinfo {volume} {100}},\ \bibinfo
  {pages} {034520} (\bibinfo {year} {2019})},\ \Eprint
  {https://arxiv.org/abs/1903.09471} {arXiv:1903.09471 [hep-lat]} \BibitemShut
  {NoStop}%
\bibitem [{\citenamefont {Bijnens}\ \emph {et~al.}(2019)\citenamefont
  {Bijnens}, \citenamefont {Hermansson-Truedsson},\ and\ \citenamefont
  {Rodr\'\i{}guez-S\'anchez}}]{Bijnens:2019ghy}%
  \BibitemOpen
  \bibfield  {author} {\bibinfo {author} {\bibfnamefont {J.}~\bibnamefont
  {Bijnens}}, \bibinfo {author} {\bibfnamefont {N.}~\bibnamefont
  {Hermansson-Truedsson}},\ and\ \bibinfo {author} {\bibfnamefont
  {A.}~\bibnamefont {Rodr\'\i{}guez-S\'anchez}},\ }\bibfield  {title} {\bibinfo
  {title} {{Short-distance constraints for the HLbL contribution to the muon
  anomalous magnetic moment}},\ }\href
  {https://doi.org/10.1016/j.physletb.2019.134994} {\bibfield  {journal}
  {\bibinfo  {journal} {Phys. Lett. B}\ }\textbf {\bibinfo {volume} {798}},\
  \bibinfo {pages} {134994} (\bibinfo {year} {2019})},\ \Eprint
  {https://arxiv.org/abs/1908.03331} {arXiv:1908.03331 [hep-ph]} \BibitemShut
  {NoStop}%
\bibitem [{\citenamefont {Colangelo}\ \emph {et~al.}(2020)\citenamefont
  {Colangelo}, \citenamefont {Hagelstein}, \citenamefont {Hoferichter},
  \citenamefont {Laub},\ and\ \citenamefont {Stoffer}}]{Colangelo:2019uex}%
  \BibitemOpen
  \bibfield  {author} {\bibinfo {author} {\bibfnamefont {G.}~\bibnamefont
  {Colangelo}}, \bibinfo {author} {\bibfnamefont {F.}~\bibnamefont
  {Hagelstein}}, \bibinfo {author} {\bibfnamefont {M.}~\bibnamefont
  {Hoferichter}}, \bibinfo {author} {\bibfnamefont {L.}~\bibnamefont {Laub}},\
  and\ \bibinfo {author} {\bibfnamefont {P.}~\bibnamefont {Stoffer}},\
  }\bibfield  {title} {\bibinfo {title} {{Longitudinal short-distance
  constraints for the hadronic light-by-light contribution to $(g-2)_\mu$ with
  large-$N_c$ Regge models}},\ }\href {https://doi.org/10.1007/JHEP03(2020)101}
  {\bibfield  {journal} {\bibinfo  {journal} {JHEP}\ }\textbf {\bibinfo
  {volume} {03}},\ \bibinfo {pages} {101}},\ \Eprint
  {https://arxiv.org/abs/1910.13432} {arXiv:1910.13432 [hep-ph]} \BibitemShut
  {NoStop}%
\bibitem [{\citenamefont {Blum}\ \emph {et~al.}(2020)\citenamefont {Blum},
  \citenamefont {Christ}, \citenamefont {Hayakawa}, \citenamefont {Izubuchi},
  \citenamefont {Jin}, \citenamefont {Jung},\ and\ \citenamefont
  {Lehner}}]{Blum:2019ugy}%
  \BibitemOpen
  \bibfield  {author} {\bibinfo {author} {\bibfnamefont {T.}~\bibnamefont
  {Blum}}, \bibinfo {author} {\bibfnamefont {N.}~\bibnamefont {Christ}},
  \bibinfo {author} {\bibfnamefont {M.}~\bibnamefont {Hayakawa}}, \bibinfo
  {author} {\bibfnamefont {T.}~\bibnamefont {Izubuchi}}, \bibinfo {author}
  {\bibfnamefont {L.}~\bibnamefont {Jin}}, \bibinfo {author} {\bibfnamefont
  {C.}~\bibnamefont {Jung}},\ and\ \bibinfo {author} {\bibfnamefont
  {C.}~\bibnamefont {Lehner}},\ }\bibfield  {title} {\bibinfo {title}
  {{Hadronic Light-by-Light Scattering Contribution to the Muon Anomalous
  Magnetic Moment from Lattice QCD}},\ }\href
  {https://doi.org/10.1103/PhysRevLett.124.132002} {\bibfield  {journal}
  {\bibinfo  {journal} {Phys. Rev. Lett.}\ }\textbf {\bibinfo {volume} {124}},\
  \bibinfo {pages} {132002} (\bibinfo {year} {2020})},\ \Eprint
  {https://arxiv.org/abs/1911.08123} {arXiv:1911.08123 [hep-lat]} \BibitemShut
  {NoStop}%
\bibitem [{\citenamefont {Colangelo}\ \emph {et~al.}(2014)\citenamefont
  {Colangelo}, \citenamefont {Hoferichter}, \citenamefont {Nyffeler},
  \citenamefont {Passera},\ and\ \citenamefont {Stoffer}}]{Colangelo:2014qya}%
  \BibitemOpen
  \bibfield  {author} {\bibinfo {author} {\bibfnamefont {G.}~\bibnamefont
  {Colangelo}}, \bibinfo {author} {\bibfnamefont {M.}~\bibnamefont
  {Hoferichter}}, \bibinfo {author} {\bibfnamefont {A.}~\bibnamefont
  {Nyffeler}}, \bibinfo {author} {\bibfnamefont {M.}~\bibnamefont {Passera}},\
  and\ \bibinfo {author} {\bibfnamefont {P.}~\bibnamefont {Stoffer}},\
  }\bibfield  {title} {\bibinfo {title} {{Remarks on higher-order hadronic
  corrections to the muon g\ensuremath{-}2}},\ }\href
  {https://doi.org/10.1016/j.physletb.2014.06.012} {\bibfield  {journal}
  {\bibinfo  {journal} {Phys. Lett. B}\ }\textbf {\bibinfo {volume} {735}},\
  \bibinfo {pages} {90} (\bibinfo {year} {2014})},\ \Eprint
  {https://arxiv.org/abs/1403.7512} {arXiv:1403.7512 [hep-ph]} \BibitemShut
  {NoStop}%
\bibitem [{\citenamefont {Aoyama}\ \emph {et~al.}(2020)\citenamefont {Aoyama}
  \emph {et~al.}}]{Aoyama:2020ynm}%
  \BibitemOpen
  \bibfield  {author} {\bibinfo {author} {\bibfnamefont {T.}~\bibnamefont
  {Aoyama}} \emph {et~al.},\ }\bibfield  {title} {\bibinfo {title} {{The
  anomalous magnetic moment of the muon in the Standard Model}},\ }\href
  {https://doi.org/10.1016/j.physrep.2020.07.006} {\bibfield  {journal}
  {\bibinfo  {journal} {Phys. Rept.}\ }\textbf {\bibinfo {volume} {887}},\
  \bibinfo {pages} {1} (\bibinfo {year} {2020})},\ \Eprint
  {https://arxiv.org/abs/2006.04822} {arXiv:2006.04822 [hep-ph]} \BibitemShut
  {NoStop}%
\bibitem [{\citenamefont {Iinuma}(2011)}]{Iinuma:2011zz}%
  \BibitemOpen
  \bibfield  {author} {\bibinfo {author} {\bibfnamefont {H.}~\bibnamefont
  {Iinuma}} (\bibinfo {collaboration} {J-PARC muon g-2/EDM}),\ }\bibfield
  {title} {\bibinfo {title} {{New approach to the muon g-2 and EDM experiment
  at J-PARC}},\ }\href {https://doi.org/10.1088/1742-6596/295/1/012032}
  {\bibfield  {journal} {\bibinfo  {journal} {J. Phys. Conf. Ser.}\ }\textbf
  {\bibinfo {volume} {295}},\ \bibinfo {pages} {012032} (\bibinfo {year}
  {2011})}\BibitemShut {NoStop}%
\bibitem [{\citenamefont {Cowan}(2021)}]{Cowan:2021sdy}%
  \BibitemOpen
  \bibfield  {author} {\bibinfo {author} {\bibfnamefont {G.}~\bibnamefont
  {Cowan}},\ }\bibfield  {title} {\bibinfo {title} {{Effect of Systematic
  Uncertainty Estimation on the Muon $g-2$ Anomaly}},\ }\href@noop {} {\
  (\bibinfo {year} {2021})},\ \Eprint {https://arxiv.org/abs/2107.02652}
  {arXiv:2107.02652 [hep-ph]} \BibitemShut {NoStop}%
\bibitem [{\citenamefont {Czarnecki}\ and\ \citenamefont
  {Marciano}(2001)}]{Czarnecki:2001pv}%
  \BibitemOpen
  \bibfield  {author} {\bibinfo {author} {\bibfnamefont {A.}~\bibnamefont
  {Czarnecki}}\ and\ \bibinfo {author} {\bibfnamefont {W.~J.}\ \bibnamefont
  {Marciano}},\ }\bibfield  {title} {\bibinfo {title} {{The Muon anomalous
  magnetic moment: A Harbinger for 'new physics'}},\ }\href
  {https://doi.org/10.1103/PhysRevD.64.013014} {\bibfield  {journal} {\bibinfo
  {journal} {Phys. Rev. D}\ }\textbf {\bibinfo {volume} {64}},\ \bibinfo
  {pages} {013014} (\bibinfo {year} {2001})},\ \Eprint
  {https://arxiv.org/abs/hep-ph/0102122} {arXiv:hep-ph/0102122} \BibitemShut
  {NoStop}%
\bibitem [{\citenamefont {Queiroz}\ and\ \citenamefont
  {Shepherd}(2014)}]{Queiroz:2014zfa}%
  \BibitemOpen
  \bibfield  {author} {\bibinfo {author} {\bibfnamefont {F.~S.}\ \bibnamefont
  {Queiroz}}\ and\ \bibinfo {author} {\bibfnamefont {W.}~\bibnamefont
  {Shepherd}},\ }\bibfield  {title} {\bibinfo {title} {{New Physics
  Contributions to the Muon Anomalous Magnetic Moment: A Numerical Code}},\
  }\href {https://doi.org/10.1103/PhysRevD.89.095024} {\bibfield  {journal}
  {\bibinfo  {journal} {Phys. Rev. D}\ }\textbf {\bibinfo {volume} {89}},\
  \bibinfo {pages} {095024} (\bibinfo {year} {2014})},\ \Eprint
  {https://arxiv.org/abs/1403.2309} {arXiv:1403.2309 [hep-ph]} \BibitemShut
  {NoStop}%
\bibitem [{\citenamefont {Lindner}\ \emph {et~al.}(2018)\citenamefont
  {Lindner}, \citenamefont {Platscher},\ and\ \citenamefont
  {Queiroz}}]{Lindner:2016bgg}%
  \BibitemOpen
  \bibfield  {author} {\bibinfo {author} {\bibfnamefont {M.}~\bibnamefont
  {Lindner}}, \bibinfo {author} {\bibfnamefont {M.}~\bibnamefont {Platscher}},\
  and\ \bibinfo {author} {\bibfnamefont {F.~S.}\ \bibnamefont {Queiroz}},\
  }\bibfield  {title} {\bibinfo {title} {{A Call for New Physics : The Muon
  Anomalous Magnetic Moment and Lepton Flavor Violation}},\ }\href
  {https://doi.org/10.1016/j.physrep.2017.12.001} {\bibfield  {journal}
  {\bibinfo  {journal} {Phys. Rept.}\ }\textbf {\bibinfo {volume} {731}},\
  \bibinfo {pages} {1} (\bibinfo {year} {2018})},\ \Eprint
  {https://arxiv.org/abs/1610.06587} {arXiv:1610.06587 [hep-ph]} \BibitemShut
  {NoStop}%
%%CITATION = ARXIV:1610.06587;%%
\bibitem [{\citenamefont {Athron}\ \emph {et~al.}(2021)\citenamefont {Athron},
  \citenamefont {Bal\'azs}, \citenamefont {Jacob}, \citenamefont {Kotlarski},
  \citenamefont {St\"ockinger},\ and\ \citenamefont
  {St\"ockinger-Kim}}]{Athron:2021iuf}%
  \BibitemOpen
  \bibfield  {author} {\bibinfo {author} {\bibfnamefont {P.}~\bibnamefont
  {Athron}}, \bibinfo {author} {\bibfnamefont {C.}~\bibnamefont {Bal\'azs}},
  \bibinfo {author} {\bibfnamefont {D.~H.}\ \bibnamefont {Jacob}}, \bibinfo
  {author} {\bibfnamefont {W.}~\bibnamefont {Kotlarski}}, \bibinfo {author}
  {\bibfnamefont {D.}~\bibnamefont {St\"ockinger}},\ and\ \bibinfo {author}
  {\bibfnamefont {H.}~\bibnamefont {St\"ockinger-Kim}},\ }\bibfield  {title}
  {\bibinfo {title} {{New physics explanations of $a_\mu$ in light of the FNAL
  muon $g-2$ measurement}},\ }\href {https://doi.org/10.1007/JHEP09(2021)080}
  {\bibfield  {journal} {\bibinfo  {journal} {JHEP}\ }\textbf {\bibinfo
  {volume} {09}},\ \bibinfo {pages} {080}},\ \Eprint
  {https://arxiv.org/abs/2104.03691} {arXiv:2104.03691 [hep-ph]} \BibitemShut
  {NoStop}%
\bibitem [{\citenamefont {Moroi}(1996)}]{Moroi:1995yh}%
  \BibitemOpen
  \bibfield  {author} {\bibinfo {author} {\bibfnamefont {T.}~\bibnamefont
  {Moroi}},\ }\bibfield  {title} {\bibinfo {title} {{The Muon anomalous
  magnetic dipole moment in the minimal supersymmetric standard model}},\
  }\href {https://doi.org/10.1103/PhysRevD.53.6565} {\bibfield  {journal}
  {\bibinfo  {journal} {Phys. Rev. D}\ }\textbf {\bibinfo {volume} {53}},\
  \bibinfo {pages} {6565} (\bibinfo {year} {1996})},\ \bibinfo {note}
  {[Erratum: Phys.Rev.D 56, 4424 (1997)]},\ \Eprint
  {https://arxiv.org/abs/hep-ph/9512396} {arXiv:hep-ph/9512396} \BibitemShut
  {NoStop}%
\bibitem [{\citenamefont {Martin}\ and\ \citenamefont
  {Wells}(2001)}]{Martin:2001st}%
  \BibitemOpen
  \bibfield  {author} {\bibinfo {author} {\bibfnamefont {S.~P.}\ \bibnamefont
  {Martin}}\ and\ \bibinfo {author} {\bibfnamefont {J.~D.}\ \bibnamefont
  {Wells}},\ }\bibfield  {title} {\bibinfo {title} {{Muon Anomalous Magnetic
  Dipole Moment in Supersymmetric Theories}},\ }\href
  {https://doi.org/10.1103/PhysRevD.64.035003} {\bibfield  {journal} {\bibinfo
  {journal} {Phys. Rev. D}\ }\textbf {\bibinfo {volume} {64}},\ \bibinfo
  {pages} {035003} (\bibinfo {year} {2001})},\ \Eprint
  {https://arxiv.org/abs/hep-ph/0103067} {arXiv:hep-ph/0103067} \BibitemShut
  {NoStop}%
\bibitem [{\citenamefont {Heinemeyer}\ \emph {et~al.}(2004)\citenamefont
  {Heinemeyer}, \citenamefont {Stockinger},\ and\ \citenamefont
  {Weiglein}}]{Heinemeyer:2004yq}%
  \BibitemOpen
  \bibfield  {author} {\bibinfo {author} {\bibfnamefont {S.}~\bibnamefont
  {Heinemeyer}}, \bibinfo {author} {\bibfnamefont {D.}~\bibnamefont
  {Stockinger}},\ and\ \bibinfo {author} {\bibfnamefont {G.}~\bibnamefont
  {Weiglein}},\ }\bibfield  {title} {\bibinfo {title} {{Electroweak and
  supersymmetric two-loop corrections to $(g-2)_{\mu}$}},\ }\href
  {https://doi.org/10.1016/j.nuclphysb.2004.08.014} {\bibfield  {journal}
  {\bibinfo  {journal} {Nucl. Phys. B}\ }\textbf {\bibinfo {volume} {699}},\
  \bibinfo {pages} {103} (\bibinfo {year} {2004})},\ \Eprint
  {https://arxiv.org/abs/hep-ph/0405255} {arXiv:hep-ph/0405255} \BibitemShut
  {NoStop}%
\bibitem [{\citenamefont {Stockinger}(2007)}]{Stockinger:2006zn}%
  \BibitemOpen
  \bibfield  {author} {\bibinfo {author} {\bibfnamefont {D.}~\bibnamefont
  {Stockinger}},\ }\bibfield  {title} {\bibinfo {title} {{The Muon Magnetic
  Moment and Supersymmetry}},\ }\href
  {https://doi.org/10.1088/0954-3899/34/2/R01} {\bibfield  {journal} {\bibinfo
  {journal} {J. Phys. G}\ }\textbf {\bibinfo {volume} {34}},\ \bibinfo {pages}
  {R45} (\bibinfo {year} {2007})},\ \Eprint
  {https://arxiv.org/abs/hep-ph/0609168} {arXiv:hep-ph/0609168} \BibitemShut
  {NoStop}%
\bibitem [{\citenamefont {Yin}(2021)}]{Yin:2021yqy}%
  \BibitemOpen
  \bibfield  {author} {\bibinfo {author} {\bibfnamefont {W.}~\bibnamefont
  {Yin}},\ }\bibfield  {title} {\bibinfo {title} {{Radiative lepton mass and
  muon $g-2$ with suppressed lepton flavor and CP violations}}\ }\href
  {https://doi.org/10.1007/JHEP08(2021)043} {10.1007/JHEP08(2021)043} (\bibinfo
  {year} {2021}),\ \Eprint {https://arxiv.org/abs/2103.14234} {arXiv:2103.14234
  [hep-ph]} \BibitemShut {NoStop}%
\bibitem [{\citenamefont {Ilisie}(2015)}]{Ilisie:2015tra}%
  \BibitemOpen
  \bibfield  {author} {\bibinfo {author} {\bibfnamefont {V.}~\bibnamefont
  {Ilisie}},\ }\bibfield  {title} {\bibinfo {title} {{New Barr-Zee
  contributions to $\mathbf{(g-2)_\mu}$ in two-Higgs-doublet models}},\ }\href
  {https://doi.org/10.1007/JHEP04(2015)077} {\bibfield  {journal} {\bibinfo
  {journal} {JHEP}\ }\textbf {\bibinfo {volume} {04}},\ \bibinfo {pages}
  {077}},\ \Eprint {https://arxiv.org/abs/1502.04199} {arXiv:1502.04199
  [hep-ph]} \BibitemShut {NoStop}%
\bibitem [{\citenamefont {Cherchiglia}\ \emph {et~al.}(2017)\citenamefont
  {Cherchiglia}, \citenamefont {Kneschke}, \citenamefont {St\"ockinger},\ and\
  \citenamefont {St\"ockinger-Kim}}]{Cherchiglia:2016eui}%
  \BibitemOpen
  \bibfield  {author} {\bibinfo {author} {\bibfnamefont {A.}~\bibnamefont
  {Cherchiglia}}, \bibinfo {author} {\bibfnamefont {P.}~\bibnamefont
  {Kneschke}}, \bibinfo {author} {\bibfnamefont {D.}~\bibnamefont
  {St\"ockinger}},\ and\ \bibinfo {author} {\bibfnamefont {H.}~\bibnamefont
  {St\"ockinger-Kim}},\ }\bibfield  {title} {\bibinfo {title} {{The muon
  magnetic moment in the 2HDM: complete two-loop result}},\ }\href
  {https://doi.org/10.1007/JHEP01(2017)007} {\bibfield  {journal} {\bibinfo
  {journal} {JHEP}\ }\textbf {\bibinfo {volume} {01}},\ \bibinfo {pages}
  {007}},\ \Eprint {https://arxiv.org/abs/1607.06292} {arXiv:1607.06292
  [hep-ph]} \BibitemShut {NoStop}%
\bibitem [{\citenamefont {Davoudiasl}\ and\ \citenamefont
  {Marciano}(2018)}]{Davoudiasl:2018fbb}%
  \BibitemOpen
  \bibfield  {author} {\bibinfo {author} {\bibfnamefont {H.}~\bibnamefont
  {Davoudiasl}}\ and\ \bibinfo {author} {\bibfnamefont {W.~J.}\ \bibnamefont
  {Marciano}},\ }\bibfield  {title} {\bibinfo {title} {{Tale of two
  anomalies}},\ }\href {https://doi.org/10.1103/PhysRevD.98.075011} {\bibfield
  {journal} {\bibinfo  {journal} {Phys. Rev. D}\ }\textbf {\bibinfo {volume}
  {98}},\ \bibinfo {pages} {075011} (\bibinfo {year} {2018})},\ \Eprint
  {https://arxiv.org/abs/1806.10252} {arXiv:1806.10252 [hep-ph]} \BibitemShut
  {NoStop}%
\bibitem [{\citenamefont {Sabatta}\ \emph {et~al.}(2020)\citenamefont
  {Sabatta}, \citenamefont {Cornell}, \citenamefont {Goyal}, \citenamefont
  {Kumar}, \citenamefont {Mellado},\ and\ \citenamefont
  {Ruan}}]{Sabatta:2019nfg}%
  \BibitemOpen
  \bibfield  {author} {\bibinfo {author} {\bibfnamefont {D.}~\bibnamefont
  {Sabatta}}, \bibinfo {author} {\bibfnamefont {A.~S.}\ \bibnamefont
  {Cornell}}, \bibinfo {author} {\bibfnamefont {A.}~\bibnamefont {Goyal}},
  \bibinfo {author} {\bibfnamefont {M.}~\bibnamefont {Kumar}}, \bibinfo
  {author} {\bibfnamefont {B.}~\bibnamefont {Mellado}},\ and\ \bibinfo {author}
  {\bibfnamefont {X.}~\bibnamefont {Ruan}},\ }\bibfield  {title} {\bibinfo
  {title} {{Connecting muon anomalous magnetic moment and multi-lepton
  anomalies at LHC}},\ }\href {https://doi.org/10.1088/1674-1137/44/6/063103}
  {\bibfield  {journal} {\bibinfo  {journal} {Chin. Phys. C}\ }\textbf
  {\bibinfo {volume} {44}},\ \bibinfo {pages} {063103} (\bibinfo {year}
  {2020})},\ \Eprint {https://arxiv.org/abs/1909.03969} {arXiv:1909.03969
  [hep-ph]} \BibitemShut {NoStop}%
\bibitem [{\citenamefont {Jana}\ \emph {et~al.}(2020)\citenamefont {Jana},
  \citenamefont {K.},\ and\ \citenamefont {Saad}}]{Jana:2020pxx}%
  \BibitemOpen
  \bibfield  {author} {\bibinfo {author} {\bibfnamefont {S.}~\bibnamefont
  {Jana}}, \bibinfo {author} {\bibfnamefont {V.~P.}\ \bibnamefont {K.}},\ and\
  \bibinfo {author} {\bibfnamefont {S.}~\bibnamefont {Saad}},\ }\bibfield
  {title} {\bibinfo {title} {{Resolving electron and muon $g-2$ within the
  2HDM}},\ }\href {https://doi.org/10.1103/PhysRevD.101.115037} {\bibfield
  {journal} {\bibinfo  {journal} {Phys. Rev. D}\ }\textbf {\bibinfo {volume}
  {101}},\ \bibinfo {pages} {115037} (\bibinfo {year} {2020})},\ \Eprint
  {https://arxiv.org/abs/2003.03386} {arXiv:2003.03386 [hep-ph]} \BibitemShut
  {NoStop}%
\bibitem [{\citenamefont {Botella}\ \emph {et~al.}(2020)\citenamefont
  {Botella}, \citenamefont {Cornet-Gomez},\ and\ \citenamefont
  {Nebot}}]{Botella:2020xzf}%
  \BibitemOpen
  \bibfield  {author} {\bibinfo {author} {\bibfnamefont {F.~J.}\ \bibnamefont
  {Botella}}, \bibinfo {author} {\bibfnamefont {F.}~\bibnamefont
  {Cornet-Gomez}},\ and\ \bibinfo {author} {\bibfnamefont {M.}~\bibnamefont
  {Nebot}},\ }\bibfield  {title} {\bibinfo {title} {{Electron and muon $g-2$
  anomalies in general flavour conserving two Higgs doublets models}},\ }\href
  {https://doi.org/10.1103/PhysRevD.102.035023} {\bibfield  {journal} {\bibinfo
   {journal} {Phys. Rev. D}\ }\textbf {\bibinfo {volume} {102}},\ \bibinfo
  {pages} {035023} (\bibinfo {year} {2020})},\ \Eprint
  {https://arxiv.org/abs/2006.01934} {arXiv:2006.01934 [hep-ph]} \BibitemShut
  {NoStop}%
\bibitem [{\citenamefont {Barman}\ \emph {et~al.}(2021)\citenamefont {Barman},
  \citenamefont {Dcruz},\ and\ \citenamefont {Thapa}}]{Barman:2021xeq}%
  \BibitemOpen
  \bibfield  {author} {\bibinfo {author} {\bibfnamefont {R.~K.}\ \bibnamefont
  {Barman}}, \bibinfo {author} {\bibfnamefont {R.}~\bibnamefont {Dcruz}},\ and\
  \bibinfo {author} {\bibfnamefont {A.}~\bibnamefont {Thapa}},\ }\bibfield
  {title} {\bibinfo {title} {{Neutrino masses and magnetic moments of electron
  and muon in the Zee Model}},\ }\href@noop {} {\  (\bibinfo {year} {2021})},\
  \Eprint {https://arxiv.org/abs/2112.04523} {arXiv:2112.04523 [hep-ph]}
  \BibitemShut {NoStop}%
\bibitem [{\citenamefont {Raby}\ and\ \citenamefont
  {Trautner}(2018)}]{Raby:2017igl}%
  \BibitemOpen
  \bibfield  {author} {\bibinfo {author} {\bibfnamefont {S.}~\bibnamefont
  {Raby}}\ and\ \bibinfo {author} {\bibfnamefont {A.}~\bibnamefont
  {Trautner}},\ }\bibfield  {title} {\bibinfo {title} {{Vectorlike chiral
  fourth family to explain muon anomalies}},\ }\href
  {https://doi.org/10.1103/PhysRevD.97.095006} {\bibfield  {journal} {\bibinfo
  {journal} {Phys. Rev. D}\ }\textbf {\bibinfo {volume} {97}},\ \bibinfo
  {pages} {095006} (\bibinfo {year} {2018})},\ \Eprint
  {https://arxiv.org/abs/1712.09360} {arXiv:1712.09360 [hep-ph]} \BibitemShut
  {NoStop}%
\bibitem [{\citenamefont {Crivellin}\ and\ \citenamefont
  {Hoferichter}(2021)}]{Crivellin:2021rbq}%
  \BibitemOpen
  \bibfield  {author} {\bibinfo {author} {\bibfnamefont {A.}~\bibnamefont
  {Crivellin}}\ and\ \bibinfo {author} {\bibfnamefont {M.}~\bibnamefont
  {Hoferichter}},\ }\bibfield  {title} {\bibinfo {title} {{Consequences of
  chirally enhanced explanations of $(g-2)_\mu$ for $h\to \mu\mu$ and $Z\to
  \mu\mu$}},\ }\href {https://doi.org/10.1007/JHEP07(2021)135} {\bibfield
  {journal} {\bibinfo  {journal} {JHEP}\ }\textbf {\bibinfo {volume} {07}},\
  \bibinfo {pages} {135}},\ \Eprint {https://arxiv.org/abs/2104.03202}
  {arXiv:2104.03202 [hep-ph]} \BibitemShut {NoStop}%
\bibitem [{\citenamefont {Kannike}\ \emph {et~al.}(2012)\citenamefont
  {Kannike}, \citenamefont {Raidal}, \citenamefont {Straub},\ and\
  \citenamefont {Strumia}}]{Kannike:2011ng}%
  \BibitemOpen
  \bibfield  {author} {\bibinfo {author} {\bibfnamefont {K.}~\bibnamefont
  {Kannike}}, \bibinfo {author} {\bibfnamefont {M.}~\bibnamefont {Raidal}},
  \bibinfo {author} {\bibfnamefont {D.~M.}\ \bibnamefont {Straub}},\ and\
  \bibinfo {author} {\bibfnamefont {A.}~\bibnamefont {Strumia}},\ }\bibfield
  {title} {\bibinfo {title} {{Anthropic solution to the magnetic muon anomaly:
  the charged see-saw}},\ }\href {https://doi.org/10.1007/JHEP02(2012)106}
  {\bibfield  {journal} {\bibinfo  {journal} {JHEP}\ }\textbf {\bibinfo
  {volume} {02}},\ \bibinfo {pages} {106}},\ \bibinfo {note} {[Erratum: JHEP
  10, 136 (2012)]},\ \Eprint {https://arxiv.org/abs/1111.2551} {arXiv:1111.2551
  [hep-ph]} \BibitemShut {NoStop}%
\bibitem [{\citenamefont {Freitas}\ \emph {et~al.}(2014)\citenamefont
  {Freitas}, \citenamefont {Lykken}, \citenamefont {Kell},\ and\ \citenamefont
  {Westhoff}}]{Freitas:2014pua}%
  \BibitemOpen
  \bibfield  {author} {\bibinfo {author} {\bibfnamefont {A.}~\bibnamefont
  {Freitas}}, \bibinfo {author} {\bibfnamefont {J.}~\bibnamefont {Lykken}},
  \bibinfo {author} {\bibfnamefont {S.}~\bibnamefont {Kell}},\ and\ \bibinfo
  {author} {\bibfnamefont {S.}~\bibnamefont {Westhoff}},\ }\bibfield  {title}
  {\bibinfo {title} {{Testing the Muon g-2 Anomaly at the LHC}},\ }\href
  {https://doi.org/10.1007/JHEP09(2014)155} {\bibfield  {journal} {\bibinfo
  {journal} {JHEP}\ }\textbf {\bibinfo {volume} {05}},\ \bibinfo {pages}
  {145}},\ \bibinfo {note} {[Erratum: JHEP 09, 155 (2014)]},\ \Eprint
  {https://arxiv.org/abs/1402.7065} {arXiv:1402.7065 [hep-ph]} \BibitemShut
  {NoStop}%
\bibitem [{\citenamefont {Dermisek}\ and\ \citenamefont
  {Raval}(2013)}]{Dermisek:2013gta}%
  \BibitemOpen
  \bibfield  {author} {\bibinfo {author} {\bibfnamefont {R.}~\bibnamefont
  {Dermisek}}\ and\ \bibinfo {author} {\bibfnamefont {A.}~\bibnamefont
  {Raval}},\ }\bibfield  {title} {\bibinfo {title} {{Explanation of the Muon
  g-2 Anomaly with Vectorlike Leptons and its Implications for Higgs Decays}},\
  }\href {https://doi.org/10.1103/PhysRevD.88.013017} {\bibfield  {journal}
  {\bibinfo  {journal} {Phys. Rev. D}\ }\textbf {\bibinfo {volume} {88}},\
  \bibinfo {pages} {013017} (\bibinfo {year} {2013})},\ \Eprint
  {https://arxiv.org/abs/1305.3522} {arXiv:1305.3522 [hep-ph]} \BibitemShut
  {NoStop}%
\bibitem [{\citenamefont {Crivellin}\ \emph {et~al.}(2018)\citenamefont
  {Crivellin}, \citenamefont {Hoferichter},\ and\ \citenamefont
  {Schmidt-Wellenburg}}]{Crivellin:2018qmi}%
  \BibitemOpen
  \bibfield  {author} {\bibinfo {author} {\bibfnamefont {A.}~\bibnamefont
  {Crivellin}}, \bibinfo {author} {\bibfnamefont {M.}~\bibnamefont
  {Hoferichter}},\ and\ \bibinfo {author} {\bibfnamefont {P.}~\bibnamefont
  {Schmidt-Wellenburg}},\ }\bibfield  {title} {\bibinfo {title} {{Combined
  explanations of $(g-2)_{\mu,e}$ and implications for a large muon EDM}},\
  }\href {https://doi.org/10.1103/PhysRevD.98.113002} {\bibfield  {journal}
  {\bibinfo  {journal} {Phys. Rev. D}\ }\textbf {\bibinfo {volume} {98}},\
  \bibinfo {pages} {113002} (\bibinfo {year} {2018})},\ \Eprint
  {https://arxiv.org/abs/1807.11484} {arXiv:1807.11484 [hep-ph]} \BibitemShut
  {NoStop}%
\bibitem [{\citenamefont {Frank}\ and\ \citenamefont
  {Saha}(2020)}]{Frank:2020smf}%
  \BibitemOpen
  \bibfield  {author} {\bibinfo {author} {\bibfnamefont {M.}~\bibnamefont
  {Frank}}\ and\ \bibinfo {author} {\bibfnamefont {I.}~\bibnamefont {Saha}},\
  }\bibfield  {title} {\bibinfo {title} {{Muon anomalous magnetic moment in
  two-Higgs-doublet models with vectorlike leptons}},\ }\href
  {https://doi.org/10.1103/PhysRevD.102.115034} {\bibfield  {journal} {\bibinfo
   {journal} {Phys. Rev. D}\ }\textbf {\bibinfo {volume} {102}},\ \bibinfo
  {pages} {115034} (\bibinfo {year} {2020})},\ \Eprint
  {https://arxiv.org/abs/2008.11909} {arXiv:2008.11909 [hep-ph]} \BibitemShut
  {NoStop}%
\bibitem [{\citenamefont {Dermisek}\ \emph {et~al.}(2021)\citenamefont
  {Dermisek}, \citenamefont {Hermanek},\ and\ \citenamefont
  {McGinnis}}]{Dermisek:2021ajd}%
  \BibitemOpen
  \bibfield  {author} {\bibinfo {author} {\bibfnamefont {R.}~\bibnamefont
  {Dermisek}}, \bibinfo {author} {\bibfnamefont {K.}~\bibnamefont {Hermanek}},\
  and\ \bibinfo {author} {\bibfnamefont {N.}~\bibnamefont {McGinnis}},\
  }\bibfield  {title} {\bibinfo {title} {{Muon g-2 in two-Higgs-doublet models
  with vectorlike leptons}},\ }\href
  {https://doi.org/10.1103/PhysRevD.104.055033} {\bibfield  {journal} {\bibinfo
   {journal} {Phys. Rev. D}\ }\textbf {\bibinfo {volume} {104}},\ \bibinfo
  {pages} {055033} (\bibinfo {year} {2021})},\ \Eprint
  {https://arxiv.org/abs/2103.05645} {arXiv:2103.05645 [hep-ph]} \BibitemShut
  {NoStop}%
\bibitem [{\citenamefont {C\'arcamo~Hern\'andez}\ \emph
  {et~al.}(2020)\citenamefont {C\'arcamo~Hern\'andez}, \citenamefont {King},
  \citenamefont {Lee},\ and\ \citenamefont
  {Rowley}}]{CarcamoHernandez:2019ydc}%
  \BibitemOpen
  \bibfield  {author} {\bibinfo {author} {\bibfnamefont {A.~E.}\ \bibnamefont
  {C\'arcamo~Hern\'andez}}, \bibinfo {author} {\bibfnamefont {S.~F.}\
  \bibnamefont {King}}, \bibinfo {author} {\bibfnamefont {H.}~\bibnamefont
  {Lee}},\ and\ \bibinfo {author} {\bibfnamefont {S.~J.}\ \bibnamefont
  {Rowley}},\ }\bibfield  {title} {\bibinfo {title} {{Is it possible to explain
  the muon and electron $g-2$ in a $Z'$ model?}},\ }\href
  {https://doi.org/10.1103/PhysRevD.101.115016} {\bibfield  {journal} {\bibinfo
   {journal} {Phys. Rev. D}\ }\textbf {\bibinfo {volume} {101}},\ \bibinfo
  {pages} {115016} (\bibinfo {year} {2020})},\ \Eprint
  {https://arxiv.org/abs/1910.10734} {arXiv:1910.10734 [hep-ph]} \BibitemShut
  {NoStop}%
\bibitem [{\citenamefont {Navarro}\ and\ \citenamefont
  {King}(2021)}]{Navarro:2021sfb}%
  \BibitemOpen
  \bibfield  {author} {\bibinfo {author} {\bibfnamefont {M.~F.}\ \bibnamefont
  {Navarro}}\ and\ \bibinfo {author} {\bibfnamefont {S.~F.}\ \bibnamefont
  {King}},\ }\bibfield  {title} {\bibinfo {title} {{Fermiophobic $Z'$ model for
  simultaneously explaining the muon anomalies $R_{K^{(*)}}$ and
  $(g-2)_{\mu}$}},\ }\href@noop {} {\  (\bibinfo {year} {2021})},\ \Eprint
  {https://arxiv.org/abs/2109.08729} {arXiv:2109.08729 [hep-ph]} \BibitemShut
  {NoStop}%
\bibitem [{\citenamefont {Ko}\ \emph {et~al.}(2021)\citenamefont {Ko},
  \citenamefont {Nomura},\ and\ \citenamefont {Okada}}]{Ko:2021lpx}%
  \BibitemOpen
  \bibfield  {author} {\bibinfo {author} {\bibfnamefont {P.}~\bibnamefont
  {Ko}}, \bibinfo {author} {\bibfnamefont {T.}~\bibnamefont {Nomura}},\ and\
  \bibinfo {author} {\bibfnamefont {H.}~\bibnamefont {Okada}},\ }\bibfield
  {title} {\bibinfo {title} {{Muon $g-2$, $B\to K^{(*)}\mu^+ \mu^-$ anomalies,
  and leptophilic dark matter in $U(1)_{\mu-\tau}$ gauge symmetry}},\
  }\href@noop {} {\  (\bibinfo {year} {2021})},\ \Eprint
  {https://arxiv.org/abs/2110.10513} {arXiv:2110.10513 [hep-ph]} \BibitemShut
  {NoStop}%
\bibitem [{\citenamefont {Pati}\ and\ \citenamefont
  {Salam}(1974)}]{Pati:1974yy}%
  \BibitemOpen
  \bibfield  {author} {\bibinfo {author} {\bibfnamefont {J.~C.}\ \bibnamefont
  {Pati}}\ and\ \bibinfo {author} {\bibfnamefont {A.}~\bibnamefont {Salam}},\
  }\bibfield  {title} {\bibinfo {title} {{Lepton Number as the Fourth Color}},\
  }\href {https://doi.org/10.1103/PhysRevD.10.275} {\bibfield  {journal}
  {\bibinfo  {journal} {Phys. Rev. D}\ }\textbf {\bibinfo {volume} {10}},\
  \bibinfo {pages} {275} (\bibinfo {year} {1974})},\ \bibinfo {note} {[Erratum:
  Phys.Rev.D 11, 703--703 (1975)]}\BibitemShut {NoStop}%
\bibitem [{\citenamefont {Georgi}\ and\ \citenamefont
  {Glashow}(1974)}]{Georgi:1974sy}%
  \BibitemOpen
  \bibfield  {author} {\bibinfo {author} {\bibfnamefont {H.}~\bibnamefont
  {Georgi}}\ and\ \bibinfo {author} {\bibfnamefont {S.~L.}\ \bibnamefont
  {Glashow}},\ }\bibfield  {title} {\bibinfo {title} {{Unity of All Elementary
  Particle Forces}},\ }\href {https://doi.org/10.1103/PhysRevLett.32.438}
  {\bibfield  {journal} {\bibinfo  {journal} {Phys. Rev. Lett.}\ }\textbf
  {\bibinfo {volume} {32}},\ \bibinfo {pages} {438} (\bibinfo {year}
  {1974})}\BibitemShut {NoStop}%
\bibitem [{\citenamefont {Fritzsch}\ and\ \citenamefont
  {Minkowski}(1975)}]{Fritzsch:1974nn}%
  \BibitemOpen
  \bibfield  {author} {\bibinfo {author} {\bibfnamefont {H.}~\bibnamefont
  {Fritzsch}}\ and\ \bibinfo {author} {\bibfnamefont {P.}~\bibnamefont
  {Minkowski}},\ }\bibfield  {title} {\bibinfo {title} {{Unified Interactions
  of Leptons and Hadrons}},\ }\href
  {https://doi.org/10.1016/0003-4916(75)90211-0} {\bibfield  {journal}
  {\bibinfo  {journal} {Annals Phys.}\ }\textbf {\bibinfo {volume} {93}},\
  \bibinfo {pages} {193} (\bibinfo {year} {1975})}\BibitemShut {NoStop}%
\bibitem [{\citenamefont {Buchmuller}\ \emph {et~al.}(1987)\citenamefont
  {Buchmuller}, \citenamefont {Ruckl},\ and\ \citenamefont
  {Wyler}}]{Buchmuller:1986zs}%
  \BibitemOpen
  \bibfield  {author} {\bibinfo {author} {\bibfnamefont {W.}~\bibnamefont
  {Buchmuller}}, \bibinfo {author} {\bibfnamefont {R.}~\bibnamefont {Ruckl}},\
  and\ \bibinfo {author} {\bibfnamefont {D.}~\bibnamefont {Wyler}},\ }\bibfield
   {title} {\bibinfo {title} {{Leptoquarks in Lepton - Quark Collisions}},\
  }\href {https://doi.org/10.1016/0370-2693(87)90637-X} {\bibfield  {journal}
  {\bibinfo  {journal} {Phys. Lett. B}\ }\textbf {\bibinfo {volume} {191}},\
  \bibinfo {pages} {442} (\bibinfo {year} {1987})},\ \bibinfo {note} {[Erratum:
  Phys.Lett.B 448, 320--320 (1999)]}\BibitemShut {NoStop}%
\bibitem [{\citenamefont {Djouadi}\ \emph {et~al.}(1990)\citenamefont
  {Djouadi}, \citenamefont {Kohler}, \citenamefont {Spira},\ and\ \citenamefont
  {Tutas}}]{Djouadi:1989md}%
  \BibitemOpen
  \bibfield  {author} {\bibinfo {author} {\bibfnamefont {A.}~\bibnamefont
  {Djouadi}}, \bibinfo {author} {\bibfnamefont {T.}~\bibnamefont {Kohler}},
  \bibinfo {author} {\bibfnamefont {M.}~\bibnamefont {Spira}},\ and\ \bibinfo
  {author} {\bibfnamefont {J.}~\bibnamefont {Tutas}},\ }\bibfield  {title}
  {\bibinfo {title} {{$(e b)$, $(e t)$ type leptoquarks at $ep$ colliders}},\
  }\href {https://doi.org/10.1007/BF01560270} {\bibfield  {journal} {\bibinfo
  {journal} {Z. Phys. C}\ }\textbf {\bibinfo {volume} {46}},\ \bibinfo {pages}
  {679} (\bibinfo {year} {1990})}\BibitemShut {NoStop}%
\bibitem [{\citenamefont {Dor\v{s}ner}\ \emph {et~al.}(2016)\citenamefont
  {Dor\v{s}ner}, \citenamefont {Fajfer}, \citenamefont {Greljo}, \citenamefont
  {Kamenik},\ and\ \citenamefont {Ko\v{s}nik}}]{Dorsner:2016wpm}%
  \BibitemOpen
  \bibfield  {author} {\bibinfo {author} {\bibfnamefont {I.}~\bibnamefont
  {Dor\v{s}ner}}, \bibinfo {author} {\bibfnamefont {S.}~\bibnamefont {Fajfer}},
  \bibinfo {author} {\bibfnamefont {A.}~\bibnamefont {Greljo}}, \bibinfo
  {author} {\bibfnamefont {J.~F.}\ \bibnamefont {Kamenik}},\ and\ \bibinfo
  {author} {\bibfnamefont {N.}~\bibnamefont {Ko\v{s}nik}},\ }\bibfield  {title}
  {\bibinfo {title} {{Physics of leptoquarks in precision experiments and at
  particle colliders}},\ }\href {https://doi.org/10.1016/j.physrep.2016.06.001}
  {\bibfield  {journal} {\bibinfo  {journal} {Phys. Rept.}\ }\textbf {\bibinfo
  {volume} {641}},\ \bibinfo {pages} {1} (\bibinfo {year} {2016})},\ \Eprint
  {https://arxiv.org/abs/1603.04993} {arXiv:1603.04993 [hep-ph]} \BibitemShut
  {NoStop}%
\bibitem [{\citenamefont {Crivellin}\ \emph {et~al.}(2021)\citenamefont
  {Crivellin}, \citenamefont {Mueller},\ and\ \citenamefont
  {Saturnino}}]{Crivellin:2020tsz}%
  \BibitemOpen
  \bibfield  {author} {\bibinfo {author} {\bibfnamefont {A.}~\bibnamefont
  {Crivellin}}, \bibinfo {author} {\bibfnamefont {D.}~\bibnamefont {Mueller}},\
  and\ \bibinfo {author} {\bibfnamefont {F.}~\bibnamefont {Saturnino}},\
  }\bibfield  {title} {\bibinfo {title} {{Correlating
  h\textrightarrow{}\ensuremath{\mu}+\ensuremath{\mu}- to the Anomalous
  Magnetic Moment of the Muon via Leptoquarks}},\ }\href
  {https://doi.org/10.1103/PhysRevLett.127.021801} {\bibfield  {journal}
  {\bibinfo  {journal} {Phys. Rev. Lett.}\ }\textbf {\bibinfo {volume} {127}},\
  \bibinfo {pages} {021801} (\bibinfo {year} {2021})},\ \Eprint
  {https://arxiv.org/abs/2008.02643} {arXiv:2008.02643 [hep-ph]} \BibitemShut
  {NoStop}%
\bibitem [{\citenamefont {Dor\v{s}ner}\ \emph
  {et~al.}(2020{\natexlab{a}})\citenamefont {Dor\v{s}ner}, \citenamefont
  {Fajfer},\ and\ \citenamefont {Sumensari}}]{Dorsner:2019itg}%
  \BibitemOpen
  \bibfield  {author} {\bibinfo {author} {\bibfnamefont {I.}~\bibnamefont
  {Dor\v{s}ner}}, \bibinfo {author} {\bibfnamefont {S.}~\bibnamefont
  {Fajfer}},\ and\ \bibinfo {author} {\bibfnamefont {O.}~\bibnamefont
  {Sumensari}},\ }\bibfield  {title} {\bibinfo {title} {{Muon $g-2$ and scalar
  leptoquark mixing}},\ }\href {https://doi.org/10.1007/JHEP06(2020)089}
  {\bibfield  {journal} {\bibinfo  {journal} {JHEP}\ }\textbf {\bibinfo
  {volume} {06}},\ \bibinfo {pages} {089}},\ \Eprint
  {https://arxiv.org/abs/1910.03877} {arXiv:1910.03877 [hep-ph]} \BibitemShut
  {NoStop}%
\bibitem [{\citenamefont {Dor\v{s}ner}\ \emph
  {et~al.}(2020{\natexlab{b}})\citenamefont {Dor\v{s}ner}, \citenamefont
  {Fajfer},\ and\ \citenamefont {Saad}}]{Dorsner:2020aaz}%
  \BibitemOpen
  \bibfield  {author} {\bibinfo {author} {\bibfnamefont {I.}~\bibnamefont
  {Dor\v{s}ner}}, \bibinfo {author} {\bibfnamefont {S.}~\bibnamefont
  {Fajfer}},\ and\ \bibinfo {author} {\bibfnamefont {S.}~\bibnamefont {Saad}},\
  }\bibfield  {title} {\bibinfo {title} {{$\mu \to e \gamma$ selecting scalar
  leptoquark solutions for the $(g-2)_{e,\mu}$ puzzles}},\ }\href
  {https://doi.org/10.1103/PhysRevD.102.075007} {\bibfield  {journal} {\bibinfo
   {journal} {Phys. Rev. D}\ }\textbf {\bibinfo {volume} {102}},\ \bibinfo
  {pages} {075007} (\bibinfo {year} {2020}{\natexlab{b}})},\ \Eprint
  {https://arxiv.org/abs/2006.11624} {arXiv:2006.11624 [hep-ph]} \BibitemShut
  {NoStop}%
\bibitem [{\citenamefont {Zhang}(2021)}]{Zhang:2021dgl}%
  \BibitemOpen
  \bibfield  {author} {\bibinfo {author} {\bibfnamefont {D.}~\bibnamefont
  {Zhang}},\ }\bibfield  {title} {\bibinfo {title} {{Radiative neutrino masses,
  lepton flavor mixing and muon g \ensuremath{-} 2 in a leptoquark model}},\
  }\href {https://doi.org/10.1007/JHEP07(2021)069} {\bibfield  {journal}
  {\bibinfo  {journal} {JHEP}\ }\textbf {\bibinfo {volume} {07}},\ \bibinfo
  {pages} {069}},\ \Eprint {https://arxiv.org/abs/2105.08670} {arXiv:2105.08670
  [hep-ph]} \BibitemShut {NoStop}%
\bibitem [{\citenamefont {Coluccio~Leskow}\ \emph {et~al.}(2017)\citenamefont
  {Coluccio~Leskow}, \citenamefont {D'Ambrosio}, \citenamefont {Crivellin},\
  and\ \citenamefont {M\"uller}}]{ColuccioLeskow:2016dox}%
  \BibitemOpen
  \bibfield  {author} {\bibinfo {author} {\bibfnamefont {E.}~\bibnamefont
  {Coluccio~Leskow}}, \bibinfo {author} {\bibfnamefont {G.}~\bibnamefont
  {D'Ambrosio}}, \bibinfo {author} {\bibfnamefont {A.}~\bibnamefont
  {Crivellin}},\ and\ \bibinfo {author} {\bibfnamefont {D.}~\bibnamefont
  {M\"uller}},\ }\bibfield  {title} {\bibinfo {title} {{$(g-2)_\mu$, lepton
  flavor violation, and $Z$ decays with leptoquarks: Correlations and future
  prospects}},\ }\href {https://doi.org/10.1103/PhysRevD.95.055018} {\bibfield
  {journal} {\bibinfo  {journal} {Phys. Rev. D}\ }\textbf {\bibinfo {volume}
  {95}},\ \bibinfo {pages} {055018} (\bibinfo {year} {2017})},\ \Eprint
  {https://arxiv.org/abs/1612.06858} {arXiv:1612.06858 [hep-ph]} \BibitemShut
  {NoStop}%
\bibitem [{\citenamefont {Calibbi}\ \emph {et~al.}(2018)\citenamefont
  {Calibbi}, \citenamefont {Crivellin},\ and\ \citenamefont
  {Li}}]{Calibbi:2017qbu}%
  \BibitemOpen
  \bibfield  {author} {\bibinfo {author} {\bibfnamefont {L.}~\bibnamefont
  {Calibbi}}, \bibinfo {author} {\bibfnamefont {A.}~\bibnamefont {Crivellin}},\
  and\ \bibinfo {author} {\bibfnamefont {T.}~\bibnamefont {Li}},\ }\bibfield
  {title} {\bibinfo {title} {{Model of vector leptoquarks in view of the
  $B$-physics anomalies}},\ }\href {https://doi.org/10.1103/PhysRevD.98.115002}
  {\bibfield  {journal} {\bibinfo  {journal} {Phys. Rev. D}\ }\textbf {\bibinfo
  {volume} {98}},\ \bibinfo {pages} {115002} (\bibinfo {year} {2018})},\
  \Eprint {https://arxiv.org/abs/1709.00692} {arXiv:1709.00692 [hep-ph]}
  \BibitemShut {NoStop}%
\bibitem [{\citenamefont {Saad}(2020)}]{Saad:2020ihm}%
  \BibitemOpen
  \bibfield  {author} {\bibinfo {author} {\bibfnamefont {S.}~\bibnamefont
  {Saad}},\ }\bibfield  {title} {\bibinfo {title} {{Combined explanations of
  $(g-2)_{\mu}$, $R_{D^{(*)}}$, $R_{K^{(*)}}$ anomalies in a two-loop radiative
  neutrino mass model}},\ }\href {https://doi.org/10.1103/PhysRevD.102.015019}
  {\bibfield  {journal} {\bibinfo  {journal} {Phys. Rev. D}\ }\textbf {\bibinfo
  {volume} {102}},\ \bibinfo {pages} {015019} (\bibinfo {year} {2020})},\
  \Eprint {https://arxiv.org/abs/2005.04352} {arXiv:2005.04352 [hep-ph]}
  \BibitemShut {NoStop}%
\bibitem [{\citenamefont {Babu}\ \emph {et~al.}(2021)\citenamefont {Babu},
  \citenamefont {Dev}, \citenamefont {Jana},\ and\ \citenamefont
  {Thapa}}]{Babu:2020hun}%
  \BibitemOpen
  \bibfield  {author} {\bibinfo {author} {\bibfnamefont {K.~S.}\ \bibnamefont
  {Babu}}, \bibinfo {author} {\bibfnamefont {P.~S.~B.}\ \bibnamefont {Dev}},
  \bibinfo {author} {\bibfnamefont {S.}~\bibnamefont {Jana}},\ and\ \bibinfo
  {author} {\bibfnamefont {A.}~\bibnamefont {Thapa}},\ }\bibfield  {title}
  {\bibinfo {title} {{Unified framework for $B$-anomalies, muon $g-2$ and
  neutrino masses}},\ }\href {https://doi.org/10.1007/JHEP03(2021)179}
  {\bibfield  {journal} {\bibinfo  {journal} {JHEP}\ }\textbf {\bibinfo
  {volume} {03}},\ \bibinfo {pages} {179}},\ \Eprint
  {https://arxiv.org/abs/2009.01771} {arXiv:2009.01771 [hep-ph]} \BibitemShut
  {NoStop}%
\bibitem [{\citenamefont {Marzocca}\ and\ \citenamefont
  {Trifinopoulos}(2021)}]{Marzocca:2021azj}%
  \BibitemOpen
  \bibfield  {author} {\bibinfo {author} {\bibfnamefont {D.}~\bibnamefont
  {Marzocca}}\ and\ \bibinfo {author} {\bibfnamefont {S.}~\bibnamefont
  {Trifinopoulos}},\ }\bibfield  {title} {\bibinfo {title} {{Minimal
  Explanation of Flavor Anomalies: B-Meson Decays, Muon Magnetic Moment, and
  the Cabibbo Angle}},\ }\href {https://doi.org/10.1103/PhysRevLett.127.061803}
  {\bibfield  {journal} {\bibinfo  {journal} {Phys. Rev. Lett.}\ }\textbf
  {\bibinfo {volume} {127}},\ \bibinfo {pages} {061803} (\bibinfo {year}
  {2021})},\ \Eprint {https://arxiv.org/abs/2104.05730} {arXiv:2104.05730
  [hep-ph]} \BibitemShut {NoStop}%
\bibitem [{\citenamefont {Panico}\ and\ \citenamefont
  {Wulzer}(2016)}]{Panico:2015jxa}%
  \BibitemOpen
  \bibfield  {author} {\bibinfo {author} {\bibfnamefont {G.}~\bibnamefont
  {Panico}}\ and\ \bibinfo {author} {\bibfnamefont {A.}~\bibnamefont
  {Wulzer}},\ }\href {https://doi.org/10.1007/978-3-319-22617-0} {\emph
  {\bibinfo {title} {{The Composite Nambu-Goldstone Higgs}}}},\ Vol.\ \bibinfo
  {volume} {913}\ (\bibinfo  {publisher} {Springer},\ \bibinfo {year} {2016})\
  \Eprint {https://arxiv.org/abs/1506.01961} {arXiv:1506.01961 [hep-ph]}
  \BibitemShut {NoStop}%
\bibitem [{\citenamefont {Schmaltz}\ and\ \citenamefont
  {Tucker-Smith}(2005)}]{Schmaltz:2005ky}%
  \BibitemOpen
  \bibfield  {author} {\bibinfo {author} {\bibfnamefont {M.}~\bibnamefont
  {Schmaltz}}\ and\ \bibinfo {author} {\bibfnamefont {D.}~\bibnamefont
  {Tucker-Smith}},\ }\bibfield  {title} {\bibinfo {title} {{Little Higgs
  review}},\ }\href {https://doi.org/10.1146/annurev.nucl.55.090704.151502}
  {\bibfield  {journal} {\bibinfo  {journal} {Ann. Rev. Nucl. Part. Sci.}\
  }\textbf {\bibinfo {volume} {55}},\ \bibinfo {pages} {229} (\bibinfo {year}
  {2005})},\ \Eprint {https://arxiv.org/abs/hep-ph/0502182}
  {arXiv:hep-ph/0502182} \BibitemShut {NoStop}%
\bibitem [{\citenamefont {Hewett}\ and\ \citenamefont
  {Rizzo}(1989)}]{Hewett:1988xc}%
  \BibitemOpen
  \bibfield  {author} {\bibinfo {author} {\bibfnamefont {J.~L.}\ \bibnamefont
  {Hewett}}\ and\ \bibinfo {author} {\bibfnamefont {T.~G.}\ \bibnamefont
  {Rizzo}},\ }\bibfield  {title} {\bibinfo {title} {{Low-Energy Phenomenology
  of Superstring Inspired E(6) Models}},\ }\href
  {https://doi.org/10.1016/0370-1573(89)90071-9} {\bibfield  {journal}
  {\bibinfo  {journal} {Phys. Rept.}\ }\textbf {\bibinfo {volume} {183}},\
  \bibinfo {pages} {193} (\bibinfo {year} {1989})}\BibitemShut {NoStop}%
\bibitem [{\citenamefont {Randall}\ and\ \citenamefont
  {Sundrum}(1999)}]{Randall:1999ee}%
  \BibitemOpen
  \bibfield  {author} {\bibinfo {author} {\bibfnamefont {L.}~\bibnamefont
  {Randall}}\ and\ \bibinfo {author} {\bibfnamefont {R.}~\bibnamefont
  {Sundrum}},\ }\bibfield  {title} {\bibinfo {title} {{A Large mass hierarchy
  from a small extra dimension}},\ }\href
  {https://doi.org/10.1103/PhysRevLett.83.3370} {\bibfield  {journal} {\bibinfo
   {journal} {Phys. Rev. Lett.}\ }\textbf {\bibinfo {volume} {83}},\ \bibinfo
  {pages} {3370} (\bibinfo {year} {1999})},\ \Eprint
  {https://arxiv.org/abs/hep-ph/9905221} {arXiv:hep-ph/9905221} \BibitemShut
  {NoStop}%
\bibitem [{\citenamefont {Chakraverty}\ \emph {et~al.}(2001)\citenamefont
  {Chakraverty}, \citenamefont {Choudhury},\ and\ \citenamefont
  {Datta}}]{Chakraverty:2001yg}%
  \BibitemOpen
  \bibfield  {author} {\bibinfo {author} {\bibfnamefont {D.}~\bibnamefont
  {Chakraverty}}, \bibinfo {author} {\bibfnamefont {D.}~\bibnamefont
  {Choudhury}},\ and\ \bibinfo {author} {\bibfnamefont {A.}~\bibnamefont
  {Datta}},\ }\bibfield  {title} {\bibinfo {title} {{A Nonsupersymmetric
  resolution of the anomalous muon magnetic moment}},\ }\href
  {https://doi.org/10.1016/S0370-2693(01)00419-1} {\bibfield  {journal}
  {\bibinfo  {journal} {Phys. Lett. B}\ }\textbf {\bibinfo {volume} {506}},\
  \bibinfo {pages} {103} (\bibinfo {year} {2001})},\ \Eprint
  {https://arxiv.org/abs/hep-ph/0102180} {arXiv:hep-ph/0102180} \BibitemShut
  {NoStop}%
\bibitem [{\citenamefont {Cheung}(2001)}]{Cheung:2001ip}%
  \BibitemOpen
  \bibfield  {author} {\bibinfo {author} {\bibfnamefont {K.-m.}\ \bibnamefont
  {Cheung}},\ }\bibfield  {title} {\bibinfo {title} {{Muon anomalous magnetic
  moment and leptoquark solutions}},\ }\href
  {https://doi.org/10.1103/PhysRevD.64.033001} {\bibfield  {journal} {\bibinfo
  {journal} {Phys. Rev. D}\ }\textbf {\bibinfo {volume} {64}},\ \bibinfo
  {pages} {033001} (\bibinfo {year} {2001})},\ \Eprint
  {https://arxiv.org/abs/hep-ph/0102238} {arXiv:hep-ph/0102238} \BibitemShut
  {NoStop}%
\bibitem [{\citenamefont {Leveille}(1978)}]{Leveille:1977rc}%
  \BibitemOpen
  \bibfield  {author} {\bibinfo {author} {\bibfnamefont {J.~P.}\ \bibnamefont
  {Leveille}},\ }\bibfield  {title} {\bibinfo {title} {{The Second Order Weak
  Correction to (G-2) of the Muon in Arbitrary Gauge Models}},\ }\href
  {https://doi.org/10.1016/0550-3213(78)90051-2} {\bibfield  {journal}
  {\bibinfo  {journal} {Nucl. Phys. B}\ }\textbf {\bibinfo {volume} {137}},\
  \bibinfo {pages} {63} (\bibinfo {year} {1978})}\BibitemShut {NoStop}%
\bibitem [{\citenamefont {Aebischer}\ \emph {et~al.}(2021)\citenamefont
  {Aebischer}, \citenamefont {Dekens}, \citenamefont {Jenkins}, \citenamefont
  {Manohar}, \citenamefont {Sengupta},\ and\ \citenamefont
  {Stoffer}}]{Aebischer:2021uvt}%
  \BibitemOpen
  \bibfield  {author} {\bibinfo {author} {\bibfnamefont {J.}~\bibnamefont
  {Aebischer}}, \bibinfo {author} {\bibfnamefont {W.}~\bibnamefont {Dekens}},
  \bibinfo {author} {\bibfnamefont {E.~E.}\ \bibnamefont {Jenkins}}, \bibinfo
  {author} {\bibfnamefont {A.~V.}\ \bibnamefont {Manohar}}, \bibinfo {author}
  {\bibfnamefont {D.}~\bibnamefont {Sengupta}},\ and\ \bibinfo {author}
  {\bibfnamefont {P.}~\bibnamefont {Stoffer}},\ }\bibfield  {title} {\bibinfo
  {title} {{Effective field theory interpretation of lepton magnetic and
  electric dipole moments}},\ }\href {https://doi.org/10.1007/JHEP07(2021)107}
  {\bibfield  {journal} {\bibinfo  {journal} {JHEP}\ }\textbf {\bibinfo
  {volume} {07}},\ \bibinfo {pages} {107}},\ \Eprint
  {https://arxiv.org/abs/2102.08954} {arXiv:2102.08954 [hep-ph]} \BibitemShut
  {NoStop}%
\bibitem [{\citenamefont {Aguilar-Saavedra}\ \emph {et~al.}(2013)\citenamefont
  {Aguilar-Saavedra}, \citenamefont {Benbrik}, \citenamefont {Heinemeyer},\
  and\ \citenamefont {P\'erez-Victoria}}]{Aguilar-Saavedra:2013qpa}%
  \BibitemOpen
  \bibfield  {author} {\bibinfo {author} {\bibfnamefont {J.~A.}\ \bibnamefont
  {Aguilar-Saavedra}}, \bibinfo {author} {\bibfnamefont {R.}~\bibnamefont
  {Benbrik}}, \bibinfo {author} {\bibfnamefont {S.}~\bibnamefont
  {Heinemeyer}},\ and\ \bibinfo {author} {\bibfnamefont {M.}~\bibnamefont
  {P\'erez-Victoria}},\ }\bibfield  {title} {\bibinfo {title} {{Handbook of
  vectorlike quarks: Mixing and single production}},\ }\href
  {https://doi.org/10.1103/PhysRevD.88.094010} {\bibfield  {journal} {\bibinfo
  {journal} {Phys. Rev. D}\ }\textbf {\bibinfo {volume} {88}},\ \bibinfo
  {pages} {094010} (\bibinfo {year} {2013})},\ \Eprint
  {https://arxiv.org/abs/1306.0572} {arXiv:1306.0572 [hep-ph]} \BibitemShut
  {NoStop}%
\bibitem [{\citenamefont {He}(2020)}]{He:2020suf}%
  \BibitemOpen
  \bibfield  {author} {\bibinfo {author} {\bibfnamefont {S.-P.}\ \bibnamefont
  {He}},\ }\bibfield  {title} {\bibinfo {title} {{Higgs boson to $\gamma Z$
  decay as a probe of flavor-changing neutral Yukawa couplings}},\ }\href
  {https://doi.org/10.1103/PhysRevD.102.075035} {\bibfield  {journal} {\bibinfo
   {journal} {Phys. Rev. D}\ }\textbf {\bibinfo {volume} {102}},\ \bibinfo
  {pages} {075035} (\bibinfo {year} {2020})},\ \Eprint
  {https://arxiv.org/abs/2004.12155} {arXiv:2004.12155 [hep-ph]} \BibitemShut
  {NoStop}%
\bibitem [{\citenamefont {He}(2021)}]{He:2020fqj}%
  \BibitemOpen
  \bibfield  {author} {\bibinfo {author} {\bibfnamefont {S.-P.}\ \bibnamefont
  {He}},\ }\bibfield  {title} {\bibinfo {title} {{Di-Higgs production as a
  probe of flavor changing neutral Yukawa couplings}},\ }\href
  {https://doi.org/10.1088/1674-1137/abfb50} {\bibfield  {journal} {\bibinfo
  {journal} {Chin. Phys. C}\ }\textbf {\bibinfo {volume} {45}},\ \bibinfo
  {pages} {073108} (\bibinfo {year} {2021})},\ \Eprint
  {https://arxiv.org/abs/2011.11949} {arXiv:2011.11949 [hep-ph]} \BibitemShut
  {NoStop}%
\bibitem [{\citenamefont {Zyla}\ \emph {et~al.}(2020)\citenamefont {Zyla} \emph
  {et~al.}}]{ParticleDataGroup:2020ssz}%
  \BibitemOpen
  \bibfield  {author} {\bibinfo {author} {\bibfnamefont {P.~A.}\ \bibnamefont
  {Zyla}} \emph {et~al.} (\bibinfo {collaboration} {Particle Data Group}),\
  }\bibfield  {title} {\bibinfo {title} {{Review of Particle Physics}},\ }\href
  {https://doi.org/10.1093/ptep/ptaa104} {\bibfield  {journal} {\bibinfo
  {journal} {PTEP}\ }\textbf {\bibinfo {volume} {2020}},\ \bibinfo {pages}
  {083C01} (\bibinfo {year} {2020})}\BibitemShut {NoStop}%
\bibitem [{\citenamefont {Sirunyan}\ \emph
  {et~al.}(2019{\natexlab{a}})\citenamefont {Sirunyan} \emph
  {et~al.}}]{CMS:2018wpl}%
  \BibitemOpen
  \bibfield  {author} {\bibinfo {author} {\bibfnamefont {A.~M.}\ \bibnamefont
  {Sirunyan}} \emph {et~al.} (\bibinfo {collaboration} {CMS}),\ }\bibfield
  {title} {\bibinfo {title} {{Search for vector-like quarks in events with two
  oppositely charged leptons and jets in proton-proton collisions at $\sqrt{s}
  =$ 13 TeV}},\ }\href {https://doi.org/10.1140/epjc/s10052-019-6855-8}
  {\bibfield  {journal} {\bibinfo  {journal} {Eur. Phys. J. C}\ }\textbf
  {\bibinfo {volume} {79}},\ \bibinfo {pages} {364} (\bibinfo {year}
  {2019}{\natexlab{a}})},\ \Eprint {https://arxiv.org/abs/1812.09768}
  {arXiv:1812.09768 [hep-ex]} \BibitemShut {NoStop}%
\bibitem [{\citenamefont {Sirunyan}\ \emph
  {et~al.}(2019{\natexlab{b}})\citenamefont {Sirunyan} \emph
  {et~al.}}]{CMS:2019eqb}%
  \BibitemOpen
  \bibfield  {author} {\bibinfo {author} {\bibfnamefont {A.~M.}\ \bibnamefont
  {Sirunyan}} \emph {et~al.} (\bibinfo {collaboration} {CMS}),\ }\bibfield
  {title} {\bibinfo {title} {{Search for pair production of vectorlike quarks
  in the fully hadronic final state}},\ }\href
  {https://doi.org/10.1103/PhysRevD.100.072001} {\bibfield  {journal} {\bibinfo
   {journal} {Phys. Rev. D}\ }\textbf {\bibinfo {volume} {100}},\ \bibinfo
  {pages} {072001} (\bibinfo {year} {2019}{\natexlab{b}})},\ \Eprint
  {https://arxiv.org/abs/1906.11903} {arXiv:1906.11903 [hep-ex]} \BibitemShut
  {NoStop}%
\bibitem [{\citenamefont {Aaboud}\ \emph {et~al.}(2018)\citenamefont {Aaboud}
  \emph {et~al.}}]{ATLAS:2018ziw}%
  \BibitemOpen
  \bibfield  {author} {\bibinfo {author} {\bibfnamefont {M.}~\bibnamefont
  {Aaboud}} \emph {et~al.} (\bibinfo {collaboration} {ATLAS}),\ }\bibfield
  {title} {\bibinfo {title} {{Combination of the searches for pair-produced
  vector-like partners of the third-generation quarks at $\sqrt{s} =$ 13 TeV
  with the ATLAS detector}},\ }\href
  {https://doi.org/10.1103/PhysRevLett.121.211801} {\bibfield  {journal}
  {\bibinfo  {journal} {Phys. Rev. Lett.}\ }\textbf {\bibinfo {volume} {121}},\
  \bibinfo {pages} {211801} (\bibinfo {year} {2018})},\ \Eprint
  {https://arxiv.org/abs/1808.02343} {arXiv:1808.02343 [hep-ex]} \BibitemShut
  {NoStop}%
\bibitem [{\citenamefont {Aaboud}\ \emph {et~al.}(2019)\citenamefont {Aaboud}
  \emph {et~al.}}]{ATLAS:2018dyh}%
  \BibitemOpen
  \bibfield  {author} {\bibinfo {author} {\bibfnamefont {M.}~\bibnamefont
  {Aaboud}} \emph {et~al.} (\bibinfo {collaboration} {ATLAS}),\ }\bibfield
  {title} {\bibinfo {title} {{Search for single production of vector-like
  quarks decaying into $Wb$ in $pp$ collisions at $\sqrt{s} = 13$ TeV with the
  ATLAS detector}},\ }\href {https://doi.org/10.1007/JHEP05(2019)164}
  {\bibfield  {journal} {\bibinfo  {journal} {JHEP}\ }\textbf {\bibinfo
  {volume} {05}},\ \bibinfo {pages} {164}},\ \Eprint
  {https://arxiv.org/abs/1812.07343} {arXiv:1812.07343 [hep-ex]} \BibitemShut
  {NoStop}%
\bibitem [{\citenamefont {Chen}\ \emph {et~al.}(2017)\citenamefont {Chen},
  \citenamefont {Dawson},\ and\ \citenamefont {Furlan}}]{Chen:2017hak}%
  \BibitemOpen
  \bibfield  {author} {\bibinfo {author} {\bibfnamefont {C.-Y.}\ \bibnamefont
  {Chen}}, \bibinfo {author} {\bibfnamefont {S.}~\bibnamefont {Dawson}},\ and\
  \bibinfo {author} {\bibfnamefont {E.}~\bibnamefont {Furlan}},\ }\bibfield
  {title} {\bibinfo {title} {{Vectorlike fermions and Higgs effective field
  theory revisited}},\ }\href {https://doi.org/10.1103/PhysRevD.96.015006}
  {\bibfield  {journal} {\bibinfo  {journal} {Phys. Rev. D}\ }\textbf {\bibinfo
  {volume} {96}},\ \bibinfo {pages} {015006} (\bibinfo {year} {2017})},\
  \Eprint {https://arxiv.org/abs/1703.06134} {arXiv:1703.06134 [hep-ph]}
  \BibitemShut {NoStop}%
\bibitem [{\citenamefont {Sirunyan}\ \emph {et~al.}(2018)\citenamefont
  {Sirunyan} \emph {et~al.}}]{CMS:2018oaj}%
  \BibitemOpen
  \bibfield  {author} {\bibinfo {author} {\bibfnamefont {A.~M.}\ \bibnamefont
  {Sirunyan}} \emph {et~al.} (\bibinfo {collaboration} {CMS}),\ }\bibfield
  {title} {\bibinfo {title} {{Search for leptoquarks coupled to
  third-generation quarks in proton-proton collisions at $\sqrt{s}=$ 13 TeV}},\
  }\href {https://doi.org/10.1103/PhysRevLett.121.241802} {\bibfield  {journal}
  {\bibinfo  {journal} {Phys. Rev. Lett.}\ }\textbf {\bibinfo {volume} {121}},\
  \bibinfo {pages} {241802} (\bibinfo {year} {2018})},\ \Eprint
  {https://arxiv.org/abs/1809.05558} {arXiv:1809.05558 [hep-ex]} \BibitemShut
  {NoStop}%
\bibitem [{\citenamefont {Sirunyan}\ \emph {et~al.}(2021)\citenamefont
  {Sirunyan} \emph {et~al.}}]{CMS:2020wzx}%
  \BibitemOpen
  \bibfield  {author} {\bibinfo {author} {\bibfnamefont {A.~M.}\ \bibnamefont
  {Sirunyan}} \emph {et~al.} (\bibinfo {collaboration} {CMS}),\ }\bibfield
  {title} {\bibinfo {title} {{Search for singly and pair-produced leptoquarks
  coupling to third-generation fermions in proton-proton collisions at
  s=13~TeV}},\ }\href {https://doi.org/10.1016/j.physletb.2021.136446}
  {\bibfield  {journal} {\bibinfo  {journal} {Phys. Lett. B}\ }\textbf
  {\bibinfo {volume} {819}},\ \bibinfo {pages} {136446} (\bibinfo {year}
  {2021})},\ \Eprint {https://arxiv.org/abs/2012.04178} {arXiv:2012.04178
  [hep-ex]} \BibitemShut {NoStop}%
\bibitem [{\citenamefont {Aad}\ \emph {et~al.}(2021{\natexlab{a}})\citenamefont
  {Aad} \emph {et~al.}}]{ATLAS:2020xov}%
  \BibitemOpen
  \bibfield  {author} {\bibinfo {author} {\bibfnamefont {G.}~\bibnamefont
  {Aad}} \emph {et~al.} (\bibinfo {collaboration} {ATLAS}),\ }\bibfield
  {title} {\bibinfo {title} {{Search for pair production of scalar leptoquarks
  decaying into first- or second-generation leptons and top quarks in
  proton\textendash{}proton collisions at $\sqrt{s}$ = 13 TeV with the ATLAS
  detector}},\ }\href {https://doi.org/10.1140/epjc/s10052-021-09009-8}
  {\bibfield  {journal} {\bibinfo  {journal} {Eur. Phys. J. C}\ }\textbf
  {\bibinfo {volume} {81}},\ \bibinfo {pages} {313} (\bibinfo {year}
  {2021}{\natexlab{a}})},\ \Eprint {https://arxiv.org/abs/2010.02098}
  {arXiv:2010.02098 [hep-ex]} \BibitemShut {NoStop}%
\bibitem [{\citenamefont {Aad}\ \emph {et~al.}(2021{\natexlab{b}})\citenamefont
  {Aad} \emph {et~al.}}]{ATLAS:2021oiz}%
  \BibitemOpen
  \bibfield  {author} {\bibinfo {author} {\bibfnamefont {G.}~\bibnamefont
  {Aad}} \emph {et~al.} (\bibinfo {collaboration} {ATLAS}),\ }\bibfield
  {title} {\bibinfo {title} {{Search for pair production of third-generation
  scalar leptoquarks decaying into a top quark and a $\tau$-lepton in $pp$
  collisions at $ \sqrt{s} $ = 13 TeV with the ATLAS detector}},\ }\href
  {https://doi.org/10.1007/JHEP06(2021)179} {\bibfield  {journal} {\bibinfo
  {journal} {JHEP}\ }\textbf {\bibinfo {volume} {06}},\ \bibinfo {pages}
  {179}},\ \Eprint {https://arxiv.org/abs/2101.11582} {arXiv:2101.11582
  [hep-ex]} \BibitemShut {NoStop}%
\bibitem [{\citenamefont {Blumlein}\ \emph {et~al.}(1997)\citenamefont
  {Blumlein}, \citenamefont {Boos},\ and\ \citenamefont
  {Kryukov}}]{Blumlein:1996qp}%
  \BibitemOpen
  \bibfield  {author} {\bibinfo {author} {\bibfnamefont {J.}~\bibnamefont
  {Blumlein}}, \bibinfo {author} {\bibfnamefont {E.}~\bibnamefont {Boos}},\
  and\ \bibinfo {author} {\bibfnamefont {A.}~\bibnamefont {Kryukov}},\
  }\bibfield  {title} {\bibinfo {title} {{Leptoquark pair production in
  hadronic interactions}},\ }\href {https://doi.org/10.1007/s002880050538}
  {\bibfield  {journal} {\bibinfo  {journal} {Z. Phys. C}\ }\textbf {\bibinfo
  {volume} {76}},\ \bibinfo {pages} {137} (\bibinfo {year} {1997})},\ \Eprint
  {https://arxiv.org/abs/hep-ph/9610408} {arXiv:hep-ph/9610408} \BibitemShut
  {NoStop}%
\bibitem [{\citenamefont {Kramer}\ \emph {et~al.}(2005)\citenamefont {Kramer},
  \citenamefont {Plehn}, \citenamefont {Spira},\ and\ \citenamefont
  {Zerwas}}]{Kramer:2004df}%
  \BibitemOpen
  \bibfield  {author} {\bibinfo {author} {\bibfnamefont {M.}~\bibnamefont
  {Kramer}}, \bibinfo {author} {\bibfnamefont {T.}~\bibnamefont {Plehn}},
  \bibinfo {author} {\bibfnamefont {M.}~\bibnamefont {Spira}},\ and\ \bibinfo
  {author} {\bibfnamefont {P.~M.}\ \bibnamefont {Zerwas}},\ }\bibfield  {title}
  {\bibinfo {title} {{Pair production of scalar leptoquarks at the CERN LHC}},\
  }\href {https://doi.org/10.1103/PhysRevD.71.057503} {\bibfield  {journal}
  {\bibinfo  {journal} {Phys. Rev. D}\ }\textbf {\bibinfo {volume} {71}},\
  \bibinfo {pages} {057503} (\bibinfo {year} {2005})},\ \Eprint
  {https://arxiv.org/abs/hep-ph/0411038} {arXiv:hep-ph/0411038} \BibitemShut
  {NoStop}%
\bibitem [{\citenamefont {Diaz}\ \emph {et~al.}(2017)\citenamefont {Diaz},
  \citenamefont {Schmaltz},\ and\ \citenamefont {Zhong}}]{Diaz:2017lit}%
  \BibitemOpen
  \bibfield  {author} {\bibinfo {author} {\bibfnamefont {B.}~\bibnamefont
  {Diaz}}, \bibinfo {author} {\bibfnamefont {M.}~\bibnamefont {Schmaltz}},\
  and\ \bibinfo {author} {\bibfnamefont {Y.-M.}\ \bibnamefont {Zhong}},\
  }\bibfield  {title} {\bibinfo {title} {{The leptoquark
  Hunter\textquoteright{}s guide: Pair production}},\ }\href
  {https://doi.org/10.1007/JHEP10(2017)097} {\bibfield  {journal} {\bibinfo
  {journal} {JHEP}\ }\textbf {\bibinfo {volume} {10}},\ \bibinfo {pages}
  {097}},\ \Eprint {https://arxiv.org/abs/1706.05033} {arXiv:1706.05033
  [hep-ph]} \BibitemShut {NoStop}%
\bibitem [{\citenamefont {Dor\v{s}ner}\ and\ \citenamefont
  {Greljo}(2018)}]{Dorsner:2018ynv}%
  \BibitemOpen
  \bibfield  {author} {\bibinfo {author} {\bibfnamefont {I.}~\bibnamefont
  {Dor\v{s}ner}}\ and\ \bibinfo {author} {\bibfnamefont {A.}~\bibnamefont
  {Greljo}},\ }\bibfield  {title} {\bibinfo {title} {{Leptoquark toolbox for
  precision collider studies}},\ }\href
  {https://doi.org/10.1007/JHEP05(2018)126} {\bibfield  {journal} {\bibinfo
  {journal} {JHEP}\ }\textbf {\bibinfo {volume} {05}},\ \bibinfo {pages}
  {126}},\ \Eprint {https://arxiv.org/abs/1801.07641} {arXiv:1801.07641
  [hep-ph]} \BibitemShut {NoStop}%
\bibitem [{\citenamefont {Buonocore}\ \emph {et~al.}(2020)\citenamefont
  {Buonocore}, \citenamefont {Haisch}, \citenamefont {Nason}, \citenamefont
  {Tramontano},\ and\ \citenamefont {Zanderighi}}]{Buonocore:2020erb}%
  \BibitemOpen
  \bibfield  {author} {\bibinfo {author} {\bibfnamefont {L.}~\bibnamefont
  {Buonocore}}, \bibinfo {author} {\bibfnamefont {U.}~\bibnamefont {Haisch}},
  \bibinfo {author} {\bibfnamefont {P.}~\bibnamefont {Nason}}, \bibinfo
  {author} {\bibfnamefont {F.}~\bibnamefont {Tramontano}},\ and\ \bibinfo
  {author} {\bibfnamefont {G.}~\bibnamefont {Zanderighi}},\ }\bibfield  {title}
  {\bibinfo {title} {{Lepton-Quark Collisions at the Large Hadron Collider}},\
  }\href {https://doi.org/10.1103/PhysRevLett.125.231804} {\bibfield  {journal}
  {\bibinfo  {journal} {Phys. Rev. Lett.}\ }\textbf {\bibinfo {volume} {125}},\
  \bibinfo {pages} {231804} (\bibinfo {year} {2020})},\ \Eprint
  {https://arxiv.org/abs/2005.06475} {arXiv:2005.06475 [hep-ph]} \BibitemShut
  {NoStop}%
\bibitem [{\citenamefont {Bansal}\ \emph {et~al.}(2018)\citenamefont {Bansal},
  \citenamefont {Capdevilla}, \citenamefont {Delgado}, \citenamefont {Kolda},
  \citenamefont {Martin},\ and\ \citenamefont {Raj}}]{Bansal:2018eha}%
  \BibitemOpen
  \bibfield  {author} {\bibinfo {author} {\bibfnamefont {S.}~\bibnamefont
  {Bansal}}, \bibinfo {author} {\bibfnamefont {R.~M.}\ \bibnamefont
  {Capdevilla}}, \bibinfo {author} {\bibfnamefont {A.}~\bibnamefont {Delgado}},
  \bibinfo {author} {\bibfnamefont {C.}~\bibnamefont {Kolda}}, \bibinfo
  {author} {\bibfnamefont {A.}~\bibnamefont {Martin}},\ and\ \bibinfo {author}
  {\bibfnamefont {N.}~\bibnamefont {Raj}},\ }\bibfield  {title} {\bibinfo
  {title} {{Hunting leptoquarks in monolepton searches}},\ }\href
  {https://doi.org/10.1103/PhysRevD.98.015037} {\bibfield  {journal} {\bibinfo
  {journal} {Phys. Rev. D}\ }\textbf {\bibinfo {volume} {98}},\ \bibinfo
  {pages} {015037} (\bibinfo {year} {2018})},\ \Eprint
  {https://arxiv.org/abs/1806.02370} {arXiv:1806.02370 [hep-ph]} \BibitemShut
  {NoStop}%
\bibitem [{\citenamefont {Bigaran}\ \emph {et~al.}(2019)\citenamefont
  {Bigaran}, \citenamefont {Gargalionis},\ and\ \citenamefont
  {Volkas}}]{Bigaran:2019bqv}%
  \BibitemOpen
  \bibfield  {author} {\bibinfo {author} {\bibfnamefont {I.}~\bibnamefont
  {Bigaran}}, \bibinfo {author} {\bibfnamefont {J.}~\bibnamefont
  {Gargalionis}},\ and\ \bibinfo {author} {\bibfnamefont {R.~R.}\ \bibnamefont
  {Volkas}},\ }\bibfield  {title} {\bibinfo {title} {{A near-minimal leptoquark
  model for reconciling flavour anomalies and generating radiative neutrino
  masses}},\ }\href {https://doi.org/10.1007/JHEP10(2019)106} {\bibfield
  {journal} {\bibinfo  {journal} {JHEP}\ }\textbf {\bibinfo {volume} {10}},\
  \bibinfo {pages} {106}},\ \Eprint {https://arxiv.org/abs/1906.01870}
  {arXiv:1906.01870 [hep-ph]} \BibitemShut {NoStop}%
\bibitem [{\citenamefont {Sahoo}\ \emph {et~al.}(2021)\citenamefont {Sahoo},
  \citenamefont {Singirala},\ and\ \citenamefont {Mohanta}}]{Sahoo:2021vug}%
  \BibitemOpen
  \bibfield  {author} {\bibinfo {author} {\bibfnamefont {S.}~\bibnamefont
  {Sahoo}}, \bibinfo {author} {\bibfnamefont {S.}~\bibnamefont {Singirala}},\
  and\ \bibinfo {author} {\bibfnamefont {R.}~\bibnamefont {Mohanta}},\
  }\bibfield  {title} {\bibinfo {title} {{Dark matter and flavor anomalies in
  the light of vector-like fermions and scalar leptoquark}},\ }\href@noop {} {\
   (\bibinfo {year} {2021})},\ \Eprint {https://arxiv.org/abs/2112.04382}
  {arXiv:2112.04382 [hep-ph]} \BibitemShut {NoStop}%
\end{thebibliography}%
%%%%%%%%%%%%%%%%%%%%%%%%%%%%%%%%%%%%%%%%%%%%%%%%%%%%%%%%%%%%%%%%%%%%%
\end{document}